\documentclass[prd,twocolumn,aps,amssymb,amsmath,tightenlines,floatfix]{revtex4-1}

\usepackage{graphicx,epsfig}
\usepackage[T1]{fontenc}
\usepackage[latin1]{inputenc}
\usepackage[english]{babel}
\usepackage{amsmath}
\usepackage{amsfonts}
\usepackage{amssymb}
\usepackage{amsthm}
\usepackage{nccmath}
\usepackage{dsfont}
\usepackage{color}
\usepackage{xcolor}
\usepackage{tabularx}
\usepackage{makecell} 
   \newcommand{\PreserveBackslash}[1]{\let\temp=\\#1\let\\=\temp}
   \newcolumntype{C}[1]{>{\PreserveBackslash\centering}p{#1}}
\usepackage{float}
\usepackage{placeins}
\usepackage{afterpage}
\usepackage{slashed}
\usepackage[labelformat=parens]{subfig}
\usepackage{bbm}
\usepackage[font=footnotesize]{caption}

\DeclareCaptionJustification{justified}{\leftskip=0pt \rightskip=0pt \parfillskip=0pt plus 1fil}

\usepackage{float}
\def\cP{\mathcal P}
\def\cT{\mathcal T}
\def\cPT{\mathcal{PT}}
\def\cH{\mathcal{H}}

\begin{document}

\title{Thermodynamic properties of non-Hermitian Nambu--Jona-Lasinio models}
\author{Alexander Felski$^1$}\email{felski@thphys.uni-heidelberg.de}
\author{Alireza Beygi$^2$}\email{alireza.beygi@kgu.de}
\author{S.~P. Klevansky$^1$}\email{spk@physik.uni-heidelberg.de}

\affiliation{
$^1$Institute for Theoretical Physics, Heidelberg University,\\
Philosophenweg 12, 69120 Heidelberg, Germany}

\affiliation{
$^2$Department of Molecular Bioinformatics, Institute of Computer Science, 
Goethe University Frankfurt,
Robert-Mayer-Strasse 11-15, 60325 Frankfurt a. M., Germany}

\begin{abstract}
We investigate the impact of non-Hermiticity on the thermodynamic properties of interacting fermions by examining bilinear extensions to the $3+1$ dimensional $SU(2)$-symmetric  Nambu--Jona-Lasinio (NJL) model of quantum chromodynamics at finite temperature and chemical potential.  
The system is modified through the anti-$\cPT\!$-symmetric pseudoscalar bilinear $\bar{\psi}\gamma_5 \psi$ and the $\cPT\!$-symmetric pseudovector bilinear $iB_\nu \,\bar{\psi}\gamma_5\gamma^\nu \psi$, introduced with a coupling $g$.
Beyond the possibility of dynamical fermion mass generation at finite temperature and chemical potential, our findings establish
model-dependent changes in the position of the chiral phase transition and the critical end-point. These are tunable with respect to $g$ in the former case, and both $g$ and  $|B|/B_0$ in the latter case, for both lightlike and spacelike fields.
Moreover, the behavior of the quark number, entropy, pressure, and energy densities signal a potential fermion or antifermion excess compared to the standard NJL model, due to the pseudoscalar and pseudovector extension respectively.
In both cases regions with negative interaction measure $I = \epsilon-3p$ are found.
Future indications of such behaviors in strongly interacting fermion systems, for example in the context of neutron star physics, may point toward the presence of non-Hermitian contributions.
These trends provide a first indication of curious potential mechanisms for producing non-Hermitian baryon asymmetry.
In addition, the formalism described in this study is expected to apply more generally to other Hamiltonians with four-fermion interactions and thus the effects of the non-Hermitian bilinears are likely to be generic.
\end{abstract}


\maketitle

\section{Introduction}
\label{s1}

The concept of $\cPT$ (parity-time reflection) symmetry has, since its inception by Bender and Boettcher in 1998 \cite{bb98}, become a highly active field of research in both theoretical and experimental physics.
In general, it has overthrown the prevailing principle that physical systems must be governed by a Hermitian Hamiltonian and has demonstrated that 
rich and unexpected features are found in non-Hermitian systems with $\cPT$ symmetry.
In particular, the possible occurrence of exceptional points has illustrated consequences beyond those observed in Hermitian models.
Various experimental realizations, displaying these particular properties of $\cPT\!$-symmetric systems have firmly established $\cPT$ symmetry as an important feature of classical and quantum-mechanical systems \cite{rsm07, gsd09, rme10, zsw10, lre11, fah11, slz11, bdg12, cgb12, zhl13, bbd13, pol14, ayf17, fq20}.

On a fundamental level, however, the development of a quantum-field-theoretical approach is essential. 
In the context of 3+1 dimensional fermionic field theories, the oddness of the  time-reversal operator $\cT$, i.e., $\cT^2=-1$, becomes a core feature when discussing non-Hermitian models, centered around their behavior under combined parity reflection and time reversal \cite{jsm10, jsm14}.
In a recent study \cite{bkb19} we have shown that modifying  free Dirac fermions through the inclusion of non-Hermitian bilinears, $\cPT\!$-symmetric or otherwise, results in a {\it breakdown} of the existence of a real physical fermion mass.
In hindsight, this is due to the odd nature of the fermionic time-reversal operator that also underlies Kramer's degeneracy.
It is not ensured in the relativistic context, that both necessary conditions for a real spectrum, $[H, \cPT] =0 $ and the simultaneity of eigenfunctions to both $H$ and $\cPT\!$, are met.
However, in further studies \cite{fbk20,fk21} we demonstrated that such real-mass solutions can in fact exist, when higher-order interactions are also present in the Hamiltonian.
Then mass can be \emph{generated dynamically} through the inclusion of the non-Hermitian but $\cPT\!$-symmetric pseudovector extension $igB_\nu \,\bar{\psi}\gamma_5\gamma^\nu \psi$.

However, the existence of a real mass solution and even of dynamical mass generation is not restricted to non-Hermitian $\cPT\!$-symmetric modifications:
a pseudoscalar bilinear term $g \bar{\psi} \gamma_5 \psi$, for example, while being neither Hermitian, nor $\cPT\!$-symmetric, still generates real fermion mass dynamically when taken in combination with higher-order interactions \cite{fbk20,fk21}. 
For this reason, we choose here particularly to study these two model interactions,  $g \bar{\psi} \gamma_5 \psi$ and $igB_\nu \,\bar{\psi}\gamma_5\gamma^\nu \psi$, placed in such a context, as a function of finite temperature and density. This is most conveniently done within the 
Nambu--Jona-Lasinio (NJL) model, which provides a fermionic system with a two-body contact interaction \cite{njl61,k92}, whose results may easily be taken over for other similar systems \cite{ams20,ms20}. 
As a commonly used effective field theory of quantum chromodynamics (QCD), that models in particular the spontaneous chiral-symmetry-breaking phase transition at finite temperature and density, the use of the NJL model furthermore ties this study to the development of a general framework of $\cPT\!$-symmetric field theories containing four-point contact interactions and the possibility of non-Hermitian physics beyond the Standard Model, see for example \cite{ams20,ms20,mav22, aem20,  ccr21}.
Here, analyzing the effects of finite temperature and density on {\it non-Hermitian} and in particular $\cPT\!$-symmetric theories marks a crucial step towards developing feasible non-Hermitian approaches and $\cPT$ quantum field theories applicable to experimental realizations, such as heavy-ion collisions and astrophysical models of compact stars. The crucial task of course is to identify characteristics beyond the existence and generation of real effective fermion mass, that may both differentiate between Hermitian and non-Hermitian field theories, and examine whether new features of quantum field theories may arise, when the underlying system is generally non-Hermitian, and specifically when it is $\cPT\!$ symmetric. 

This paper is structured as follows.
In Sec.~\ref{s2} the standard $ SU(2)$ NJL model is reviewed. This discussion serves as the baseline for the examination and the modified approach used in the study of the non-Hermitian extensions of the NJL model.
The gap equation for the effective fermion mass is presented in a self-consistent Hartree approximation and within the Matsubara-formalism for finite temperature $T$ and at finite chemical potential $\mu$, introducing a three-dimensional regulator $\Lambda$. The thermodynamic (grand) potential $\Omega$ is obtained following the coupling-constant integration method. Based on this, the phase diagram of the physical fermion mass is determined and the behavior of the thermodynamic observables -- quark number, pressure, entropy, and energy density, as well 
as the interaction measure -- is established.

Section~\ref{s3} adapts the NJL formalism for the study of the model which is modified through the inclusion of a non-Hermitian non-$\cPT\!$-symmetric bilinear extension based on the pseudoscalar term $\gamma_5$, introduced with the coupling constant $g$. 
The qualitative results obtained at zero temperature and chemical potential are seen to verify the behavior found previously with an Euclidean four-momentum cutoff under similar constraints \cite{fk21}. 
The behavior of the effective fermion mass, in particular the spontaneous chiral-symmetry-breaking phase transition and its critical end-point (CEP), as well as the effect on the thermodynamic observables is analyzed in dependence of the temperature $T$ and chemical potential $\mu$, as well as $T$ and the quark number density $n$ for illustrative values of the coupling $g$.
Despite a dynamical generation of fermion mass within the spontaneously broken approximate chiral symmetry regime, the behavior of the thermodynamic observables is demonstrated to coincide with the standard NJL model behavior at low temperature and small chemical potential. In the vicinity of the phase transition and throughout the restored symmetry region, however, the pseudoscalar extension drives a fermion excess compared to the standard NJL model and exhibits interaction measures $I = \varepsilon-3p <0$.

In Sec.~\ref{s4} the NJL model is extended through the inclusion of the non-Hermitian but $\cPT\!$-symmetric pseudovector bilinear $igB_\nu \,\bar{\psi}\gamma_5\gamma^\nu \psi$. 
The influence of this modification on the effective fermion mass at finite $T$ and $\mu$ is analyzed for a spacelike, a lightlike and a timelike background field $B^\nu$.
We confirm that the results for the spacelike background field obtained in the zero-temperature and vanishing chemical potential limit coincide qualitatively with those previously found
using an Euclidean cutoff, demonstrating the robustness of the regularization procedure in this limit.
The effect on the position of the chiral phase transition, the CEP, and on the behavior of the thermodynamic observables within the $T$-$\mu$--plane is investigated,
finding a notable deviation from the standard NJL model behavior and an emphasis on the antifermionic component of the system. This contrasts with the findings within the pseudoscalar extension.

We conclude and summarize our results in Sec.~\ref{s5}.

\section{The NJL model}
\label{s2}

In the grand canonical ensemble, the two-flavor version of the standard NJL model \cite{njl61} is characterized by the Hamiltonian density
\begin{equation}
\label{njl_hamiltonian}
\cH_{\text{NJL}} - \mu N \!=\! \bar{\psi} (-i \gamma^k \partial_k +m_0 -\mu) \psi
- G [(\bar{\psi} \psi)^2 \!+ (\bar{\psi} i \gamma_5 \vec{\tau} \psi)^2 ]
,
\end{equation}
where $N$ is the quark number density operator, $\mu$ is the baryon chemical potential, $G$ is the two-body coupling strength, and $m_0$ is a bare fermion mass term. $\vec{\tau}$ denotes the isospin $SU(2)$ matrices.
The Dirac matrices $\gamma$ in $3+1$ dimensional spacetime
have the form
\begin{equation}
\label{iic1s0e4}
\gamma^0 \!=\!
\begin{pmatrix}
\mathds{1} & 0 \\
0 & -\mathds{1}
\end{pmatrix} 
, \quad \,\,
\gamma^k \!=\!
\begin{pmatrix}
0 & \sigma^k \\
-\sigma^k & 0
\end{pmatrix} 
, \quad \,\,
\gamma_5 = i \gamma^0 \gamma^1 \gamma^2 \gamma^3 ,
\end{equation}
where $\sigma^k$ with $k\in[1,3]$ are the Pauli matrices.
In the limit of vanishing bare mass $m_0$, this model can be used to study the spontaneous \emph{chiral symmetry}  breaking, which occurs through fermion-antifermion pair production,
parallelling the Bardeen-Cooper-Schrieffer mechanism of superconductivity \cite{bcs57}.
It is therefore a commonly used effective model for the study of QCD in the low-energy regime. 
Due to the non-renormalizability introduced by the contact interaction, a cutoff length of $\Lambda = 653$~MeV and the coupling strength $G \Lambda^2 = 2.14$,
determined traditionally through the quark condensate density per flavor and the pion-decay constant,
are fixed within a three-momentum cutoff regularization scheme in this context, cf.~\cite{k92}.    

Following the Feynman-Dyson perturbation theory, the gap equation for the effective fermion mass $m$ is determined in a self-consistent Hartree approximation to take the well-established form
\begin{equation}
\label{njl_gap}
m_{\text{NJL}} = m_0 - 2 GN_c N_f \!\int^\Lambda \hspace{-0.2cm} \mfrac{\mathrm{d}^3\bf{p}}{(2\pi)^3}\,
T\,\! \sum_n \,\!
\mathrm{e}^{i \omega_n \eta}\,
\text{tr}[S(p_n)]
, 
\end{equation}
where $N_c=3$, $N_f=2$, and tr denotes the spinor trace over the fermion propagator $S(p_n)= (\slashed{p}_n+\mu\gamma^0-m_{\text{NJL}})^{-1}$ with $p_n= (i\omega_n,\bf{p})$ and $\omega_n = (2n+1)\pi T$.
The effects of the finite temperature $T$ are included here through the imaginary-time (or Matsubara) formalism, cf.~\cite{m55,fw03}; the parameter $\eta$ denotes an infinitesimally small positive imaginary-time difference, kept for definiteness and ultimately taken to vanish.
Upon evaluation of the Matsubara-frequency summation and in the chiral limit $m_0 \rightarrow0$, the gap equation (\ref{njl_gap}) becomes
\begin{eqnarray}
\label{njl_gap_2}
m_{\text{NJL}} &=&\,  2 GN_c N_f\,  m_{\text{NJL}}\nonumber \\
& &\,\,\times \int^\Lambda \hspace{-0.2cm} \mfrac{\mathrm{d}^3\bf{p}}{(2\pi)^3}\,
\frac{1}{E}\,  
\Bigl[\tanh\Bigl(\mfrac{E+\mu}{2T}\Bigr) + \tanh\Bigl(\mfrac{E-\mu}{2T}\Bigr)
\Bigr]
,\nonumber \\
\end{eqnarray}
where $E^2 = {\bf p} ^2 +m_{\text{NJL}}^2$, cf.~\cite{k92}.

\begin{figure}[!t]
\includegraphics[width=0.45\textwidth]
{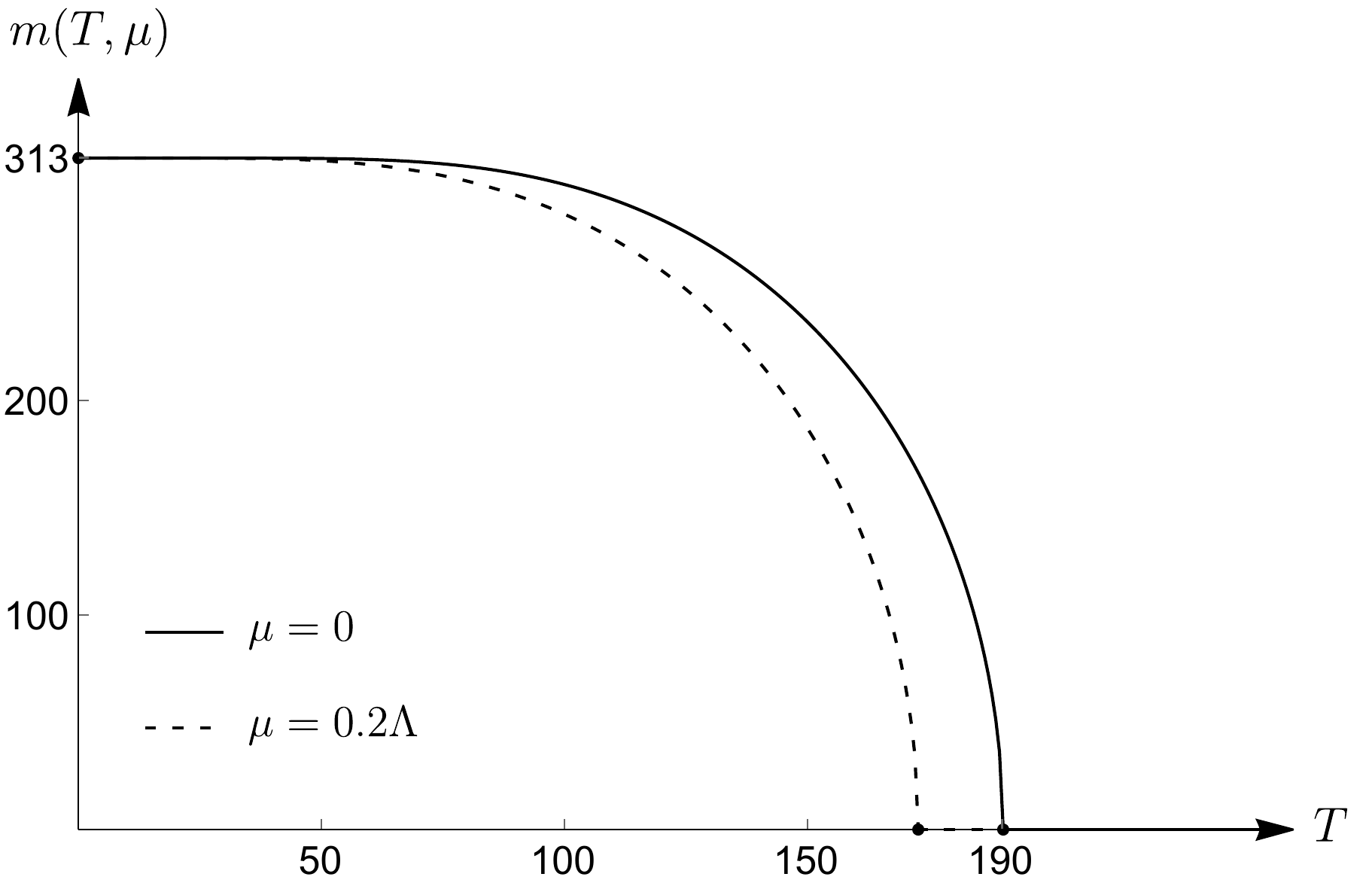}
\caption{
\label{f1a}
Behavior of the effective fermion mass $m$ within the NJL model in MeV at 
the chemical potential $\mu=0$ and $\mu=0.2 \Lambda$ as a function of the temperature $T$ in MeV.
}
\end{figure}

When evaluating the self-consistent gap equation at vanishing chemical potential $\mu$, one obtains a finite effective fermion mass solution of $m_{\text{NJL}}(0,0)\approx 313$~MeV at $T\!=\!\mu\!=\!0$,
which decreases monotonically as a function of increasing temperature $T$, until a second-order phase transition is reached at a critical value $T^c(\mu\!=\!0)$ $ \approx 190$~MeV. 
At higher temperatures the initial spontaneously broken chiral symmetry of the system is restored, and the effective fermion mass $m_{\text{NJL}}$ vanishes.
This behavior is shown in Fig.~\ref{f1a}.
A qualitatively similar second-order phase transition is found for small finite chemical potential $\mu$, differing in a decrease of the critical temperature $T^c(\mu)$ and of the mass $m_{\text{NJL}}(T,\mu)$ in the spontaneously broken chiral-symmetry phase. 

When evaluating the gap equation (\ref{njl_gap_2}) at vanishing (or small) temperature $T$ as a function of the chemical potential $\mu$, however, a parametric region is reached in which the gap equation admits multiple real mass solutions. 
The stable physical mass solution in this region can then be determined as the global minimum of the thermodynamic potential 
\begin{equation}
\label{thermodyn_pot}
\Omega_{\text{NJL}}(T,\mu) \!=\! -T \ln[\, \mathcal{Z} ]
\!=\! -T \ln\bigl( \text{tr}\bigr[\mathrm{e}^{-(\mathcal{H}_{\text{NJL}}-\mu N)/T} \bigl]\! \bigr)
\end{equation}
under variation of $m$, where $\mathcal{Z}$ denotes the (grand canonical) partition function.
Using the coupling-constant integration method, see, e.g., \cite{k92,fw03}, $\Omega_{\text{NJL}}$ can be determined as follows:
By considering the Hamiltonian density $\mathcal{H}_\lambda = \mathcal{H}_0+\lambda \mathcal{H}_\text{int}$, with $\mathcal{H}_\text{int}$ denoting the two-body contact interaction term in (\ref{njl_hamiltonian}), eq.~(\ref{thermodyn_pot}) implies that 
\begin{equation}
\frac{\mathrm{d} \Omega_\lambda}{\mathrm{d}\lambda}
=
\frac{1}{\lambda\, \mathcal{Z}_\lambda}\, \text{tr}( \lambda \mathcal{H}_\text{int} \, \mathcal{Z}_\lambda )
= 
\mfrac{1}{\lambda}  \langle\,\mathcal{H}_\text{int}\,\rangle
.
\end{equation}
Accordingly, the thermodynamic potential $\Omega_{\text{NJL}}$ of the system (\ref{njl_hamiltonian}), associated with $\lambda=1$, can be determined from the thermodynamic average of the interaction energy to be
\begin{equation}
\label{coupling_integration_method}
\Omega_{\text{NJL}}-\Omega_0=
\mfrac{1}{4G} \Bigl[
(m_{\text{NJL}}-m_0)^2 - 2\!\int_0^1 \hspace{-0.1cm} \mfrac{\mathrm{d}\lambda}{\lambda} \,(m_\lambda-m_0) \mfrac{\mathrm{d}m_\lambda}{\mathrm{d}\lambda}
\Bigr]
,
\end{equation}
cf.~\cite{k92}.
By substituting the $\lambda$-dependent equivalent of the gap equation (\ref{njl_gap}) for $m_\lambda-m_0$, the coupling-constant integration thus results in the expression 

\begin{figure}[]
\centering
\begin{minipage}[t]{0.45\textwidth}
\centering
\includegraphics[width=1\textwidth]
{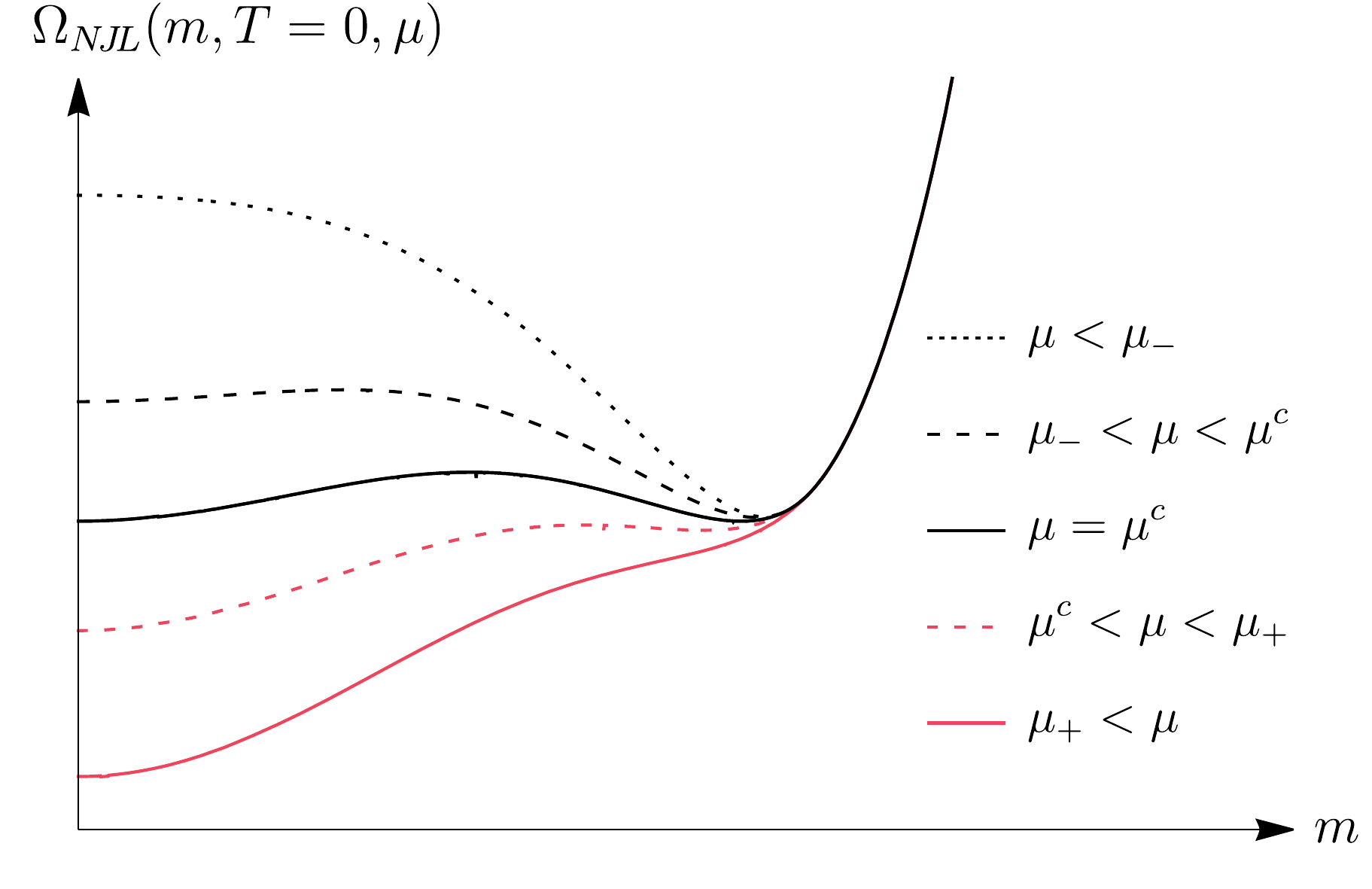}
\caption{
\label{f1b}
Qualitative behavior of the thermodynamic potential $\Omega_{\text{NJL}}$ at vanishing temperature $T$ as a function of the effective mass $m$ for various chemical potentials $\mu$.
The different cases illustrate the possible existence of extrema at vanishing and finite mass.
}
\end{minipage}
\hfill
\begin{minipage}[t]{0.45\textwidth}
\centering
\includegraphics[width=1\textwidth]
{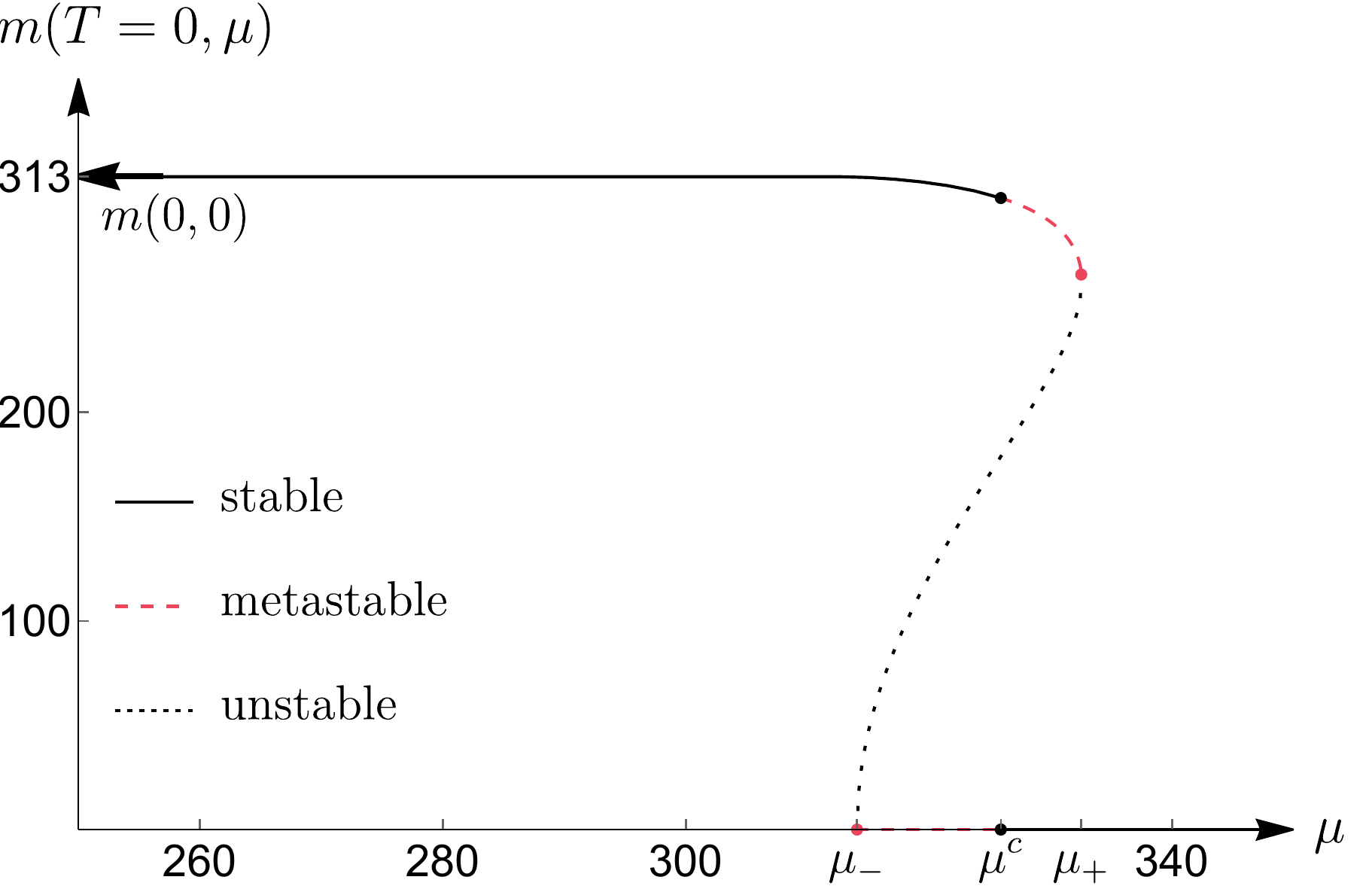}
\caption{
\label{f1c}
Behavior of the effective fermion mass $m$ within the NJL model in MeV at 
vanishing temperature $T$ as a function of the chemical potential $\mu$  in MeV. The stable physical mass solution associated with the global minimum of 
$\Omega_{\text{NJL}}$ is shown as a solid black line, undergoing a first-order phase transition at $\mu^c$.
}
\end{minipage}
\end{figure}

\begin{eqnarray}
\label{njl_potential_coupling_integrated}
\Omega_{\text{NJL}}&-&\Omega_0=
\frac{(m_{\text{NJL}}-m_0)^2}{4G} \nonumber \\
&-&2\,T\,  N_c N_f
\int^\Lambda \hspace{-0.1cm} \mfrac{\mathrm{d}^3{\bf p}}{(2\pi)^3}\,
\ln \Bigg[\,
\frac{\cosh(\frac{E+\mu}{2T}) \cosh(\frac{E-\mu}{2T})}{
\cosh(\frac{E_0+\mu}{2T}) \cosh(\frac{E_0-\mu}{2T})}
\,\Bigg]
,\nonumber\\
\end{eqnarray}
with $E_0^2 = {\bf p} ^2 +m_{0}^2$.
Subtracting the contribution of the denominator in the logarithm, which is associated with the thermodynamic potential $\Omega_0$ of the free theory obtained at $\lambda=0$, one thus finds 
\begin{align}
\label{njl_potential}
\begin{split}
&\Omega_{\text{NJL}}(T,\mu)=\,
\mfrac{(m_{\text{NJL}}-m_0)^2}{4G} 
-2 N_c N_f
\int^\Lambda \hspace{-0.2cm} \mfrac{\mathrm{d}^3{\bf p}}{(2\pi)^3}\,
E
\\
&-2\,T\, N_c N_f
\int^\Lambda \hspace{-0.2cm} \mfrac{\mathrm{d}^3 {\bf p}}{(2\pi)^3}\,
\ln\Bigl(
\bigl[1+\mathrm{e}^{-(E+\mu)/T}\,\bigr]\,
\bigl[1+\mathrm{e}^{-(E-\mu)/T}\, \bigr]
\Bigr)
.
\end{split}
\end{align}
The gap equation (\ref{njl_gap_2}) can be regained from the extremal condition $\mathrm{d}\Omega/\mathrm{d}m = 0$ in the chiral limit.

In Fig.~\ref{f1b} the behavior of the thermodynamic potential is visualized for vanishing temperature $T$ and various chemical potentials $\mu$ as a function of the effective mass $m$. 
For small chemical potentials $\mu < \mu_- \approx 314$~MeV (dotted black line), the only minimum lies at a finite value of the fermion mass, which identifies the physical solution in this region of spontaneously broken chiral symmetry. 
For $\mu_-<\mu<\mu_+\approx 333$~MeV the thermodynamic potential admits a second minimum at vanishing mass, which for $\mu_-<\mu<\mu^c\approx 326$~MeV (dashed black line) is only a local, not a global minimum of $\Omega_{\text{NJL}}(T=0,\mu)$. Therefore, the vanishing mass solution is a metastable state in this region. 
At $\mu^c$ (solid black line) both minima of $\Omega_{\text{NJL}}$ lie at the same height.
The abrupt transition from the finite fermion mass to the vanishing mass solution at $\mu^c$ thus marks a chiral-symmetry-breaking first-order phase transition.
For $\mu^c<\mu<\mu_+$ (dashed red line), the global minimum characterizing the physical solution lies at vanishing mass, while the finite-mass solution describes a local minimum associated with a metastable solution only. 
When $\mu>\mu_+$ (solid red line), the only minimum lies at vanishing mass. No additional solutions exist.
Moreover, a local maximum of $\Omega_{\text{NJL}}(T=0,\mu)$ can be found for $\mu< \mu_-$ at $m=0$ (see dotted black curve) and for $\mu_-<\mu<\mu_+$ in between the two existing minima (see dashed curves and solid black curve). While this maximum denotes a possible mass solution of the gap equation, such a solution is an unstable state.

The behavior of the stable physical fermion mass, as well as the metastable and unstable mass solutions of the gap equation, are visualized as a function of the chemical potential $\mu$ in Fig.~\ref{f1c} as solid, dashed, and dotted lines respectively.
A qualitatively similar first-order phase transition behavior is found at small finite temperatures $T$, causing a small decrease in the critical chemical potential $\mu^c(T)$ and the mass $m_{\text{NJL}}(T,\mu)$ in the spontaneously broken chiral-symmetry phase.

\begin{figure}[]
\centering
\includegraphics[width=0.45\textwidth]{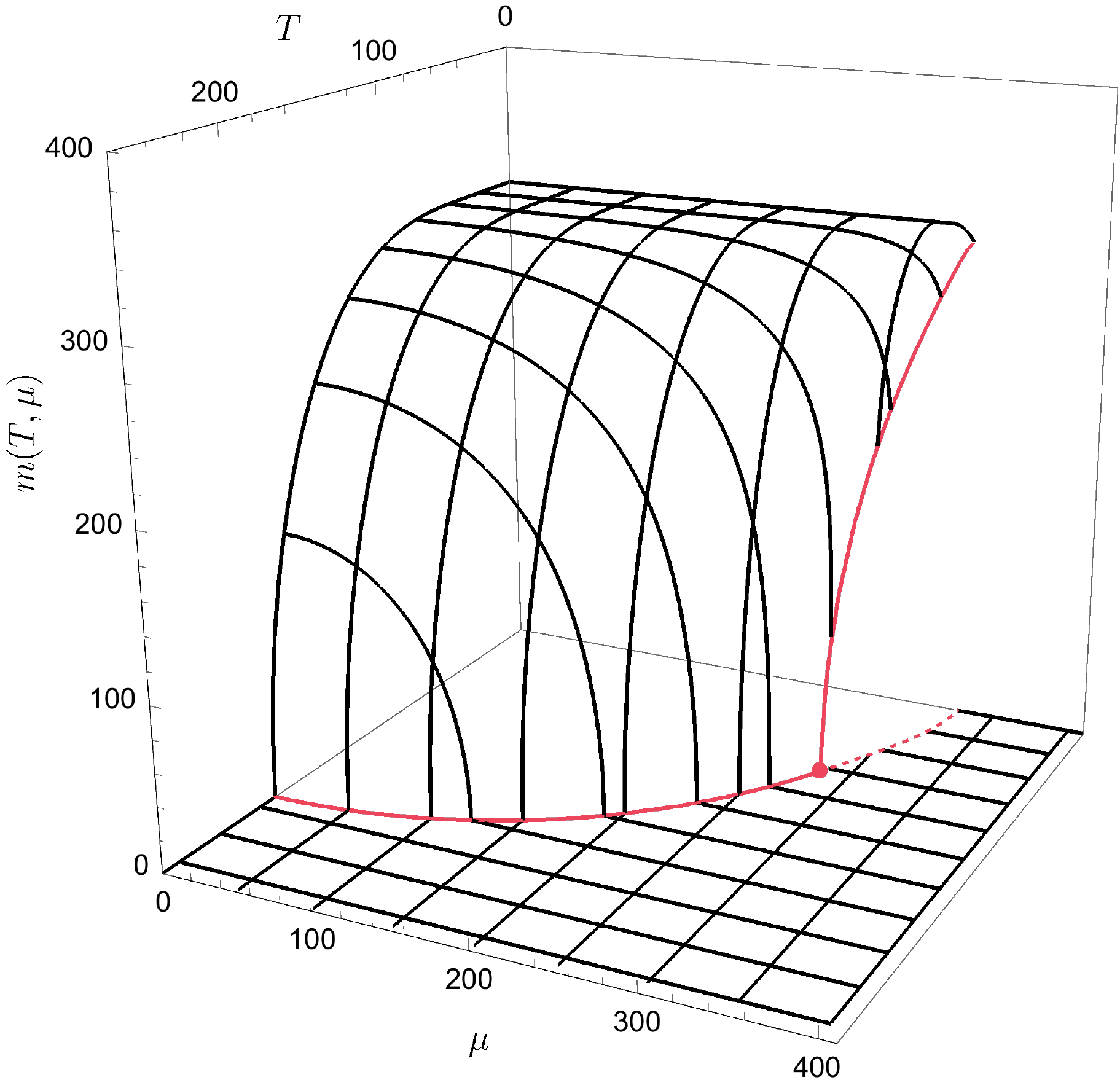}
\caption{ 
\label{f2}
Effective fermion mass $m_{\text{NJL}}$ as a function of the temperature $T$ and the chemical potential $\mu$ in MeV. The chiral phase transition is denoted in red, with a red dot indicating the CEP.
At low temperatures the mass undergoes a discontinuous first-order chiral phase transition, while the transition is of second order at small chemical potentials.
}
\vspace*{-0.3cm}
\end{figure}

In Fig.~\ref{f2} the behavior of the physical mass $m_{\text{NJL}}$ is shown as a function of both the temperature $T$ and the chemical potential $\mu$, with the chiral phase transition being denoted in red. The red dot marks the critical end-point (CEP) at which the first-order transition behavior, found at small temperatures, changes to a second-order transition, found at small chemical potentials. It lies at $T_{\text{CEP}}\approx 79$~MeV and $\mu_{\text{CEP}}\approx 281$~MeV (with a ratio of $(\mu/T)_{\text{CEP}} \approx 3.56$). 

Another instructive representation of the chiral phase transition that is commonly used in the context of heavy-ion collisions and astrophysical models of compact stars, considers the behavior of the effective fermion mass as a function of the quark number density $n(T,\mu)$ instead of the chemical potential $\mu$, which is determined through the thermodynamic potential as
\begin{eqnarray}
\label{njl_quarknumber}
 n_{\text{NJL}}(T,\mu) &=& -\frac{\partial \,\Omega_{\text{NJL}}(T,\mu)}{\partial \mu} \,\Bigr\rvert_T \nonumber \\
&=&
N_c N_f \!\!\int^\Lambda \hspace{-0.25cm} \mfrac{\mathrm{d}^3{\bf p}}{(2\pi)^3}\,
\!\!\Bigl[\tanh\Bigl(\mfrac{E+\mu}{2T}\Bigr)\! - \!\tanh\Bigl(\mfrac{E-\mu}{2T}\Bigr)
\Bigr]. \nonumber \\
\end{eqnarray}

\begin{figure}
\subfloat[]{
\includegraphics[width=0.45\textwidth]
{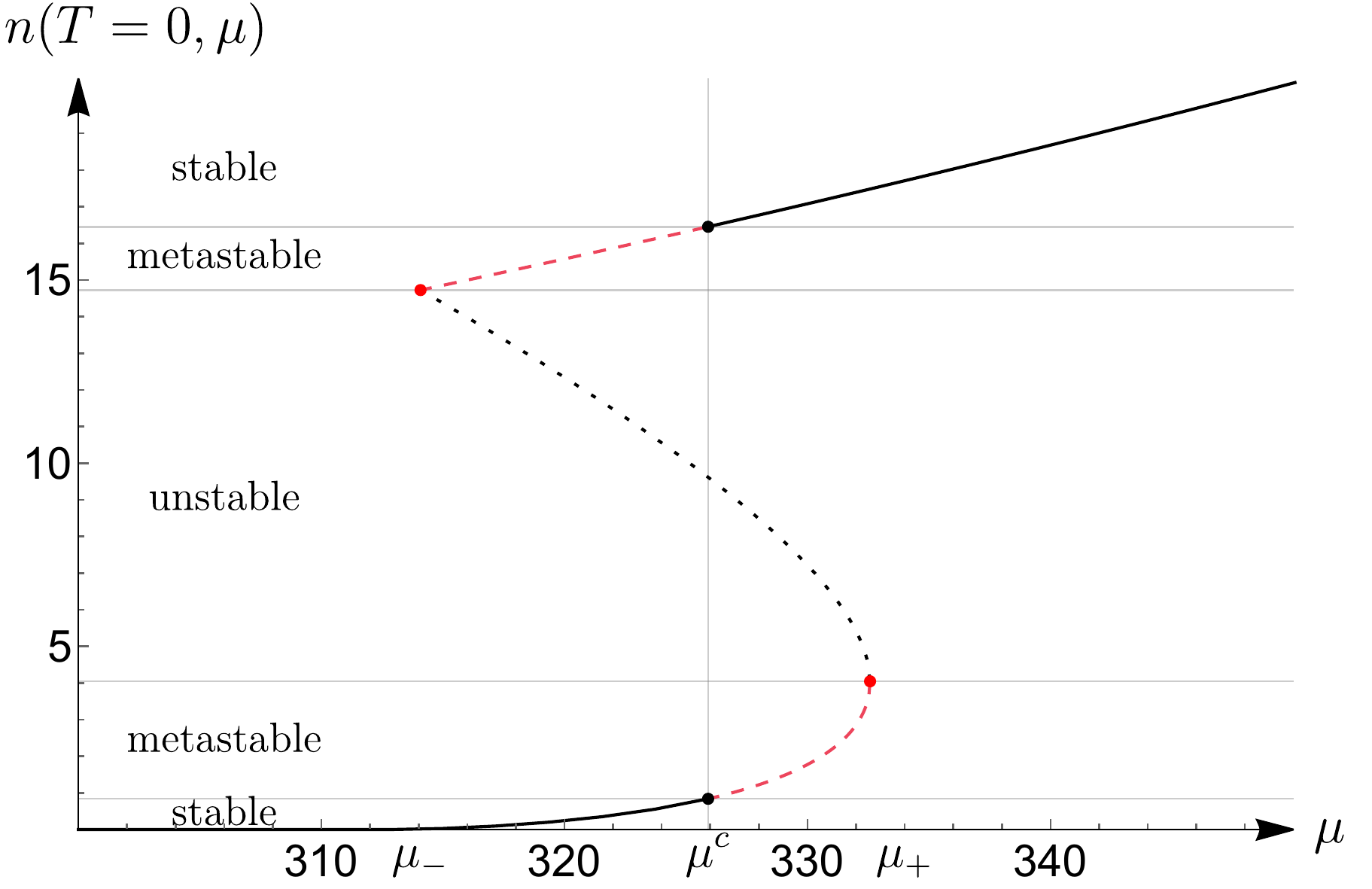}
\label{f3a}
}
\hfill
\subfloat[]{
\includegraphics[width=0.45\textwidth]
{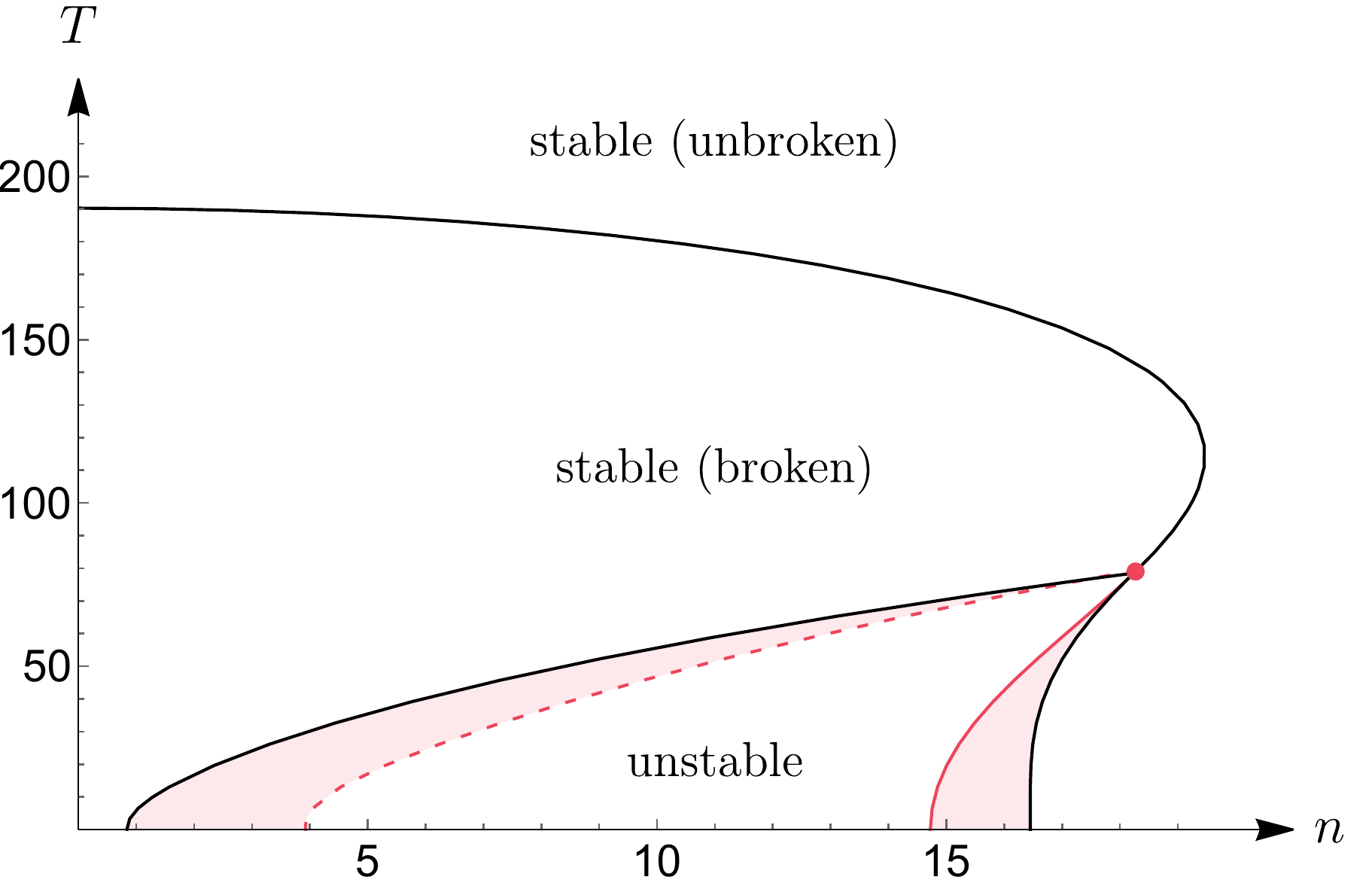}
\label{f3b}
 }
\caption{ 
(a) Quark number density at vanishing temperature as a function of the chemical potential $\mu$ in MeV, illustrating that, while the stable, metastable, and unstable mass solutions of the gap equation are a function of the chemical potential with multiple branches, they form a single-valued function of the quark number density. 
(b) Chiral phase diagram of the NJL model in the temperature-quark number density plane. The phase transition of the stable physical fermion mass is denoted as solid black line. Red lines denote the spinoidals associated with $\mu_+$ (dashed) and $\mu_-$ (solid) marking the transition from the metastable (shaded) to the unstable mass regions.
}
\end{figure}

It is the physical quantity that enters into the equations of state and which is accessible experimentally.
Such a representation also has the advantage that the stable, metastable and unstable solutions of the fermion mass found in the region with a first-order phase transition arise as separate regions of a single-valued function in the quark number density, while these solutions form overlapping branches in the chemical potential between $\mu_-$ and $\mu_+$. 
This is illustrated for the case of vanishing temperature in Fig.~\ref{f3a}, where $n_{\text{NJL}}(T=0,\mu)$ is shown as a function of the chemical potential. 
The solutions for stable, metastable, and unstable fermion masses are indicated as solid, dashed, and dotted lines respectively and the corresponding regions of the quark number density $n_{\text{NJL}}$ are indicated.
The chiral phase diagram in the $T$-$n$--plane is shown in Fig.~\ref{f3b}.
The phase transition is denoted as a solid black line. Red lines denote the spinoidals associated with $\mu_\pm$, that bind the regions of metastable solutions (shaded in red) and mark the transition to the region of only unstable results.

Beyond the identification of the physical effective fermion mass, the thermodynamic potential in (\ref{njl_potential}) allows for the study of various thermodynamic observables.
In addition to the quark number density $n_{\text{NJL}}$, the entropy density $s_{\text{NJL}}$ is determined through $\Omega_{\text{NJL}}(T,\mu)$ to be
\begin{widetext}
\begin{eqnarray}
\label{njl_entropy}
s_{\text{NJL}}(T,\mu) & =&\,  -\frac{\partial \,\Omega_{\text{NJL}}(T,\mu)}{\partial T} \,\Bigr\rvert_\mu
=\,
2 N_c N_f \int^\Lambda \hspace{-0.1cm} \mfrac{\mathrm{d}^3{\bf p}}{(2\pi)^3}\,
\Bigl\{
\ln\Bigl(
\bigl[1+\mathrm{e}^{-(E+\mu)/T}\,\bigr]\,
\bigl[1+\mathrm{e}^{-(E-\mu)/T}\, \bigr]
\Bigr)
+\mfrac{E}{T}\nonumber \\
&& \hspace*{6.0cm}
-\mfrac{E+\mu}{2T}
\tanh\Bigl(\mfrac{E+\mu}{2T}\Bigr) 
- \mfrac{E-\mu}{2T}
\tanh\Bigl(\mfrac{E-\mu}{2T}\Bigr)
\Bigr\}
.
\end{eqnarray}
\end{widetext}

The pressure density $p_{\text{NJL}}(T,\mu)$ corresponds to the thermodynamic potential (\ref{njl_potential}) up to an overall sign and relative to the physical vacuum 
at vanishing temperature and chemical potential,   
\begin{equation}
\label{njl_pressure}
p_{\text{NJL}}(T,\mu) = - \bigl[\, \Omega_{\text{NJL}}(T,\mu) - \Omega_{\text{NJL}}(0,0) \,\bigr]
.
\end{equation}
The energy density $\epsilon_{\text{NJL}}(T,\mu)$ and the interaction measure (or trace anomaly of the energy-momentum tensor) $I_{\text{NJL}}(T,\mu)$, quantifying the deviation from an ideal-gas behavior, have the respective forms
\begin{align}
\label{njl_energy}
\begin{split}
& \epsilon_{\text{NJL}}(T,\mu) =-p_{\text{NJL}} + T \,s_{\text{NJL}} +\mu\, n_{\text{NJL}}
,
\end{split} \\[5pt]
\label{njl_anomaly}
\begin{split}
& I_{\text{NJL}}(T,\mu) =\epsilon_{\text{NJL}}-3\,p_{\text{NJL}}
.
\end{split}
\end{align}

To study the behavior of these observables 
(\ref{njl_quarknumber}) -- (\ref{njl_anomaly}), their analysis as a function of the temperature $T$ along lines of constant $\mu/T$ in the $T$-$\mu$--plane is instructive. 
Furthermore, the influence of the three-momentum cutoff scale $\Lambda$ can be examined by considering the behavior of the thermodynamic observables along these lines in the limit $\Lambda \rightarrow \infty$.  
Note that, while the quark number density (\ref{njl_quarknumber}) and entropy density (\ref{njl_entropy}) remain convergent in this limit, the thermodynamic potential (\ref{njl_potential}), and consequently the pressure density (\ref{njl_pressure}), energy density (\ref{njl_energy}), and interaction measure (\ref{njl_anomaly}), show an ultra-violet divergence.
In the form (\ref{njl_potential}), however, one finds this divergence contained in the last term contributing to $\Omega_{\text{NJL}}(T,\mu)$, while the three-momentum integration of the logarithmic term remains finite in the large cutoff limit.
Since the divergent term is, in particular, independent of the temperature $T$ and the chemical potential $\mu$, it is sufficient to remove the cutoff of the finite logarithmic term in (\ref{njl_potential}) to study the behavior at high temperatures. 

Notably the algebraic behavior of $\Omega_{\text{NJL}}$ in the large temperature limit for fixed $\mu/T$ and when removing the cutoff scale $\Lambda \rightarrow \infty$ is found to be
\begin{equation}
\label{njl_SBlimit_potential}
\Omega_{\text{NJL}}(T,\mu) \sim
-T^4 \, N_c N_f\,
\Bigl[\,
\mfrac{7\pi^2}{180}
+\mfrac{1}{6}\, \bigl(\mfrac{\mu}{T}\bigr)^2
+ \mfrac{1}{12 \pi^2}\, \bigl(\mfrac{\mu}{T}\bigr)^4 \,
\Bigr],
\end{equation}
and coincides with the Stefan-Boltzmann (SB) behavior of an ideal massless fermion gas, which is expected to dominate the behavior of the NJL model in the chirally symmetric region.
Together with the corresponding expansions of (\ref{njl_quarknumber}) and (\ref{njl_entropy}),
\begin{align}
\label{njl_SBlimit_number}
n_{\text{NJL}}(T,\mu)\sim&
\,\, T^3 \, N_c N_f\,
\bigl[\,
\mfrac{1}{3}\, \bigl(\mfrac{\mu}{T}\bigr)
+ \mfrac{1}{3 \pi^2}\, \bigl(\mfrac{\mu}{T}\bigr)^3 \,
\bigr]
,
\\[5pt]
\label{njl_SBlimit_entropy}
s_{\text{NJL}}(T,\mu)\sim&
\,\, T^3 \, N_c N_f\,
\bigl[\,
\mfrac{7\pi^2}{45}
+\mfrac{1}{3}\, \bigl(\mfrac{\mu}{T}\bigr)^2 \,
\bigr]
,
\end{align}
it is thus sensible to present the scaled quantities
$n/T^3$, $s/T^3$, $p/T^4$, $\epsilon/T^4$, and $I/T^4$ when considering the behavior of these observables as functions of $T$ for fixed values $\mu/T$,
because they approach finite limits at large temperatures.
\begin{figure*}
\centering
\subfloat[]{
\includegraphics[width=0.45\textwidth]
{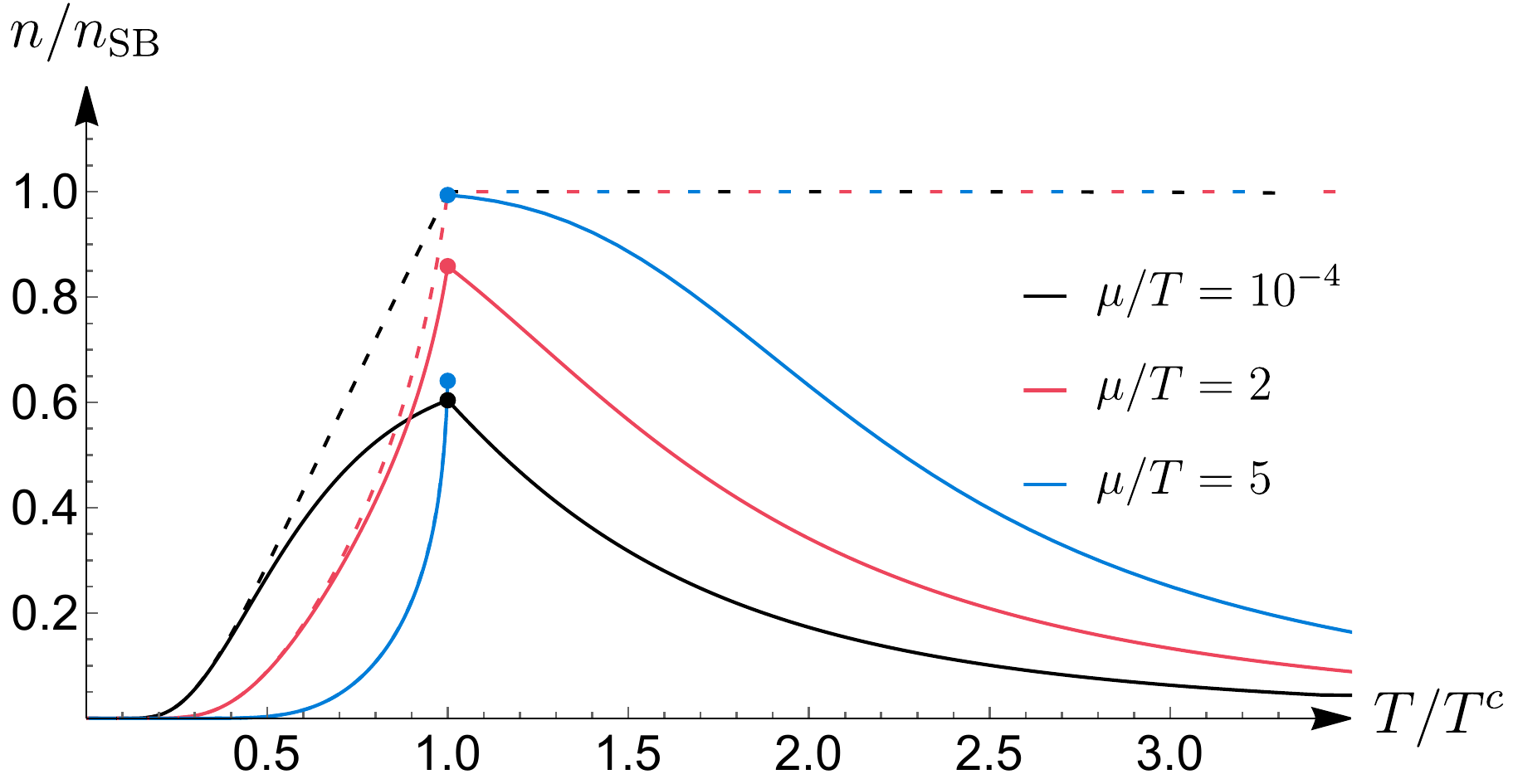}
\label{f4a}
}
\subfloat[]{
\includegraphics[width=0.45\textwidth]
{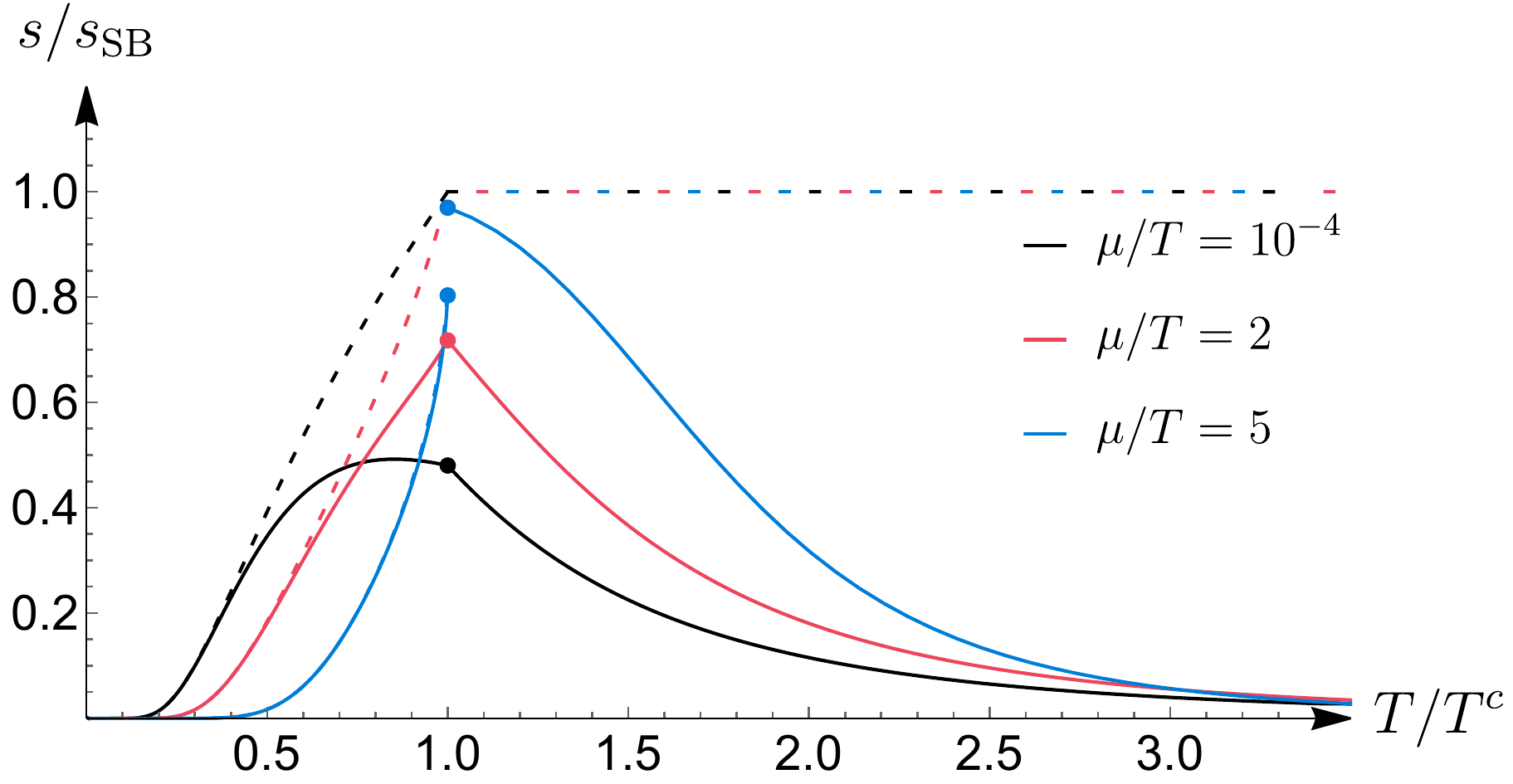}
\label{f4b}
}
\hfill\\
\subfloat[]{
\includegraphics[width=0.45\textwidth]
{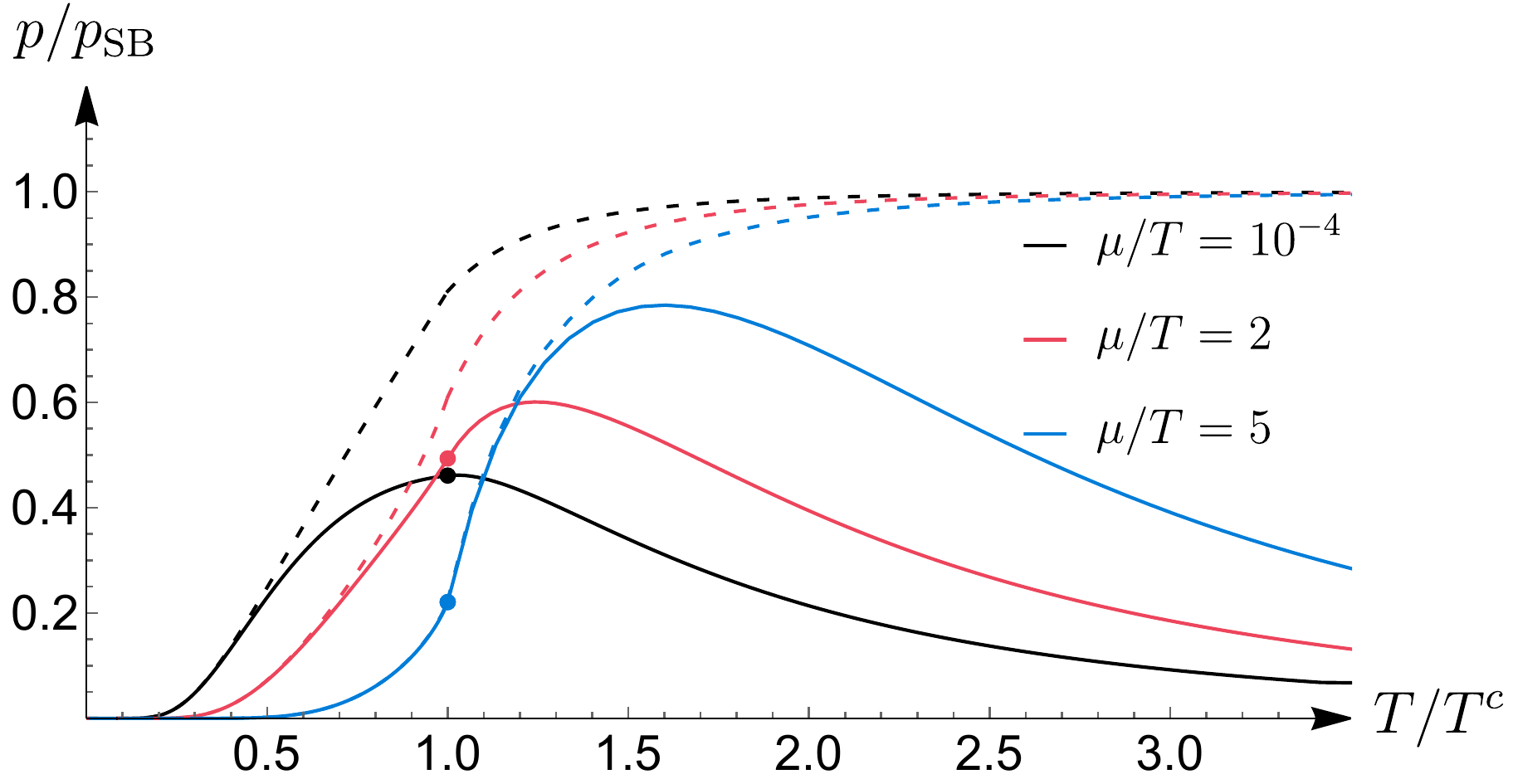}
\label{f4c}
}
\subfloat[]{
\includegraphics[width=0.45\textwidth]
{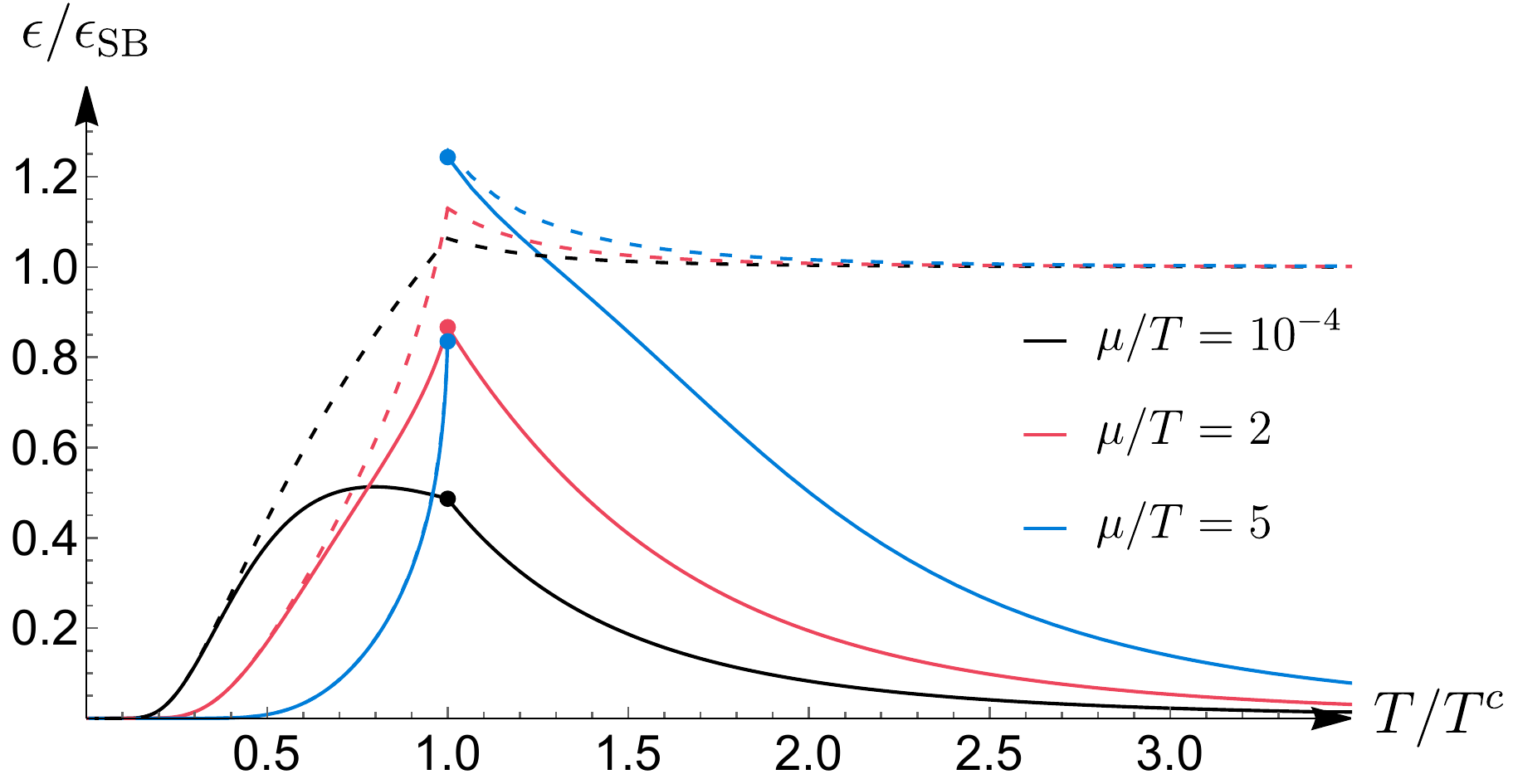}
\label{f4d}
}
\hfill\\
\subfloat[]{
\includegraphics[width=0.45\textwidth]
{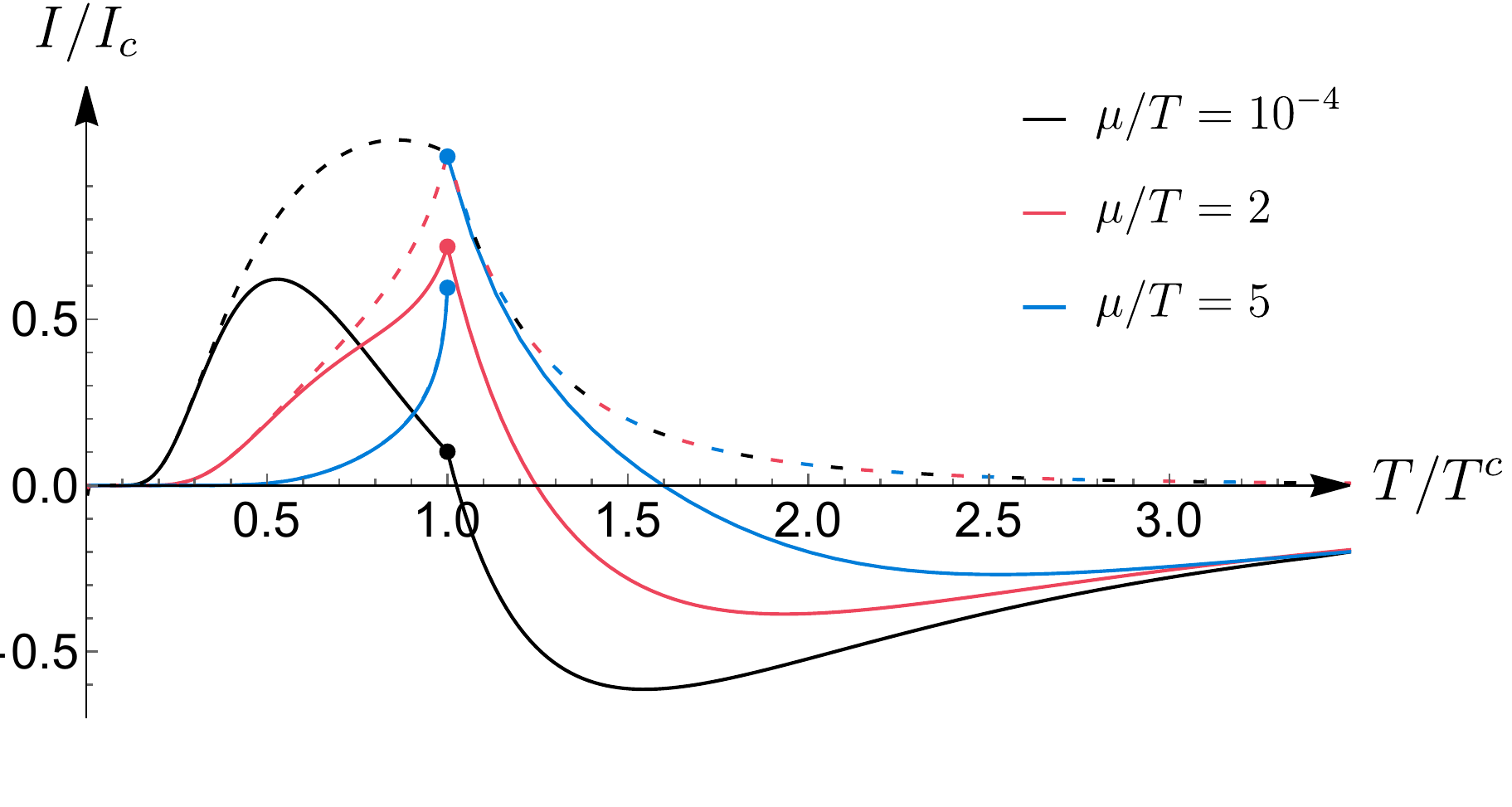}
\label{f4e}
}
\caption{
Thermodynamic observables as a function of the scaled temperature $T/T^c$ with three-momentum cutoff $\Lambda$ (solid) and in the limit $\Lambda \rightarrow \infty$ (dashed). 
}
\end{figure*}
To compare the behavior of the thermodynamic observables along different lines of fixed values $\mu/T$, they are here furthermore normalized to their respective SB limit values, as determined by (\ref{njl_SBlimit_potential}) -- (\ref{njl_SBlimit_entropy}).
An exception is the asymptotically vanishing interaction measure, which is scaled to its value at the phase transition in the $\Lambda\rightarrow\infty$ case instead. 
Their behaviors are shown in Figs.~\ref{f4a} -- \ref{f4e}.
In each case, the solid lines denote the behavior with fixed cutoff length $\Lambda$ as a function of the scaled temperature $T/T^c$ 
for $\mu/T=2$ (red), i.e., in the second-order phase-transition region, and for $\mu/T=5$ (blue), i.e., in the first-order transition region. 
For comparison with the frequently presented case along the temperature axis, the behavior for close-to vanishing chemical potential, $\mu/T=10^{-4}$, is included in black. 
This value of $\mu/T$ is not taken to vanish exactly in order to present a nontrivial reference for the quark number density $n$ as well, whose SB limit otherwise vanishes as $\mu \rightarrow 0$, see (\ref{njl_SBlimit_number}).
The respective normalizations, as well as the critical-temperature values $T^c$ of the phase transition, are listed in Table~\ref{t1}.

\begin{table*}[]
\centering
\setlength\tabcolsep{7.8pt}
\renewcommand{\arraystretch}{1.4}
\small   
\begin{tabular}{ |c||c|c|c|c|c|c| }  
 \cline{2-7}
 \multicolumn{1}{c||}{} & $T^c_\text{NJL}$ & $(n/T^3)_{\text{SB}}$ & $(s/T^3)_{\text{SB}}$  & $(p/T^4)_{\text{SB}}$  & $(\epsilon/T^4)_{\text{SB}}$ & $(I/T^4)_{c}$  \\ 
 \hline
 \hline
 $\mu/T= 10^{-4}$ & $190$~MeV & $ 0.0002$ & $9.2116$ & $2.3029$ & $6.9087$ & $1.7399 $ \\
 \hline 
 $\mu/T=2$ & $120$~MeV & $5.6211$ & $17.2116$ & $7.1135$ & $21.3404$ & $11.0912$\\
 \hline 
 $\mu/T=5$ & $59$~MeV & $35.3303$ & $59.2116$ & $58.9658$ & $176.8973$ & $182.9054$\\ 
 \hline
\end{tabular}
\caption{
Chiral phase transition temperature, Stefan-Boltzmann limits of the quark number, entropy, pressure, and energy densities, as well as the interaction measure value at the phase transition (for $\Lambda\rightarrow \infty$) along lines of constant $\mu/T$ in the $T$-$\mu$-plane for $\mu/T = 10^{-4}$ (close to the $T$ axis), $\mu/T = 2$ (second-order transition region), and $\mu/T = 5$ (first-order transition region).
}
\label{t1}
\end{table*}

In the region of restored chiral symmetry, that is at large temperatures $T> T^c$, the effective fermion mass vanishes and thus the free quarks are expected to dominate the physical behavior of the system.
Accordingly, the quark number density, as well as the pressure, entropy, and energy densities, should saturate to the behavior of an ideal gas of massless fermions. 
Instead, Figs.~\ref{f4a} -- \ref{f4d} show an asymptotic decay of these quantities at large temperatures beyond the chiral phase transition. 
The comparison to the $\Lambda\rightarrow \infty$ behavior illustrates, however, that this decay is the consequence of the momentum cutoff within the model. 
For definiteness, regard Fig.~\ref{f4c} for the pressure, which is continuous in all cases. 
While the NJL($\Lambda$) curves (solid lines) undershoot the SB limit and in fact go to zero, the NJL($\infty$) curves  (dashed lines), found by removing the cutoff, approach the SB limit quite rapidly.
The ideal gas behavior is not only reached in the asymptotic large-temperature limit, as previously indicated by the expansions (\ref{njl_SBlimit_potential})-(\ref{njl_SBlimit_entropy}), but one finds rather, that the thermodynamic observables plateau rapidly after crossing the phase transition at $T^c$ - in agreement with expectation.
Notably, the quark number density, Fig.~\ref{f4a}, and entropy density, Fig.~\ref{f4b}, reach their respective SB limits essentially upon undergoing the phase transition, while $p$, $\epsilon$, and $I$, Figs.~\ref{f4c} -- \ref{f4e}, approach the limits rapidly, but not at the phase transition point.
So while the effective mass solution $m$ vanishes immediately, resulting in the presence of a massless fermion gas, the free ideal-gas behavior is obtained only at some distance from $T^c$.
The effective degrees of freedom remain restricted around the chiral phase transition, which is reflected in the reduced pressure and higher energy densities as the fermions are still correlated, and
the interaction measure rapidly declining rather than immediately vanishing due to the absence of confinement.

At small temperatures $T<T^c$, the chiral symmetry is spontaneously broken through the formation of fermion-antifermion pairs and the fermions gain a finite effective mass. The behavior of the system is thus expected to be governed by the massless pionic Nambu-Goldstone mode.
Nevertheless, since the standard NJL model is not confining,  the thermodynamic functions reflect the presence of fermions within this region as well. 
Their increasing effect with the temperature $T$ rising toward the phase transition value $T^c$ is clearly seen in the growth of the quark number, entropy, pressure, and energy densities in Figs.~\ref{f4a} -- \ref{f4d}.
Notably, the removal of the momentum cutoff, $\Lambda\rightarrow\infty$, initially does  not result in a significant difference of behavior.
Such a deviation is only found as the transition to the massless-fermion-gas behavior at $T^c$ is approached.
This difference between the solid NJL($\Lambda$) and dashed NJL($\infty$) curves, as well as the influence of the fermions in the broken symmetry region overall, is much less pronounced at large ratios $\mu/T$, for which the system undergoes a first-order phase transition with a sudden discontinuous decrease in the effective fermion mass.
While the thermodynamic potential, and thus the pressure density, remains a continuous function in all cases, 
the characteristic discontinuity of the first-order transition is found in the quark number, entropy, and energy densities, as well as the interaction measure.

The interaction measure $I_{\text{NJL}}(T,\mu)$ in Fig.~\ref{f4e} furthermore depicts the change in the system between the spontaneously broken and restored chiral-symmetry regions clearly:
expressing it in terms of the scaled pressure density,
\begin{align}
\label{njl_interactionmeasure_dof}
\frac{I_{\text{NJL}}}{T^4} =
T\, \mfrac{\partial}{\partial T} \Bigl(\frac{p_{\text{NJL}}}{T^4}\Bigr) + 
\mu\,  \mfrac{\partial}{\partial \mu} \Bigl(\frac{p_{\text{NJL}}}{T^4}\Bigr)
,
\end{align}
illustrates that $I_{\text{NJL}}$ naively counts the change in the effective degrees of freedom of the fermions, cf.~\cite{f08}.
As indicated previously, the results at temperatures close to the phase transition and throughout the chirally symmetric phase are affected by the momentum cutoff $\Lambda$. In particular the behavior of an ideal massless fermion gas is only recovered in the limit $\Lambda\rightarrow\infty$.
Thus the behavior of $I_{\text{NJL}}$ in Fig.~\ref{f4e} accounting for a finite cutoff (solid lines) has to be considered largely artificial.
In the limit $\Lambda\rightarrow\infty$ (dashed lines), however, one finds a characteristic peaked structure, centered around $T^c$, which illustrates the increase in the fermionic effective degrees of freedom as the system transitions from a mixture of massive interacting fermions to the free massless fermion gas in the chirally symmetric region.

Overall, the NJL model can only provide schematic insight into the nature of the chiral phase transition, especially in the self-consistent first-order approximation discussed here. 
But it provides an adequate, accessible framework in the search for general characteristic signals of non-Hermitian fermionic field theories at finite temperature and chemical potential.
To this end the effects of the non-Hermitian $\cPT\!$-symmetry breaking pseudoscalar extension $g \bar{\psi}\gamma_5 \psi$ and the 
non-Hermitian $\cPT\!$-symmetric pseudovector extension $igB_\mu \,\bar{\psi}\gamma_5\gamma^\mu \psi$ are investigated in the following sections.

\section{Pseudoscalar extension}
\label{s3}

The NJL model is now extended through the inclusion of the pseudoscalar bilinear term $g \bar{\psi}\gamma_5 \psi$,
a modification that breaks the Hermiticity of the system. 
Nevertheless, an investigation of this model at vanishing temperature and chemical potential in \cite{fk21} has established the existence of real mass solutions and, moreover, dynamical mass generation due to this non-Hermitian extension term within the framework of the Euclidean four-momentum cutoff scheme.
This feature is particularly relevant in the context of $\cPT$ theory, for which the existence of real solutions to non-Hermitian models is a characteristic property. The non-Hermitian pseudoscalar extension, however,  breaks $\cPT$ symmetry: 
$\{\cPT, \gamma_5 \}=0$
for the parity-reflection and time-reversal operators
\begin{equation}
\label{PT_operators}
\begin{alignedat}{3}
&\cP: \psi(t, \mathbf{x})\, \to\,\, && \cP \psi(t, \mathbf{x}) \cP^{-1} && = 
\gamma^0 \psi(t, -\mathbf{x}), \\[3pt]
&\cT: \psi(t, \mathbf{x})\, \to\,\, && \cT \psi(t, \mathbf{x}) \cT^{-1} && = 
i\gamma^1\gamma^3 \psi^*(-t, \mathbf{x})
\end{alignedat}
\end{equation}
in $3+1$ dimensional spacetime. The extension term is thus anti-$\cPT\!$-symmetric and the Hamiltonian density 
\begin{equation}
\label{gamma5_hamiltonian}
\cH = \cH_{\text{NJL}} +g \bar{\psi} \gamma_5 \psi 
,
\end{equation}
of the full extended system is non-$\cPT\!$-symmetric overall.
Therefore, in the context of (at least) this fermionic quantum field theory, the reality of the effective fermion mass solution alone appears not to be a sufficient distinguishing feature of $\cPT$ models.
To identify such properties of non-Hermitian fermionic systems, we analyze  here the behavior of the effective fermion mass and the thermodynamic observables of the non-Hermitian NJL model with the pseudoscalar extension $g \gamma_5$ at \emph{finite} temperatures and densities. The following section then contrasts this with the behavior of a non-Hermitian but $\cPT$-symmetric pseudovector extension. 

In addition to breaking $\cPT\!$ symmetry, the bilinear extension based on 
$\gamma_5$ also explicitly breaks the chiral symmetry of the model considered, similar to including a small bare mass $m_0$ in the NJL model, 
see~(\ref{njl_hamiltonian}).
The limit of vanishing bare mass, $m_0\rightarrow 0$, is therefore not a chiral limit.
Nevertheless, the system retains an approximate chiral symmetry for small bilinear couplings $g$. 
Its effect on the behavior of the effective fermion mass in the $T$-$\mu$-plane and across the phase transition is, however, distinctly different to that of the scalar bare mass $m_0$, as shown in the following.

Since the non-Hermitian bilinear extension leaves the two-body interaction structure of the NJL model unchanged, the structure of the Feynman-Dyson perturbation approach remains applicable, cf.~\cite{fbk20,fk21}. The self-consistent Hartree approximation to the gap equation for the effective fermion mass keeps the general form (\ref{njl_gap}).
However, the full fermion propagator now depends on the extension term $g \gamma_5$ and can, for the evaluation of the gap equation, be written in the form
\begin{equation}
S(p_n)= (\slashed{p}_n+\mu\gamma^0-m-g\gamma_5)^{-1} = 
\mfrac{\slashed{p}_n+\mu\gamma^0 +m -g \gamma_5}{(i \omega_n+\mu)^2- (\mathbf{p}^2 + m^2 -g^2)}
,
\end{equation}
with $p_n= (i\omega_n,{\bf p})$, $\omega_n = (2n+1)\pi T$, and thus
\begin{widetext}
\begin{equation}
\text{tr}[S(\omega_n,{\bf p})] = 
\mfrac{4m}{(i \omega_n+\mu)^2- ({\bf p}^2 + m^2 -g^2)}
= 
\mfrac{2m}{\sqrt{{\bf p}^2+m^2-g^2}} \Bigl[
\mfrac{1}{i\omega_n -  (\sqrt{{\bf p}^2+m^2-g^2}-\mu)} -
\mfrac{1}{i\omega_n +  (\sqrt{{\bf p}^2+m^2-g^2}+\mu)}
\Bigr] ,
\end{equation}
\end{widetext}
similar to the treatment within the four-momentum cutoff scheme at vanishing temperature and chemical potential~\cite{fbk20,fk21}.
Note in particular, that the non-Hermitian coupling constant $g$ enters effectively 
as a quadratic shift of the fermion mass $m$, in the form $m^2-g^2$.
Accounting for this shift, the Matsubara-frequency summation can be performed analogously to the standard NJL model, resulting in the gap equation of the modified system:
\begin{widetext}
\begin{equation}
\label{gamma5_gap}
m =2 GN_c N_f\,  m \int^\Lambda \hspace{-0.2cm} \mfrac{\mathrm{d}^3{\bf p}}{(2\pi)^3}\,
\frac{1}{\sqrt{{\bf p}^2+m^2-g^2}}
\Bigl[\,\tanh\Bigl(\mfrac{\sqrt{{\bf p}^2+m^2-g^2}+\mu}{2T}\Bigr) 
+ \tanh\Bigl(\mfrac{\sqrt{{\bf p}^2+m^2-g^2}-\mu}{2T}\Bigr)
\,\Bigr]
.
\end{equation}
\end{widetext}

\begin{figure*}[]
\centering
\subfloat[]{
\includegraphics[width=0.45\textwidth]
{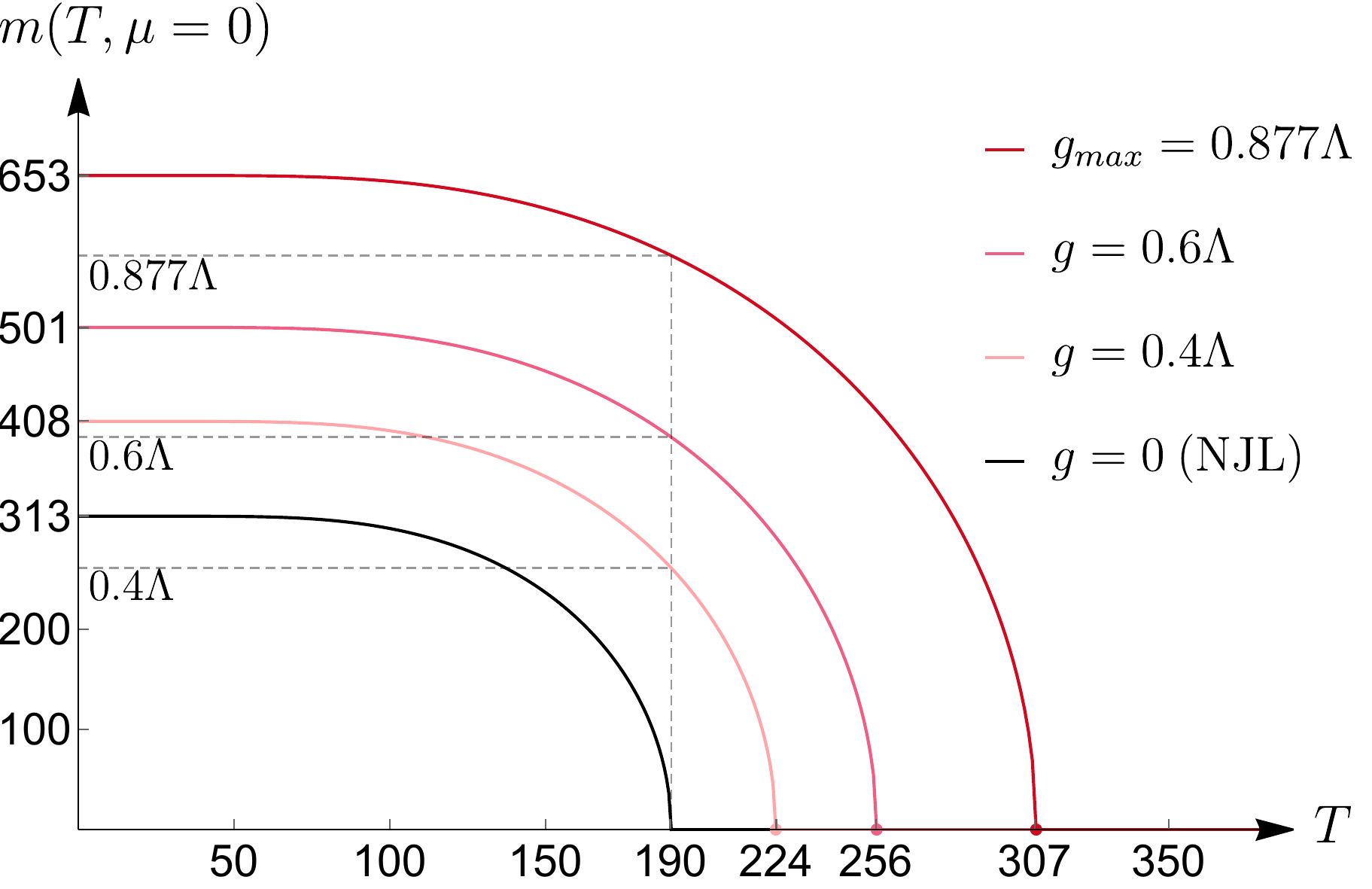}
\label{f5a}
}
\hfill
\subfloat[]{
\includegraphics[width=0.45\textwidth]
{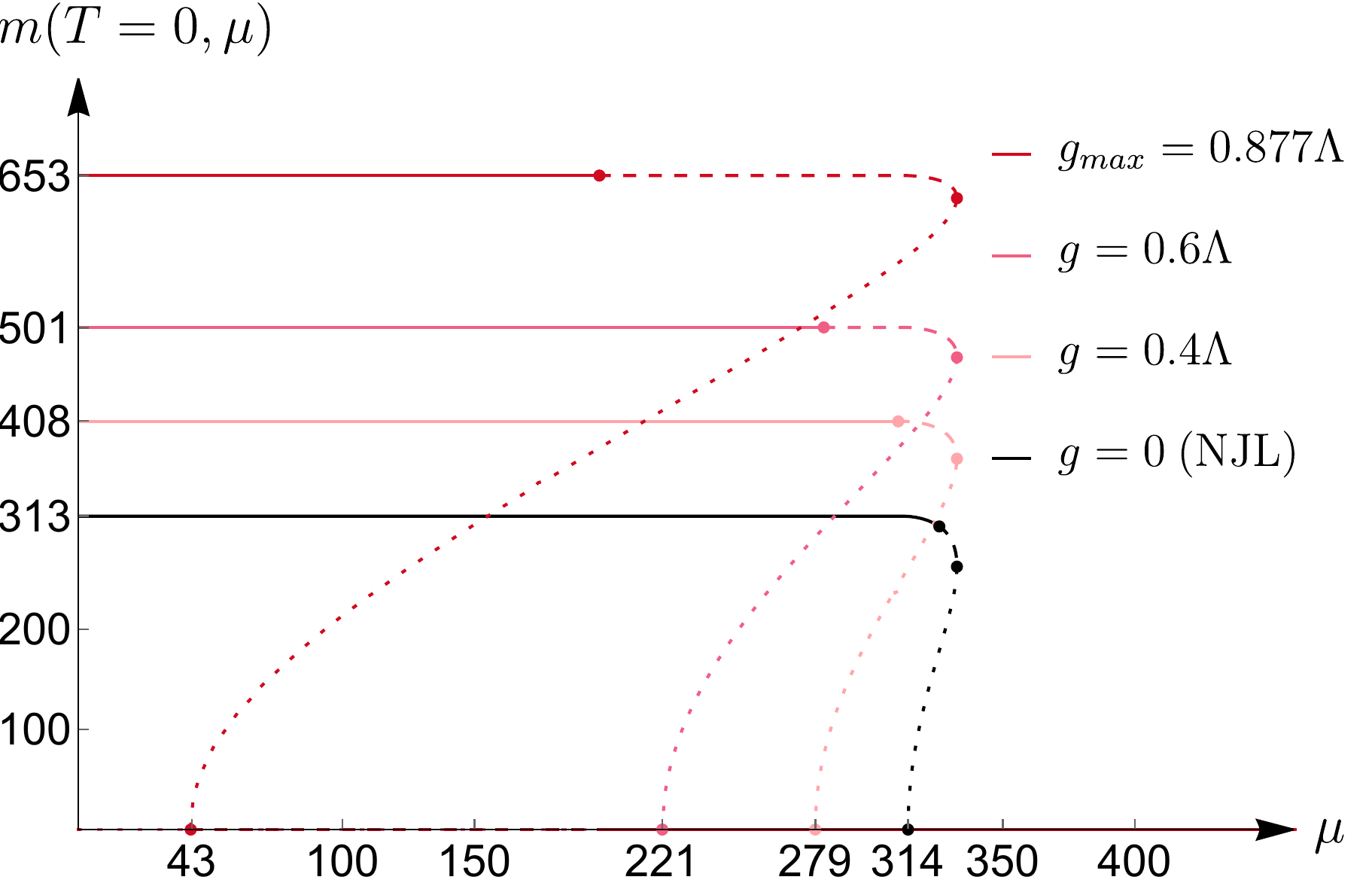}
\label{f5b}
 }
\caption{
(a) Behavior of the effective fermion mass $m$ within the pseudoscalar extension of the NJL model in MeV at vanishing chemical potential $\mu$ as a function of the temperature $T$ for various bilinear coupling values $g$.
(b)
Behavior of the effective mass $m$ at vanishing temperature $T$ as a function of the chemical potential $\mu$ for various bilinear coupling values $g$. The stable physical solutions associated with the global minimum of $\Omega$ are shown as solid lines, while metastable and unstable solutions of the gap equation are shown as dashed and dotted lines respectively. 
}
\end{figure*}

Notice that the hyperbolic tangents, in combination with the square root in the denominator, form a real-valued expression, even in the case that ${\bf p}^2+m^2-g^2 < 0$, enabling the existence of real effective mass solutions.
In fact, because of the formal similarity to the gap equation (\ref{njl_gap_2}), with $m_{\text{NJL}}^2$ replaced by $m^2-g^2$, one can directly infer the existence of a finite real fermion mass solution $m$ of the modified NJL model as long as a \emph{finite} mass solution $m_{\text{NJL}}$ in the standard NJL model exists, satisfying:
\begin{equation}
\label{gamma5_njl_mass}
m^2(T, \mu,g) = m^2_{\text{NJL}}(T, \mu)+g^2 
, \quad \text{for} \,\,\, m_{\text{NJL}} \neq 0.
\end{equation}
This generalizes the relation found in \cite{fk21} from the limit of vanishing temperature and chemical potential into the finite $T$-$\mu$--plane. 
However, the existence of such a solution $m$ does not ensure its physicality, i.e., the solution obtained from (\ref{gamma5_njl_mass}) is not guaranteed to correspond to the global minimum of the thermodynamic potential $\Omega$.
And moreover, real fermion masses in the modified NJL model are not restricted to the same $T$-$\mu$ regime as in the standard NJL model, since the relation (\ref{gamma5_njl_mass}) allows for real solutions $m$ where the corresponding solution in the standard NJL model is an unphysical imaginary result, i.e., where $-g^2 < m_{\text{NJL}}^2 < 0$. 
Nonetheless, a qualitative resemblance of the effective fermion mass behavior to the NJL-model result is found upon evaluating the self-consistent gap equation (\ref{gamma5_gap}) of the modified NJL model.

When both the temperature and chemical potential are zero, one notes that the effective fermion mass increases as a function of the bilinear coupling strength $g$, see Fig.~\ref{f5a}. That is, mass is generated on including this term, which confirms the robustness of the qualitative result previously obtained for the four-momentum Euclidean cutoff, applied to this model~\cite{fbk20,fk21}. 
At small (or vanishing) fixed chemical potential $\mu$ the real finite effective fermion mass solution decreases monotonically with increasing temperature $T$ until a second-order phase transition is reached at $T^c(g)> T^c_{\text{NJL}}$ and the spontaneously broken approximate chiral symmetry is restored, see Fig.~\ref{f5a} and Table~\ref{t0}.
Notably, the mass vanishes exactly for temperatures $T> T^c(g)$, even though the pseudoscalar extension term breaks the chiral symmetry explicitly.
Contrary to the inclusion of a bare mass $m_0$, the phase transition is not smoothed out, but remains sharp yet shifted to {\it larger} temperatures with increasing bilinear coupling strength $g$. 
While this behavior is found for arbitrarily large values of the real bilinear coupling $g$, we note that at $g_{max}\approx 0.877 \Lambda \approx 573$~MeV, the effective fermion mass at $T=\mu=0$ reaches the cutoff scale $\Lambda$. 
Therefore, no higher coupling values are considered in the following discussion.


\begin{table}
\centering
\renewcommand{\arraystretch}{1.1}
\small   
\begin{tabular}{ |c||c|c|c|c| }  
 \cline{2-5}
 \multicolumn{1}{c||}{} & NJL $ (g=0)$  & $g=0.4\Lambda$ & $g=0.6\Lambda$  &$g_{max}$ \\ 
 \hline
 \hline 
 $T^c (\mu=0, g)$ &  $190$~MeV &  $224$~MeV & $256$~MeV & $307$~MeV \\
\hline
 $\mu^c (T=0, g)$ & $326$~MeV &  $310$~MeV & $282$~MeV &
  $197$~MeV\\
\hline
\end{tabular}

\caption{
Phase transition temperatures $T^c (\mu=0, g)$ at vanishing chemical potential and transition chemical potentials $\mu^c (T=0, g)$ at vanishing temperature for various coupling strengths $g$ of the non-Hermitian extension term.
}
\label{t0}
\end{table}


When evaluating the gap equation (\ref{gamma5_gap}) at small (or vanishing) temperature $T$ as a function of the chemical potential $\mu$, again a qualitative resemblance to the standard NJL-model behavior is found, see Fig.~\ref{f5b}: real finite mass solutions exist up to a maximum value $\mu_+$ beyond which the mass solution vanishes. Contrasting the second-order transition behavior of the mass found as a function of temperature at small chemical potential, see Fig.~\ref{f5a}, the fermion mass does not decrease to vanish continuously. 
Instead a parametric region  
with multiple real finite fermion mass solutions is found.
As in the standard NJL model the thermodynamic potential $\Omega$ has to be studied to identify the stable physical mass solution in this region. 
The position of the critical chemical potential $\mu^c(g)$ of the approximate chiral phase transition, as well as the effect of the non-Hermitian extension on it, are not immediately apparent.
Note, however, the evidently decreasing lower bound $\mu_-(g)$ of the region with multiple real mass solutions as the bilinear coupling $g$  increases, while the upper bound $\mu_+\approx 333$~MeV remains unaffected. 

The thermodynamic potential $\Omega(T,\mu,g)$ of the modified NJL model can be determined from the thermodynamic average of the interaction energy, following a coupling-constant integration method that parallels the discussion for the standard NJL model, see (\ref{coupling_integration_method}). 
As such it is structurally not affected explicitly by the bilinear non-Hermitian extension term.
However, the substitution of the effective mass within the coupling-constant integral here relies on the modified gap equation (\ref{gamma5_gap}), through which the extension enters implicitly:
\begin{widetext}
\begin{eqnarray}
2\!\int_0^1 \hspace{-0.1cm} \mfrac{\mathrm{d}\lambda}{\lambda} \,(m_\lambda-m_0) \mfrac{\mathrm{d}m_\lambda}{\mathrm{d}\lambda}
&=& 
4G N_c N_f \int^\Lambda \hspace{-0.2cm} \mfrac{\mathrm{d}^3{\bf p}}{(2\pi)^3}\,
\int_0^1 \hspace{-0.2cm} \mathrm{d}\lambda \,
\mfrac{\mathrm{d} E_\lambda}{\mathrm{d}\lambda} \,
\mfrac{E_\lambda}{\sqrt{E_\lambda^2-g^2}}\,
\Bigl[\,\tanh\Bigl(\mfrac{\sqrt{E_\lambda^2-g^2}+\mu}{2T}\Bigr) + \tanh\Bigl(\mfrac{\sqrt{E_\lambda^2-g^2}-\mu}{2T}\Bigr)
\,\Bigr]\nonumber \\
&=&
8 G N_c N_f\, T\, \int^\Lambda \hspace{-0.2cm} \mfrac{\mathrm{d}^3{\bf p}}{(2\pi)^3}\,
\Bigl\{
\int_{x(0)}^{x(1)} \hspace{-0.2cm} \mathrm{d}x \, \tanh(x)
+\int_{y(0)}^{y(1)} \hspace{-0.2cm} \mathrm{d}y \, \tanh(y)
\Bigr\} \nonumber \\
&=&
8 G N_c N_f\, T\, \int^\Lambda \hspace{-0.2cm} \mfrac{\mathrm{d}^3{\bf p}}{(2\pi)^3}\,
\ln \Bigg(\,
\frac{\cosh[x(1)] \cosh[y(1)]}{\cosh[x(0)] \cosh[y(1)]}
\,\Bigg)
,
\end{eqnarray}

where $E_\lambda^2 = {\bf p}^2+m_\lambda^2$ and the variable of integration 
$\lambda$ is changed to $x(\lambda)= (\sqrt{E_\lambda^2-g^2}+\mu)/2T$ and $y(\lambda)=  (\sqrt{E_\lambda^2-g^2}-\mu)/2T$, with 
$\mathrm{d}x=\mathrm{d}y= \mathrm{d}\lambda({\mathrm{d} E_\lambda}/{\mathrm{d}\lambda} )({E_\lambda}/2T{\sqrt{E_\lambda^2-g^2}} )
$.
One thus arrives at the formal equivalent of (\ref{njl_potential_coupling_integrated}), 

\begin{equation}
\label{gamma5_potential_coupling_integrated}
\Omega-\Omega_0=
\frac{(m-m_0)^2}{4G} 
-2\,T\, N_c N_f
\int^\Lambda \hspace{-0.2cm} \mfrac{\mathrm{d}^3{\bf p}}{(2\pi)^3}\,
\ln \Bigg[\,
\frac{\cosh(\frac{\sqrt{E^2-g^2}+\mu}{2T}) \cosh(\frac{\sqrt{E^2-g^2}-\mu}{2T})}{
\cosh(\frac{\sqrt{E_0^2-g^2}+\mu}{2T}) \cosh(\frac{\sqrt{E_0^2-g^2}-\mu}{2T})}
\,\Bigg]
,
\end{equation}
which establishes the thermodynamic potential $\Omega(T,\mu,g)$ after 
subtracting the contribution associated with the thermodynamic potential $\Omega_0$ of the non-Hermitian free theory obtained at $\lambda=0$:
\begin{equation}
\label{gamma5_potential}
\Omega(T,\mu,g)=
\mfrac{(m\!-\!m_0)^2}{4G} 
-2 N_c N_f 
\int^\Lambda \hspace{-0.2cm} \mfrac{\mathrm{d}^3{\bf p}}{(2\pi)^3}\,
\sqrt{E^2\!-\!g^2}
-2\,T\, N_c N_f \!
\int^\Lambda \hspace{-0.2cm} \mfrac{\mathrm{d}^3{\bf p}}{(2\pi)^3}\,
\ln\Bigl(
\bigl[1+\mathrm{e}^{-(\sqrt{E^2-g^2}+\mu)/T}\,\bigr]
\bigl[1+\mathrm{e}^{-(\sqrt{E^2-g^2}-\mu)/T}\, \bigr]
\Bigr)
.
\end{equation}
\end{widetext}

Like the gap equation (\ref{gamma5_gap}) of the modified NJL model, which is 
recovered from the extremal condition $d\Omega/dm = 0$ in the limit of vanishing bare mass $m_0$, the thermodynamic potential (\ref{gamma5_potential})
is a real-valued expression, even in the case that $E^2-g^2 < 0$.
This property is seen clearly in (\ref{gamma5_potential_coupling_integrated}), whereas the separation of the momentum integral in (\ref{gamma5_potential}) obfuscates this slightly. 
Nevertheless, this separation of the square-root term, without explicit dependence on the temperature or the chemical potential, is advantageous for the identification of cutoff effects in the SB limit of large temperatures $T$, where it separates off the ultraviolet divergence of $\Omega(T,\mu,g)$.
Canceling imaginary contributions are in particular restricted to a fixed region of small momenta $\lvert\mathbf{p}\rvert \leq g <\Lambda$ and as such unaffected by a removal of the cutoff limit in the logarithmic integral contribution for comparison to the SB limit.

The stable physical mass solution is now determined as the global minimum of the thermodynamic potential $\Omega(T,\mu,g)$ under variation of the effective mass $m$, whereas local minima characterize metastable solutions and unstable solutions of the gap equation are maxima of $\Omega$.
These characterizations of the possible real mass results are visualized in Fig.~\ref{f5b} as solid, dashed, and dotted lines respectively. 
The chemical potential $\mu^c$ of the phase transition is denoted as a dot.
Similar to the standard NJL model, the modified system at small temperatures undergoes a first-order phase transition marked by the abrupt transition from a finite to a vanishing physical mass at $\mu^c$.
Notably, $\mu^c$ \emph{decreases} with increasing bilinear coupling  strength $g$ of the non-Hermitian extension term, see Table~\ref{t0},
contrasting the trend in the second-order transition behavior, cf. Fig.~\ref{f5a}.
This implies in particular, that while the relation (\ref{gamma5_njl_mass}) connects the possible finite real mass solutions of the standard NJL gap equation to possible real masses of the modified system, it does not preserve the physicality of this solution: finite real physical masses in the NJL model do not necessarily have corresponding finite real \emph{physical} mass solutions in the non-Hermitian model. 

\begin{figure*}[]
\centering
\subfloat[]{
\includegraphics[width=0.45\textwidth]
{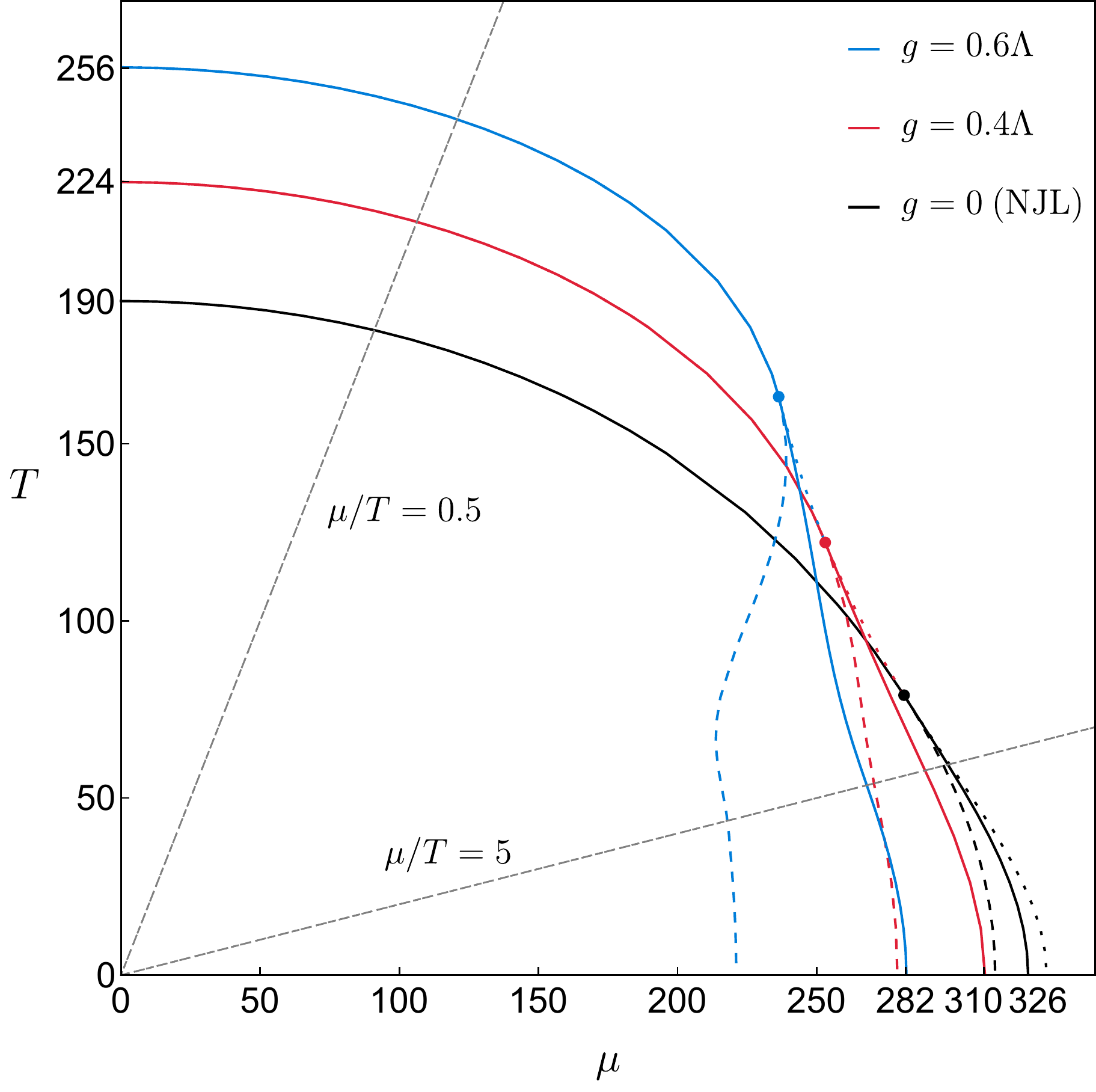}
\label{f6a}
}
\hfill
\subfloat[]{
\includegraphics[width=0.45\textwidth]
{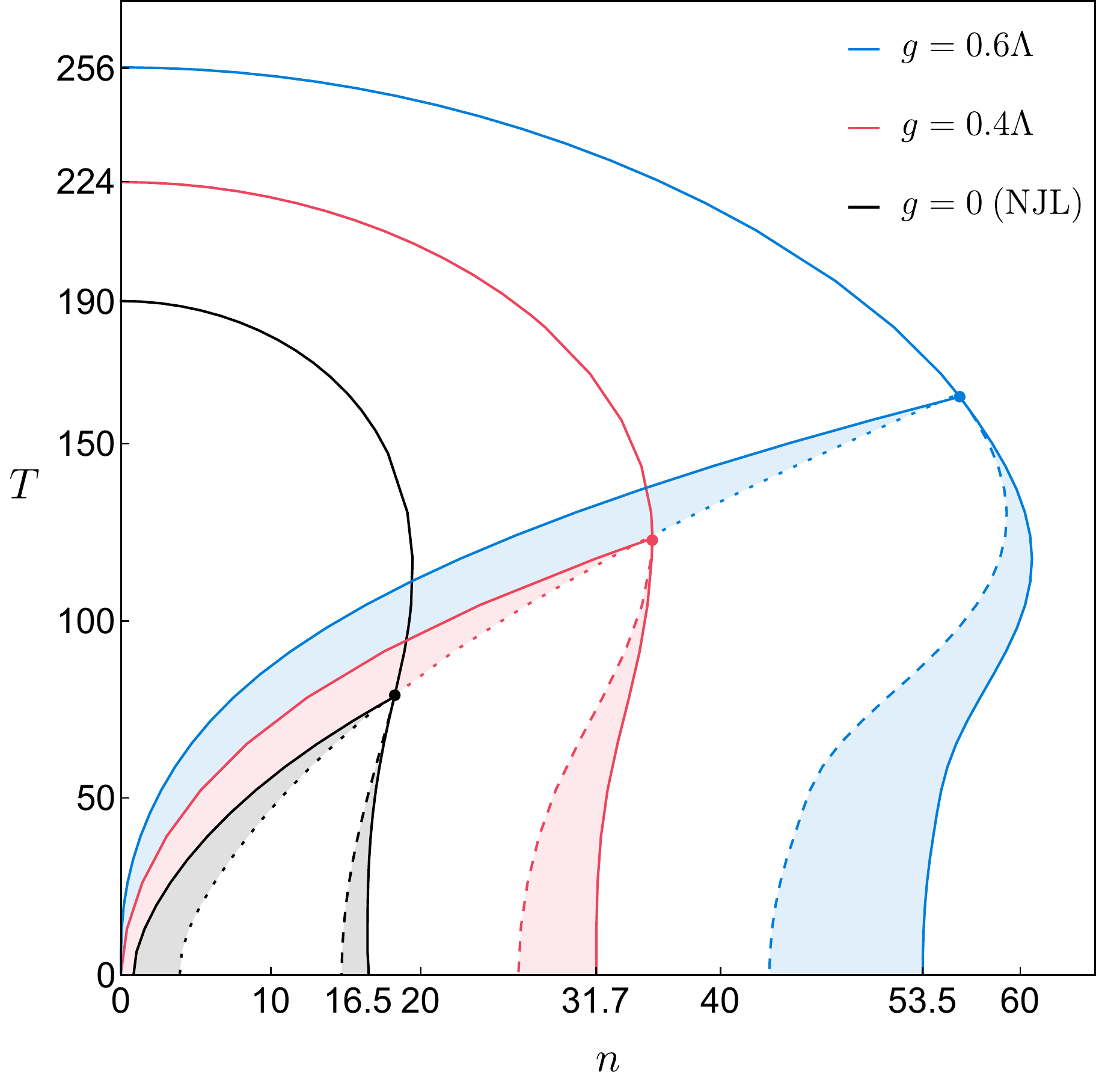}
\label{f6b}
 }
\caption{
Phase diagrams of the modified NJL model in the $T$-$\mu$--plane (a) and the $T$-$n$--plane (b) at various strengths $g$ of the non-Hermitian extension.
The phase transitions are denoted as solid lines, while the dotted and dashed lines mark the position of the spinoidals associated with $\mu_+(T,g)$ and $\mu_-(T,g)$ respectively.
The positions of the critical end-points are shown as dots.
}
\end{figure*}

\begin{table}[b]
\centering
\setlength\tabcolsep{7.8pt}
\renewcommand{\arraystretch}{1.4}
\small   
\begin{tabular}{ |c||c|c|c| } 
 \cline{2-4}
 \multicolumn{1}{c||}{} & NJL $ (g=0)$ & $g=0.4\Lambda$ & $g=0.6\Lambda$ \\ 
 \hline
 \hline 
$T_{\text{CEP}} (g)$ & $79 $~MeV & $ 122$~MeV & $163 $~MeV  \\
\hline
 $\mu_{\text{CEP}} (g)$ & $281 $~MeV &  $253 $~MeV &  $236 $~MeV \\
\hline
 $(\mu/T)_{\text{CEP}}$  &  $3.56$ &  $2.07$ & $1.45$ \\
\hline
\end{tabular}
\caption{
Temperature and chemical potential of the critical end-points of the modified NJL model for various coupling strengths $g$ of the non-Hermitian extension term.
}
\label{t00}
\end{table}

The overall behavior of the phase transition within the $T$-$\mu$-plane is visualized in Fig.~\ref{f6a} for the coupling strengths $g= 0.4\Lambda$ and $g= 0.6\Lambda$ of the non-Hermitian extension term.
The respective critical end-points, marking the change from a second-order transition behavior at low chemical potentials to a first-order behavior at low temperatures, are shown as dots and their position is listed in Table~\ref{t00}.
Note in particular, that with an increase in the bilinear coupling $g$, the position of the CEP in the modified NJL model follows the trend previously outlined of moving to higher temperature and lower chemical potential.
One observes furthermore, that the spinoidal which marks the upper bound $\mu_+(T)$ of the region with multiple finite mass solutions in the case of a first-order phase transition, cf. the special case of $T=0$ shown in Fig.~\ref{f5b}, remains generally independent of the bilinear coupling $g$.
Its position in the $T$-$\mu$-plane, shown as a dotted line in Fig.~\ref{f6a}, does not deviate from the NJL model case (in black), but merely extends beyond it as the CEP moves to higher temperatures and lower chemical potentials with increasing bilinear coupling $g$.

On the other hand, the position of the spinoidal marking the lower bound $\mu_-(T)$, shown as dashed lines in Fig.~\ref{f6a}, rapidly moves towards a small chemical potential with increasing $g$, like the phase transition itself, cf. again the case of vanishing temperature in Fig.~\ref{f5b}.

In addition to the analysis of the phase diagram in the $T$-$\mu$--plane,
the thermodynamic potential (\ref{gamma5_potential}) of the modified NJL model enables the study of the thermodynamic observables, paralleling the standard NJL model approach.
The evaluation of the quark number density, 
\begin{widetext}
\begin{equation}
\label{gamma5_quarknumber}
 n(T,\mu,g) =\,-\frac{\partial \,\Omega(T,\mu,g)}{\partial \mu} \,\Bigr\rvert_T
\\[4pt]
=\,
N_c N_f \int^\Lambda \hspace{-0.2cm} \mfrac{\mathrm{d}^3\bf{p}}{(2\pi)^3}\,
\Bigl[\tanh\Bigl(\mfrac{\sqrt{E^2-g^2}+\mu}{2T}\Bigr) - \tanh\Bigl(\mfrac{\sqrt{E^2-g^2}-\mu}{2T}\Bigr)
\Bigr]
,
\end{equation}
\end{widetext}
in particular allows for the visualization of the phase transition in the 
$T$-$n$--plane, where regions of stable, metastable, and unstable effective 
fermion mass solutions are distinguished clearly. Like the thermodynamic potential and the modified gap equation, $n(T,\mu,g)$ is a real-valued function.
The behavior of the phase transition in the $T$-$n$--plane is shown in Fig.~\ref{f6b} for the coupling values $g= 0.4\Lambda$ and $g= 0.6\Lambda$.
The transition is in each case denoted through solid lines, while
the spinoidals  associated with $\mu_+$ and $\mu_-$, which bind the shaded regions of metastable solutions and mark the transition to a region of unstable results only, are denoted as dotted and dashed lines respectively, cf. also Fig.~\ref{f3b}.

Together with the entropy density,
\begin{widetext}
\begin{align}
\label{gamma5_entropy}
\begin{split}
s(T,\mu,g) =& -\frac{\partial \,\Omega(T,\mu,g)}{\partial T} \,\Bigr\rvert_\mu
\\[4pt]
=&\,
2 N_c N_f \int^\Lambda \hspace{-0.2cm} \mfrac{\mathrm{d}^3\bf{p}}{(2\pi)^3}\,
\Bigl\{
\ln\Bigl(
\bigl[1+\mathrm{e}^{-(\sqrt{E^2-g^2}+\mu)/T}\,\bigr]\,
\bigl[1+\mathrm{e}^{-(\sqrt{E^2-g^2}-\mu)/T}\, \bigr]
\Bigr)
+\mfrac{\sqrt{E^2-g^2}}{T}\\[4pt]
&\,
-\mfrac{\sqrt{E^2-g^2}+\mu}{2T}
\tanh\Bigl(\mfrac{\sqrt{E^2-g^2}+\mu}{2T}\Bigr) 
- \mfrac{\sqrt{E^2-g^2}-\mu}{2T}
\tanh\Bigl(\mfrac{\sqrt{E^2-g^2}-\mu}{2T}\Bigr)
\Bigr\}
,
\end{split}
\end{align}
\end{widetext}
and the pressure density, determined by the thermodynamic potential (\ref{gamma5_potential}) of the modified system,
\begin{equation}
\label{gamma5_pressure}
p(T,\mu,g) = - \bigl[\, \Omega(T,\mu,g) - \Omega(0,0, g) \,\bigr],
\end{equation}
the energy density and interaction measure are found to be
\begin{align}
\label{gamma5_energy}
\begin{split}
& \epsilon(T,\mu,g) = -p(T,\mu,g) + T \,s(T,\mu,g) +\mu\, n(T,\mu,g)
,
\end{split} \\[5pt]
\label{gamma5_anomaly}
\begin{split}
& I(T,\mu, g) =\epsilon(T,\mu,g)-3\,p(T,\mu,g)
,
\end{split}
\end{align}
analogous to the standard NJL model. 
Like $\Omega(T,\mu,g)$ itself, they are entirely real-valued functions,
even when $E^2-g^2 < 0$.
Note furthermore, that the non-Hermitian bilinear coupling ultimately always enters in combination with the temperature as $g/T$ in (\ref{gamma5_quarknumber}) to (\ref{gamma5_anomaly}). 
As such, the large temperature limit remains unchanged by the inclusion of the non-Hermitian pseudoscalar bilinear extension, which is temperature suppressed, when considered for fixed $\mu/T$ and removing the cutoff scale $\Lambda \rightarrow \infty$ (with the exception of the UV-divergent term in $\Omega$ in the related pressure and energy densities as well as in the interaction measure, cf. the discussion within the standard NJL model).
One finds the same SB limits (\ref{njl_SBlimit_potential}) to (\ref{njl_SBlimit_entropy}) of an ideal massless fermion gas as in the standard NJL model.

\begin{table}[b]
\centering
\setlength\tabcolsep{7.8pt}
\renewcommand{\arraystretch}{1.4}
\small   
\begin{tabular}{ |c||c|c|c| }  
 \cline{2-4}
 \multicolumn{1}{c||}{} & $T^c_\text{NJL}$ & $T^c (g=0.4\Lambda)$ & $T^c (g=0.6\Lambda)$ \\ 
 \hline
 \hline 
$\mu/T=0.5$ & $182$~MeV & $213$~MeV& $242$~MeV \\
\hline
$\mu/T=5$ & $59$~MeV &  $58$~MeV  & $54$~MeV\\
\hline
\end{tabular}

\caption{
Phase transition temperature along lines of constant $\mu/T = 0.5$ (second-order transition region) and $\mu/T = 5$ (first-order transition region) for various coupling strengths $g$ of the non-Hermitian extension term.
}
\label{t2}
\end{table}

\begin{figure*}
\centering
\subfloat[]{
\includegraphics[width=0.45\textwidth]
{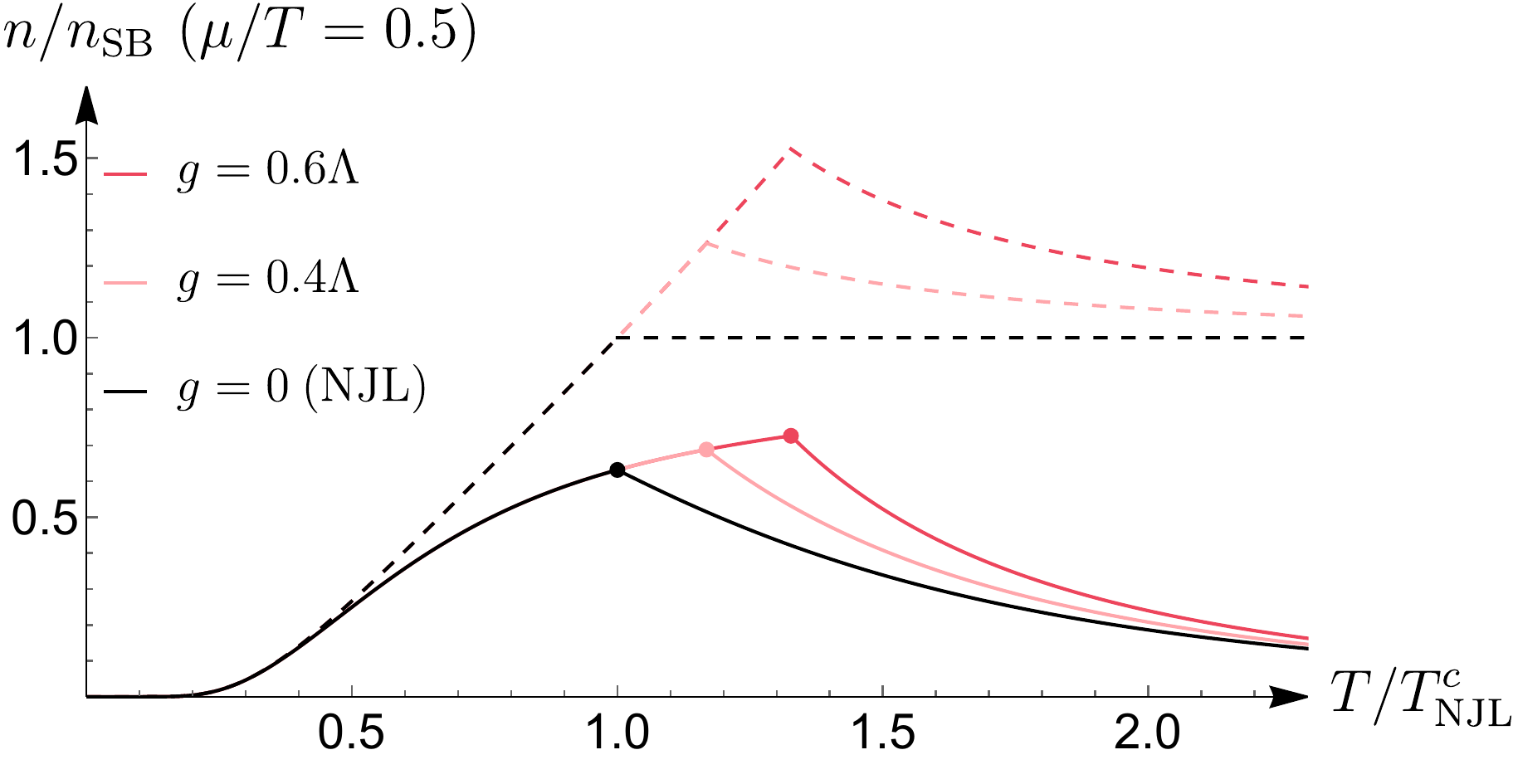}
\label{f7a}
}
\subfloat[]{
\includegraphics[width=0.45\textwidth]
{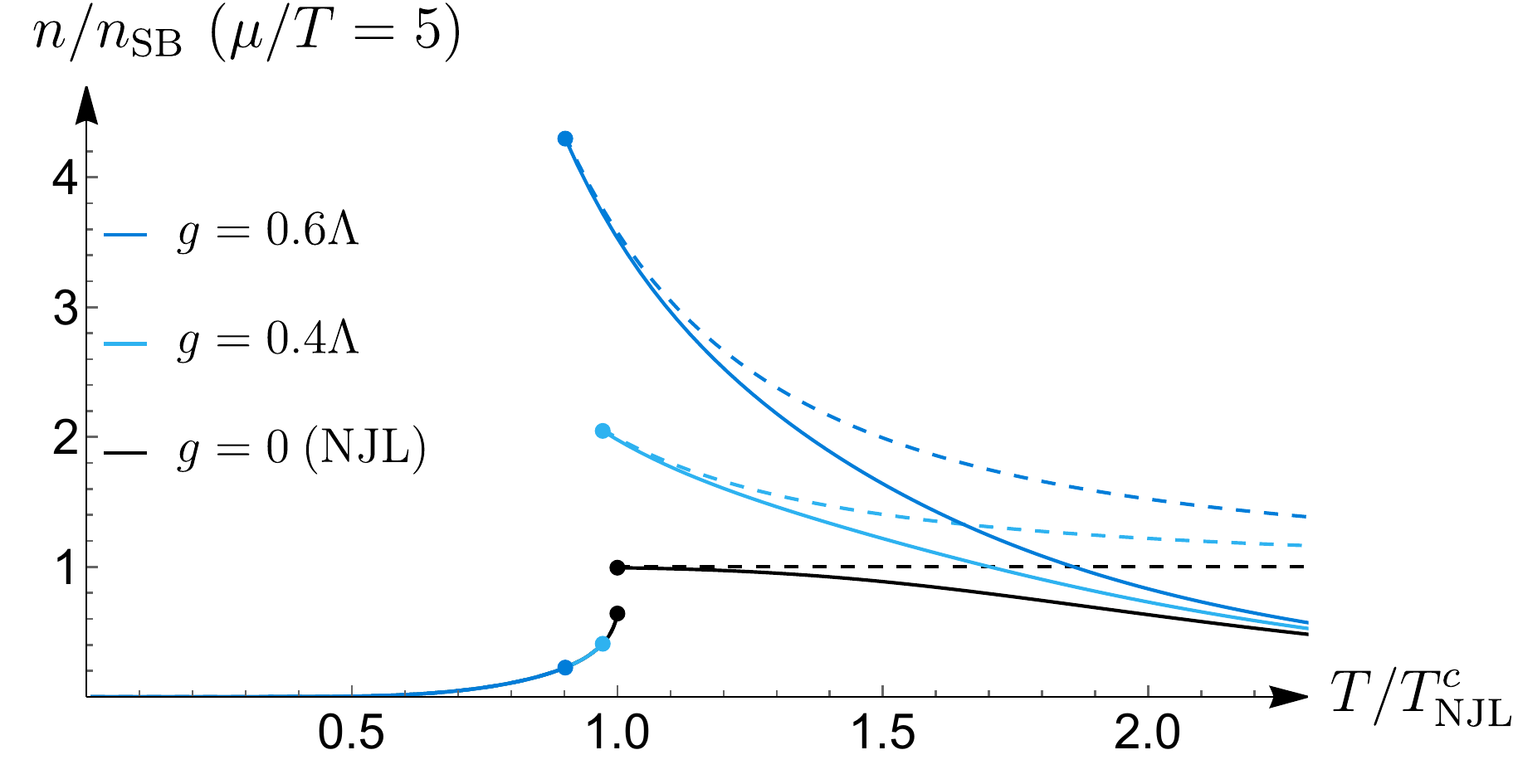}
\label{f7a2}
}
\hfill\\
\vspace*{-0.5cm}
\subfloat[]{
\includegraphics[width=0.45\textwidth]
{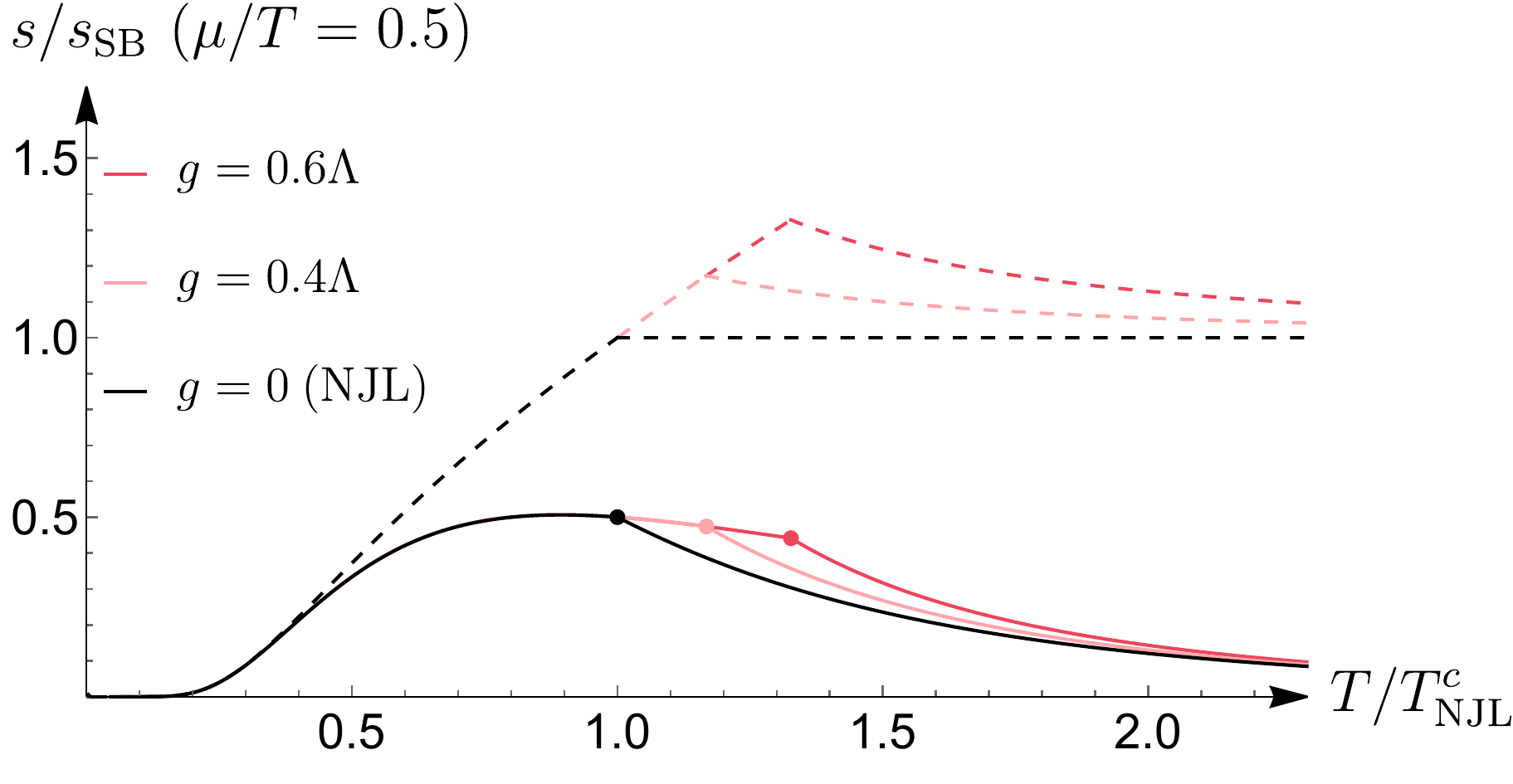}
\label{f7b}
}
\subfloat[]{
\includegraphics[width=0.45\textwidth]
{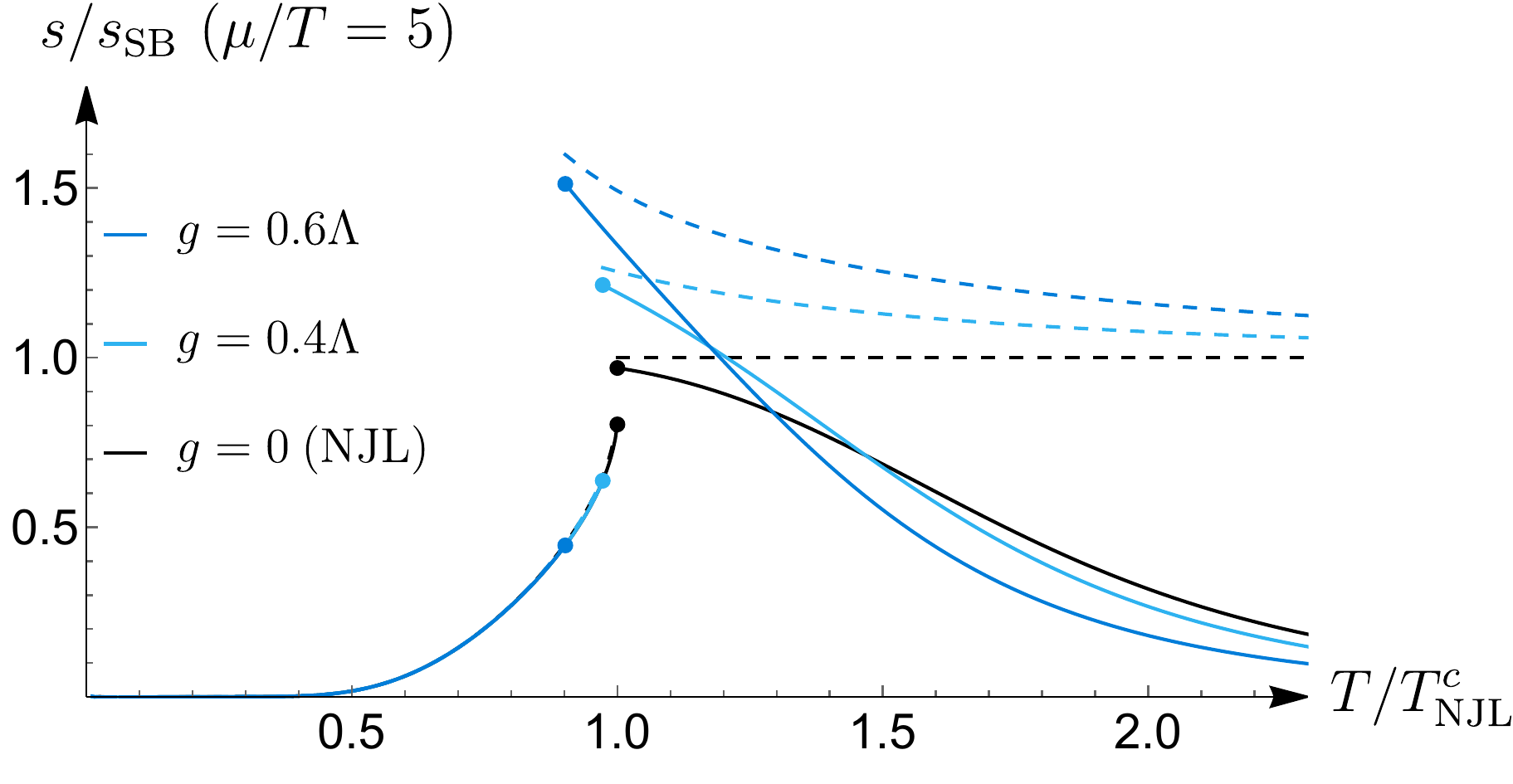}
\label{f7b2}
}
\hfill\\
\vspace*{-0.5cm}
\subfloat[]{
\includegraphics[width=0.45\textwidth]
{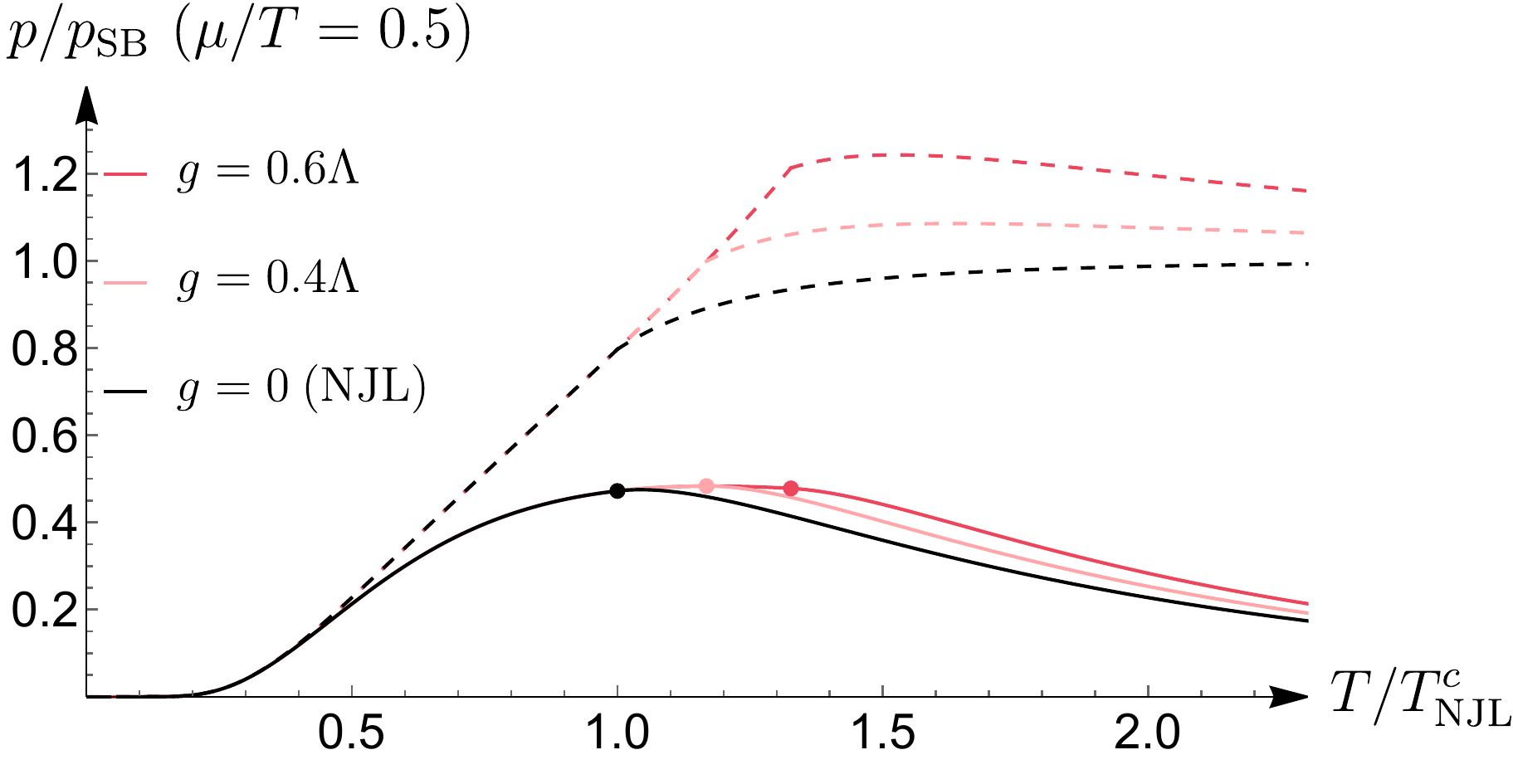}
\label{f7c}
}
\subfloat[]{
\includegraphics[width=0.45\textwidth]
{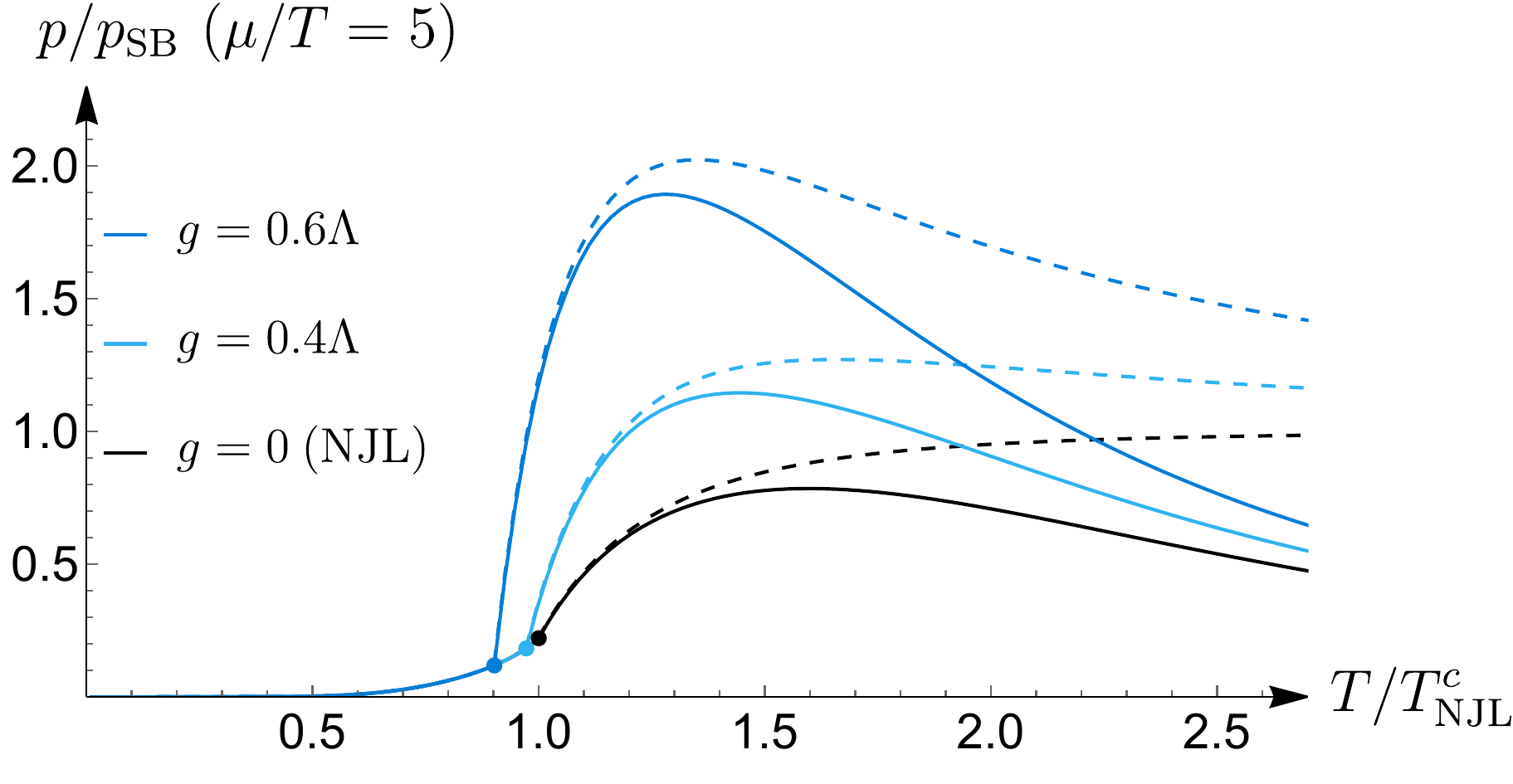}
\label{f7c2}
}
\hfill\\
\vspace*{-0.5cm}
\subfloat[]{
\includegraphics[width=0.45\textwidth]
{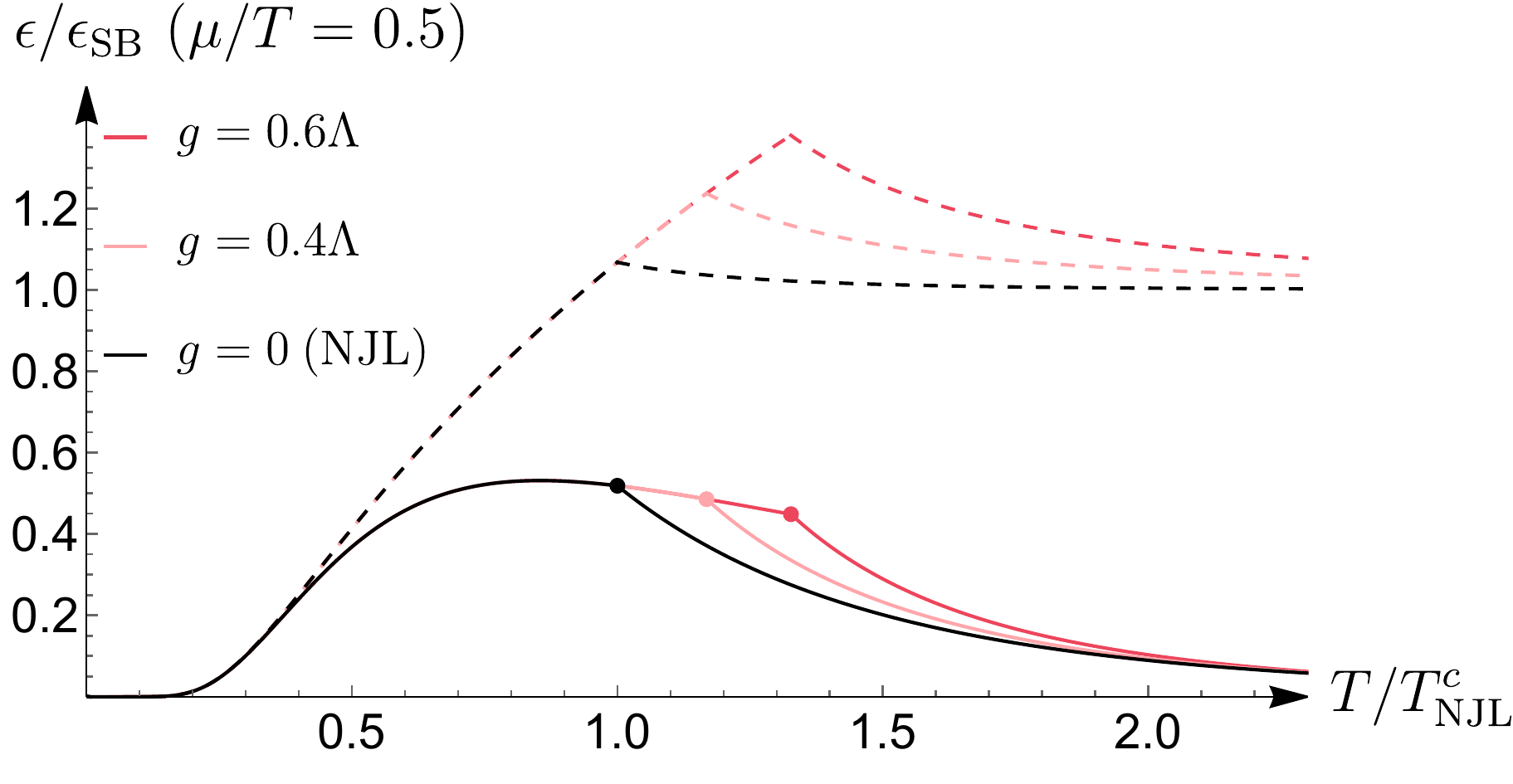}
\label{f7d}
}
\subfloat[]{
\includegraphics[width=0.45\textwidth]
{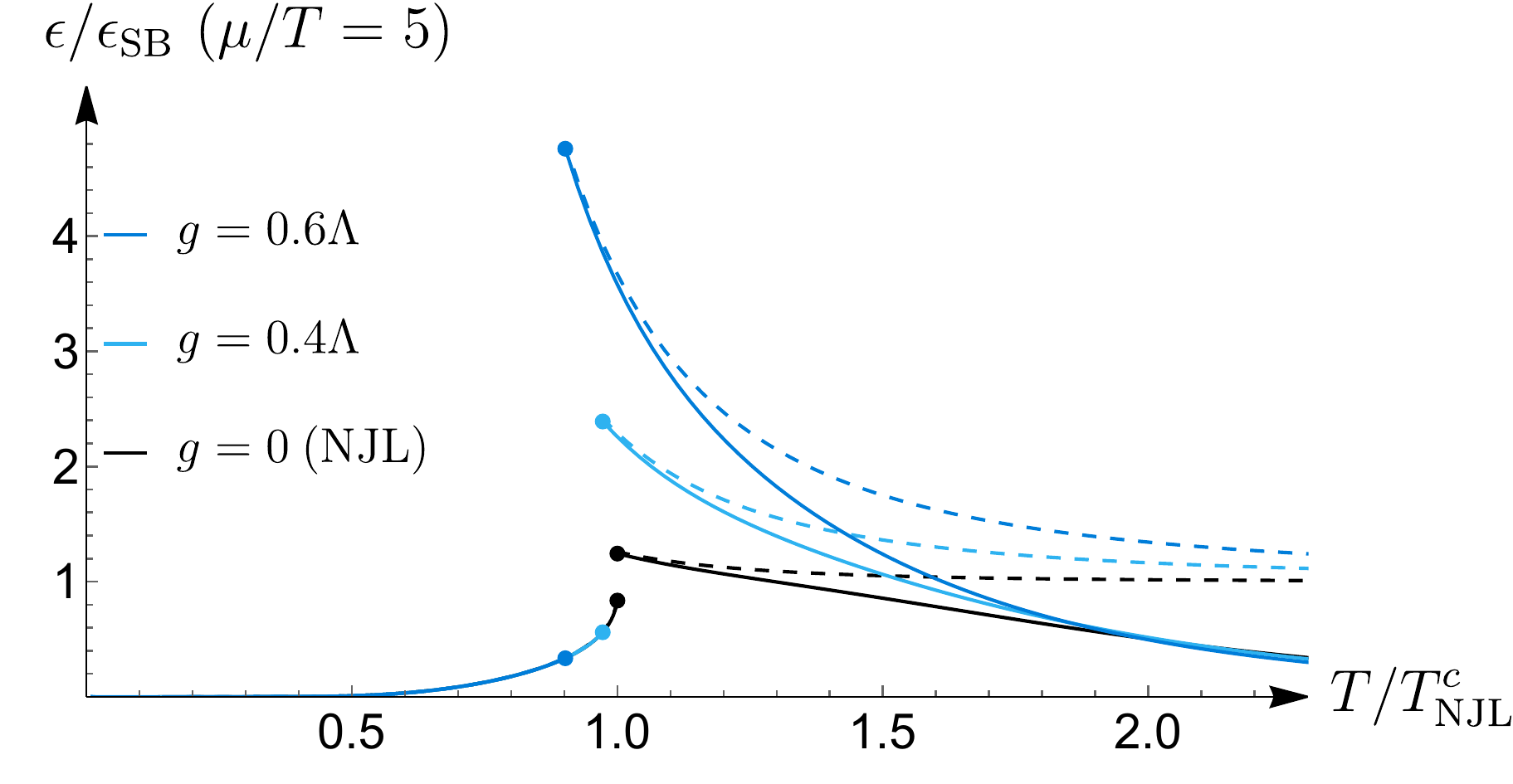}
\label{f7d2}
}
\hfill\\
\vspace*{-0.5cm}
\subfloat[]{
\includegraphics[width=0.45\textwidth]
{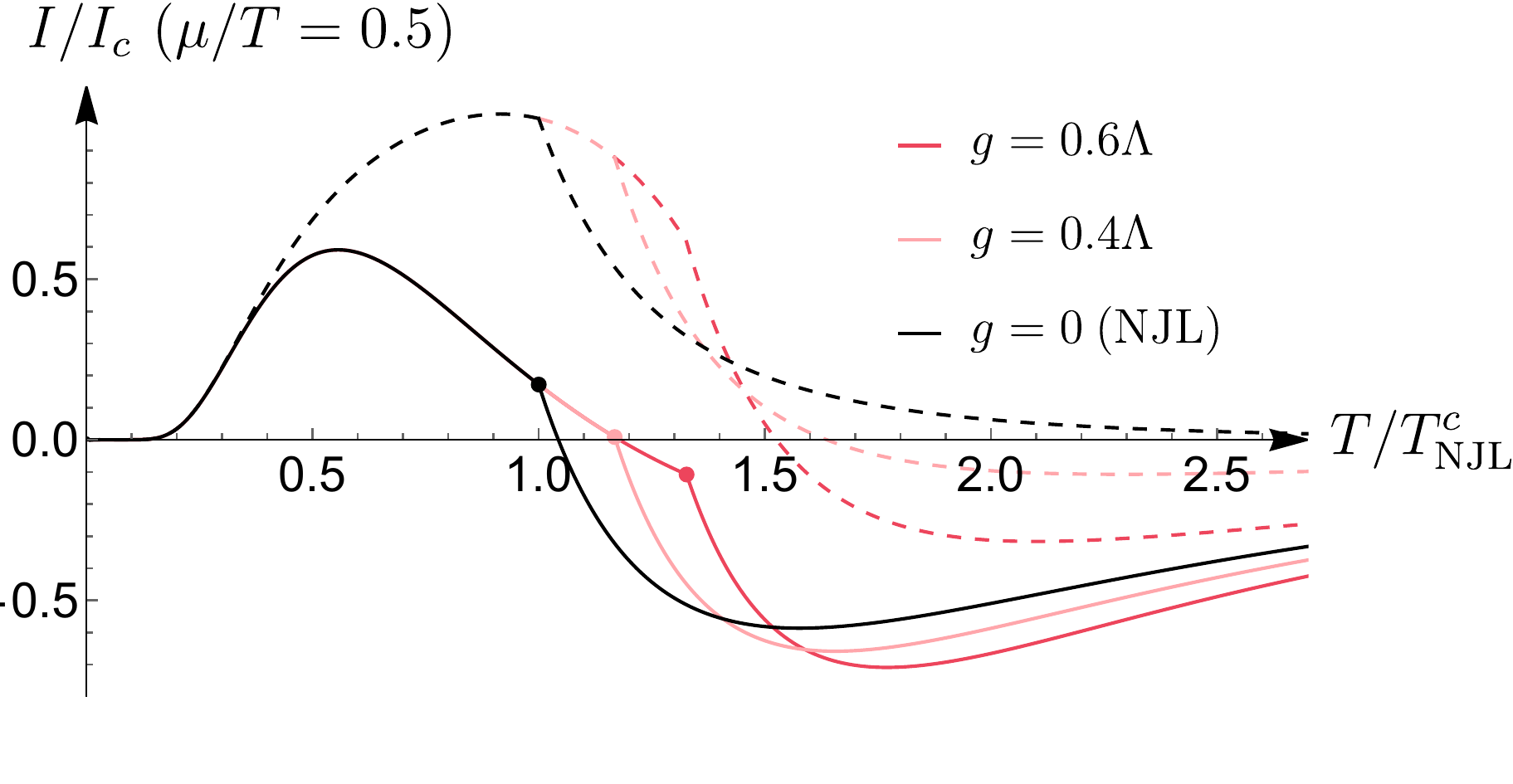}
\label{f7e}
}
\subfloat[]{
\includegraphics[width=0.45\textwidth]
{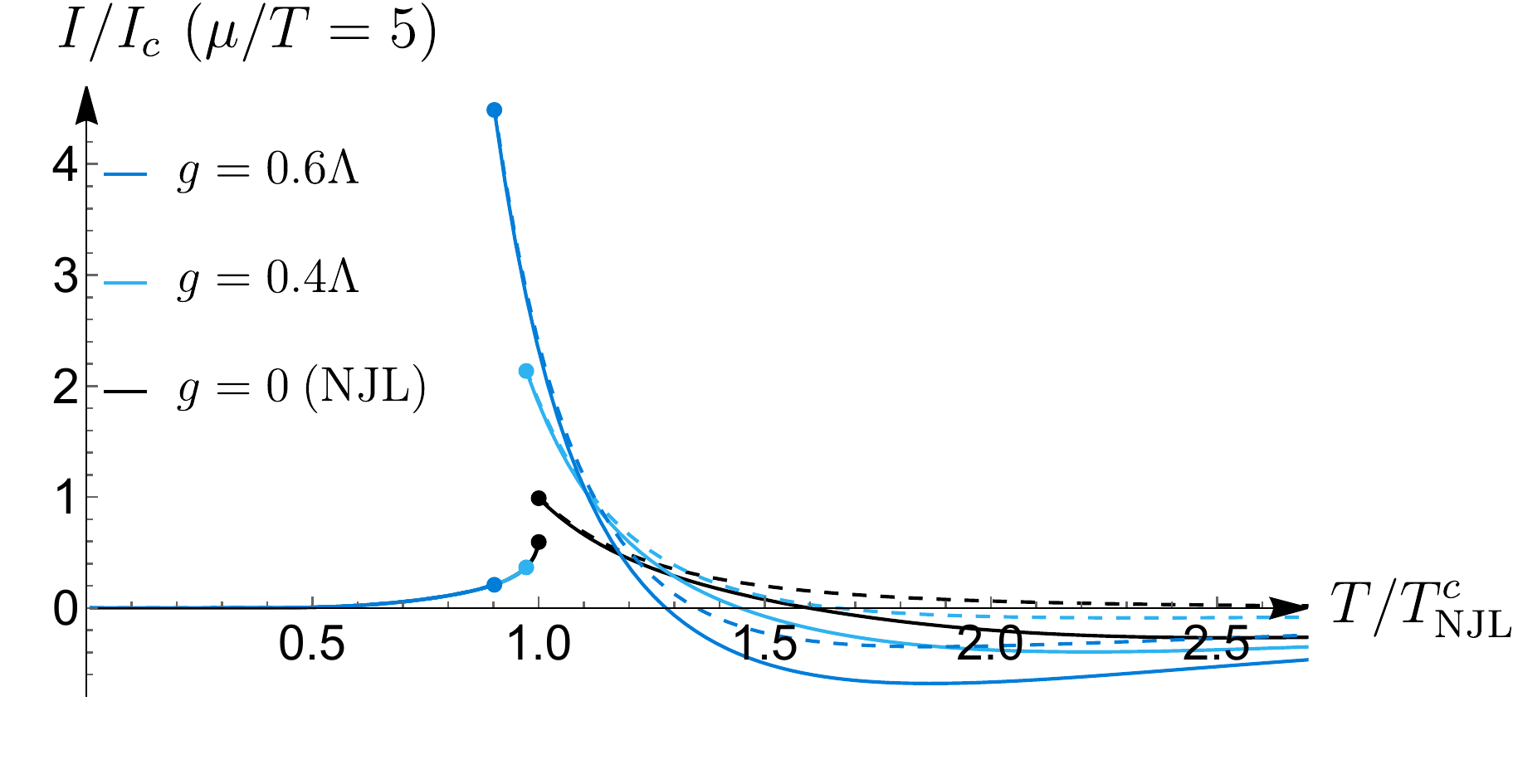}
\label{f7e2}
}
\caption{
Thermodynamic observables as a function of the scaled temperature $T/T^c_\text{NJL}$ with three-momentum cutoff $\Lambda$ (solid) and in the limit $\Lambda \rightarrow \infty$ (dashed) for fixed $\mu/T=0.5$ (second-order transition regime; in red) and $\mu/T=5$ (first-order transition regime; in blue). 
}
\end{figure*}

To illustrate the effect of the non-Hermitian extension on the thermodynamic observables at finite values of the temperature and the chemical potential, 
the behavior of the quantities $n$, $s$, $p$, and  $\epsilon$, scaled to their respective SB limits, 
is presented in Figs.~\ref{f7a} -- \ref{f7d2}.
Shown is their behavior for the non-Hermitian coupling strengths $g= 0.4\Lambda$ and $g= 0.6\Lambda$ along lines in the $T$-$\mu$--plane with constant ratio $\mu/T=0.5$ (red), in which case the phase transition remains of second order, and for the ratio $\mu/T=5$ (blue), 
for which the system undergoes a first-order phase transition, cf. Fig.~\ref{f6a}.
Solid lines denote the behavior with a fixed cutoff length $\Lambda$ as a function of the temperature scaled to the associated transition temperature within the standard NJL model, $T/T^c_{\text{NJL}}$,  
while dashed lines show the behavior when the cutoff is removed.
Table~\ref{t2} lists the corresponding critical temperatures $T^c(g)$.

In both the second-order transition case with $\mu/T=0.5$ and the first-order transition case with $\mu/T=5$, the behavior of the thermodynamic functions (\ref{gamma5_quarknumber}) -- (\ref{gamma5_anomaly}) coincides with the behavior of the standard NJL model functions within the overlapping spontaneously broken  symmetry regions, despite the dynamical generation of fermion mass.
This is to be expected because, like for the modified gap equation (\ref{gamma5_gap}), these functions show a formal equivalence to the corresponding standard NJL model observables (\ref{njl_quarknumber}) -- (\ref{njl_anomaly}) with $m_{\text{NJL}}^2$ replaced by $m^2-g^2$.  
Due to the relation (\ref{gamma5_njl_mass}) between the effective fermion masses 
any dependence on the non-Hermitian extension term thus cancels exactly at values of $T$ and $\mu$ that lie within the spontaneously broken symmetry region of both the standard NJL model \emph{and} the modified system.
However, the position of the phase transition within the $T$-$\mu$--plane is affected by the non-Hermitian extension term, resulting in some notable differences.  

Due to the increase of the transition temperature $T^c(g)$ in the second-order phase transition case with $\mu/T = 0.5$ (red), one observes a continued increase of the quark number density $n$ for $T^c_\text{NJL} < T  < T^c(g)$, see Fig.~\ref{f7a}.
When the cutoff is removed (dashed lines) this increase in particular \emph{exceeds} the SB limit. 
Beyond the phase transition, $n$ then displays an asymptotic decay toward either a vanishing limit, for finite $\Lambda$, or toward the SB limit, for $\Lambda\rightarrow\infty$. 
As in the standard NJL model the vanishing large-temperature behavior can thus be identified as a cutoff artifact.
The asymptotic decay throughout the restored approximate chiral symmetry region at $T>T^c(g)$ toward the standard SB limit of an ideal massless fermion gas for $\Lambda\rightarrow\infty$, contrasting the plateauing behavior found in the NJL($\infty$) case,
is a result of the temperature-suppressed influence of the extension term.
A notable deviation from the massless ideal gas behavior in the form of a fermion excess
remains until the temperature well exceeds the transition value $T^c(g)$.
A comparable phenomenology is found for the entropy, pressure, and energy density, cf. Figs.~\ref{f7b}, \ref{f7c}, and \ref{f7d}.

In the first-order transition case with $\mu/T=5$ (blue), see Figs.~\ref{f7a2}, \ref{f7b2}, \ref{f7c2}, and \ref{f7d2}, 
the pseudoscalar extension results in a decrease of the transition temperature $T^c(g)$ with increasing bilinear coupling $g$ instead.
As such the behavior of the modified system coincides with the NJL model behavior throughout the entire region of spontaneously broken approximate chiral symmetry at $T< T^c(g)$. 
However, after undergoing the phase transition at $T^c(g)$, with the characteristic  discontinuous jump in all thermodynamic quantities but the pressure, a notable increase beyond the SB limit is observed in all functions, followed again by an asymptotic decay toward either an artificial vanishing limit at high temperatures, for finite $\Lambda$,
or toward the SB limit of an ideal massless fermion gas, when the cutoff is removed. 
As in the second-order transition case, a notable deviation from the ideal gas behavior remains until the temperature well exceeds the transition value $T^c(g)$.



The behavior of the interaction measure is illustrated in Figs.~\ref{f7e} and \ref{f7e2}. Since $I(T,\mu,g)$ vanishes in the SB limit, it is normalized instead to the value $I_c$ at the phase transition of the NJL($\infty$) case as before.
The behavior of the pressure and the energy density close to the phase transition and beyond are affected notably by the momentum cutoff $\Lambda$,
so that again the behavior of $I/I_c$ when accounting for a finite cutoff (solid lines) has to be considered largely artificial.
In the limit $\Lambda\rightarrow\infty$ (dashed lines), the interaction measure  
behaves as follows. 

Like the corresponding energy and pressure densities, the behavior of $I/I_c$ coincides with the behavior of the standard NJL model within the overlapping spontaneously broken  symmetry regions. 
For the second-order transition case with $\mu/T=0.5$, see Fig.~\ref{f7e}, where the transition temperature increases with the bilinear coupling $g$, 
the interaction measure decreases less rapidly than in the standard NJL model for $T^c_\text{NJL} < T < T^c(g)$ until $T^c(g)$ is reached, reflecting the presence of a fermion excess in the modified system.
Beyond the phase transition $I/I_c$ continues to decrease, although - contrary to the standard NJL case - it notably does not approach the vanishing SB limit monotonically.
Instead, a decrease to \emph{negative} values is found, until a minimum is reached and the vanishing large-temperature limit is approached from below.
For the first-order transition case with $\mu/T=5$, see Fig.~\ref{f7e2},
the interaction measure of the extended system coincides with the standard NJL model all throughout the spontaneously broken symmetry region, since the transition temperature here decreases with increasing $g$.
The fermion excess becomes apparent only after undergoing the phase transition,
cf. Fig.~\ref{f7b2}; 
similar to the second-order transition case, $I/I_c$ takes on larger values than in the standard NJL model in the restored symmetry regime close to the phase transition, but rapidly decreases toward a minimum at negative values before approaching the vanishing SB limit from below. 

\begin{figure}
\centering
\subfloat[]{
\includegraphics[width=0.45\textwidth]
{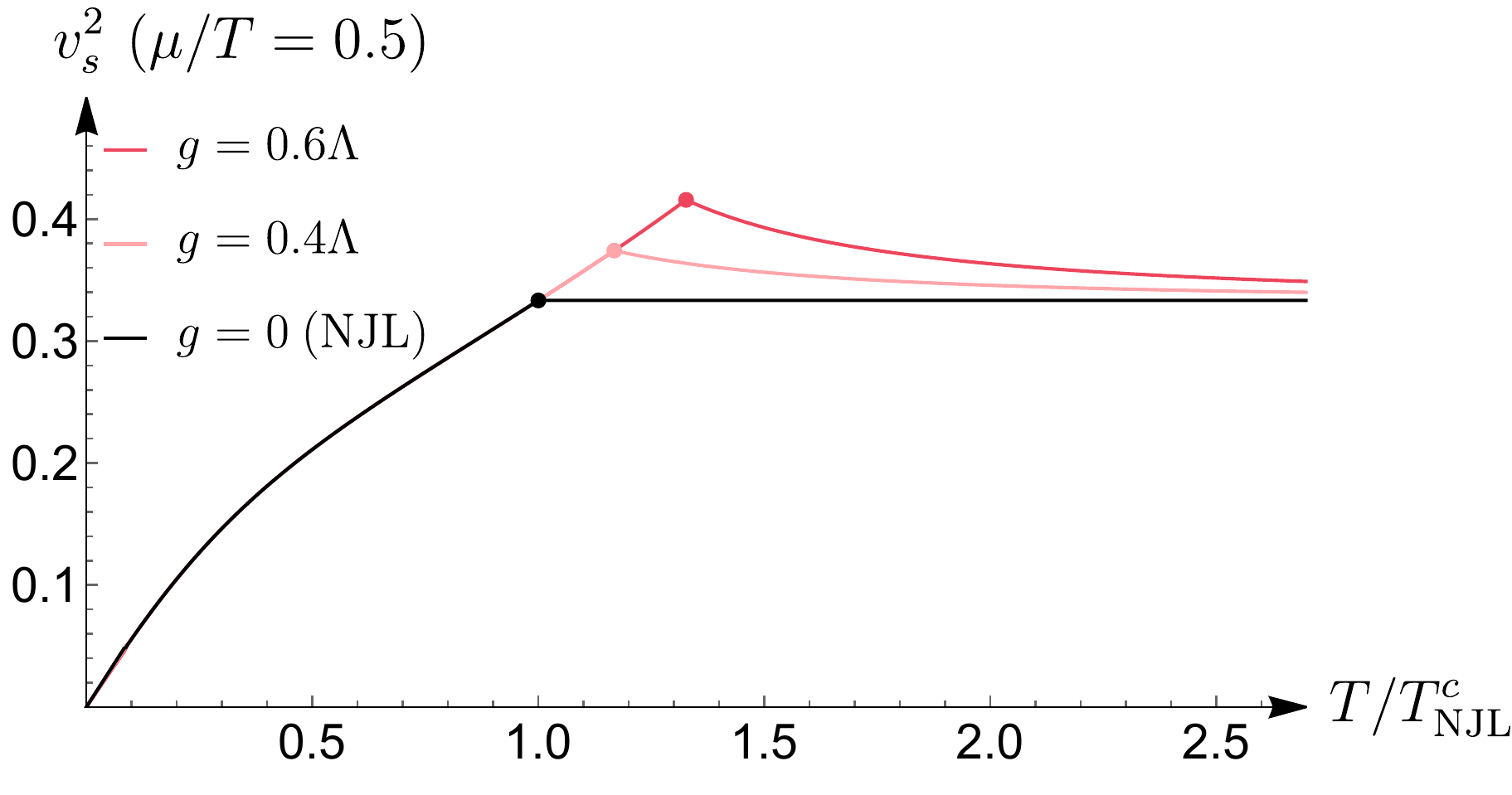}
\label{f9a}
}
\hfill
\subfloat[]{
\includegraphics[width=0.45\textwidth]
{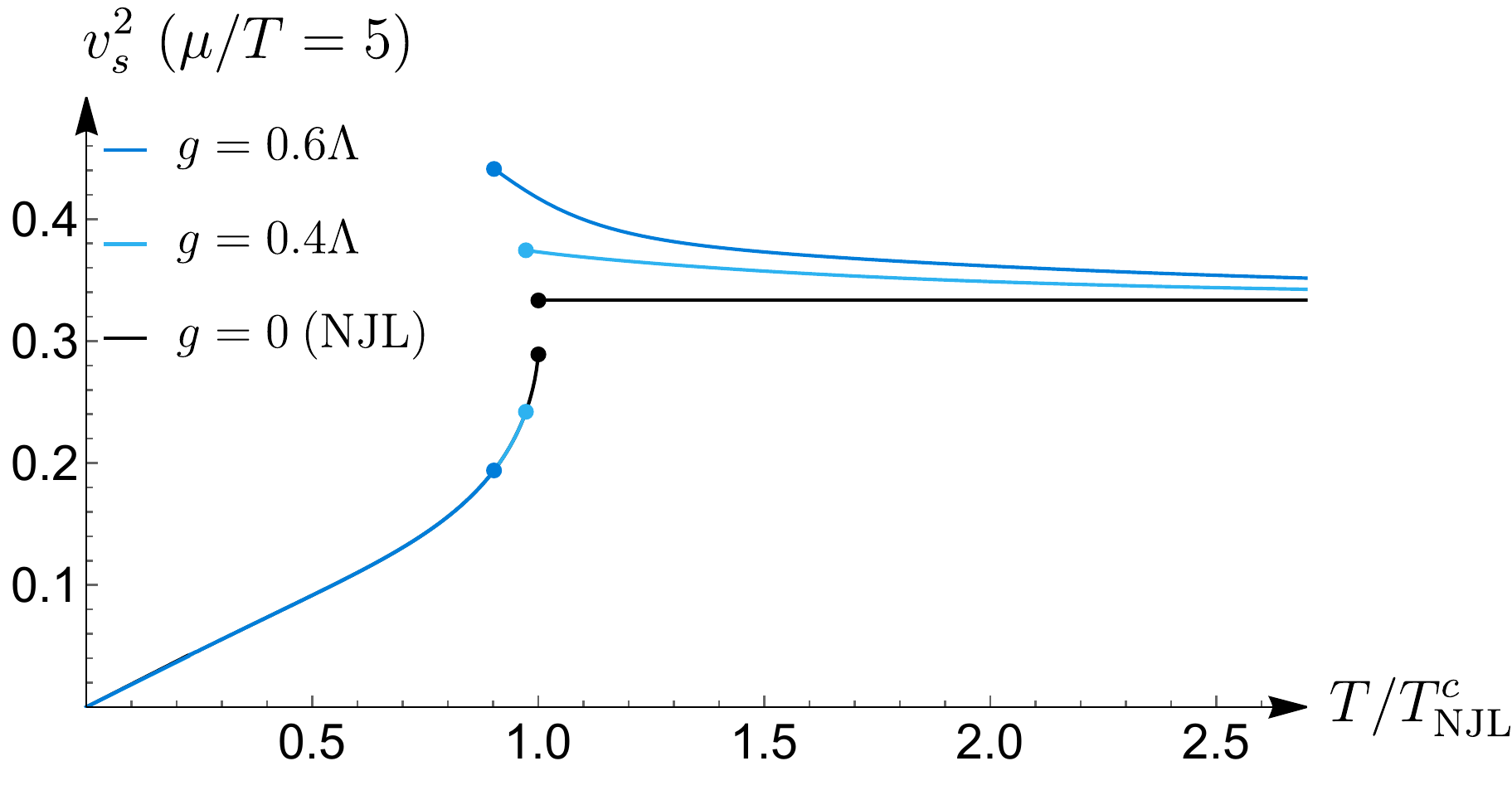}
\label{f9b}
}
\caption{\label{f9}
Behavior of the speed of sound as a function of the scaled temperature $T/T^c_\text{NJL}$ for fixed $\mu/T=0.5$ [(a); second-order transition regime; in red] and $\mu/T=5$ [(b); first-order transition regime; in blue] in the case of removed three-momentum cutoff, $\Lambda\rightarrow\infty$.
}
\end{figure}

Altogether the behavior of the interaction measure in the modified system shows the characteristic peaked structure found in the standard NJL model due to the system transitioning from a mixture of massive interacting fermions and bound states to a free massless fermion gas, combined with the effects of a fermion excess in the vicinity of the phase transition due to the non-Hermitian extension.
The presence of \emph{negative} interaction measure values within the region of restored approximate chiral symmetry marks a notable change in behavior of the modified NJL model that includes a non-Hermitian pseudoscalar bilinear term when compared to the standard NJL model case.  
While relativistic theories satisfying $\epsilon < 3 p$ were initially disregarded based on the observation that $\epsilon = 3p$ for the electromagnetic field and $\epsilon > 3p$ for massive free noninteracting particles \cite{ll75}, it has long been demonstrated that such theories are not necessarily at odds with relativistic causality \cite{z62}.
In the case of the present study, this can be clearly seen in the presence of a subluminal ($v_s<1$) speed of sound,
\begin{equation}
\label{gamma5_speed}
v_s^2(T,\mu,g) = 
\Bigl( \frac{\partial \,p(T,\mu,g)}{\partial T} \,\Bigr\rvert_\mu \,\Bigr) 
\Bigl( \frac{\partial \,\epsilon(T,\mu,g)}{\partial T} \,\Bigr\rvert_\mu \,\Bigr)^{-1} 
,
\end{equation}
shown in Fig.~\ref{f9}.
While in accordance with relativistic causality, it is noteworthy that $v_s^2(T,\mu,g) $ does exceed the conjectured speed of sound bound of $v_s^2 = 1/3$ \cite{ccn09}; this bound has, however, been called into question by examinations of the constraints of neutron star masses and radii \cite{bs15, tcgr18, hfn21} finding higher speeds of sound crucial for the existence of neutron stars above two solar masses.
Models with negative interaction measure have, furthermore, been considered recently  \cite{, m15, mo16, pmp18} in the context of scalar-tensor theories, where this property may result in a significant deviation from the theory of general relativity around neutron stars. 
Considering that the discussion of quark matter within neutron stars is frequently modeled using an equation of state taken from the NJL model, the possible occurrence of negative interaction measure values due to the non-Hermitian pseudoscalar extension provides a noteworthy feature of this system, connecting it to the discussion of extended theories of general relativity.

Overall, modifying the NJL model through the inclusion of a non-Hermitian pseudoscalar bilinear term $g \bar{\psi}\gamma_5 \psi$ results in a dynamically generated increase of the effective fermion mass in the spontaneously broken approximate chiral regime of the system at finite temperature and chemical potential.
The position of the phase transition in the $T$-$\mu$--plane is affected by this extension, moving toward higher temperatures for low chemical potential values (second-order transition region) and toward lower chemical potentials for low temperatures (first-order transition region);
the position of the CEP follows this trend, moving toward higher $T_\text{CEP}$ and lower $\mu_\text{CEP}$ with increasing coupling strength of the non-Hermitian term.
Despite the generation of fermion mass, the behavior of the quark number, entropy, pressure, and energy densities remains unchanged for small $T$ and $\mu$.
In the vicinity of the phase transition and throughout the restored symmetry regime, however, a notable fermion excess is observed, which decreases asymptotically, approaching the limit of an ideal massless fermion gas deep within the restored symmetry region.
Here, at some distance from the phase transition, the modified system is furthermore characterized by negative interaction measure values, which mark a notable deviation from the standard NJL model behavior.

\section{Pseudovector extension}
\label{s4}

Another possible non-Hermitian bilinear extension of the NJL model 
is the addition of an imaginary pseudovector extension term $igB_\nu \,\bar{\psi}\gamma_5\gamma^\nu \psi$.
As with the previous modification, an investigation at vanishing temperature and chemical potential has demonstrated the retained existence of real mass solutions and the possibility of dynamical mass generation due to the non-Hermitian extension term, within the Euclidean four-dimensional cutoff regularization scheme, and with spacelike fields $B_\nu$  \cite{fbk20,fk21}.
Unlike the previous modification, however, this pseudovector bilinear {\it preserves} the $\cPT$ symmetry in $3+1$ dimensional spacetime: $[\cPT, i\gamma_5\gamma^\nu] = 0$ for the parity-reflection and time-reversal operators in (\ref{PT_operators}).
Thus the full extended non-Hermitian system with the Hamiltonian density
\begin{equation}
\label{PT_hamiltonian}
\cH = \cH_{\text{NJL}} +igB_\nu \,\bar{\psi}\gamma_5\gamma^\nu \psi
\end{equation}
is $\cPT\!$ symmetric overall.
As is characteristic of many $\cPT$ systems, the existence of real fermion masses at $T = \mu=0$ is restricted to a finite region up to a critical value of the coupling strength $g$, cf. \cite{fbk20,fk21}, beyond which the (nontrivial) mass solutions occur in complex conjugate pairs. 
This transition is the consequence of a spontaneous breaking of the $\cPT$ symmetry of the system.

Moreover, the pseudovector extension anticommutes with $\gamma_5$, preserving the axial flavor symmetry and thus the overall chiral-symmetry properties. 
As such the limit of vanishing bare mass, $m_0 \rightarrow 0$, remains the chiral limit of this modified NJL model.

Similar to the pseudoscalar extension discussed in the previous section, the two-body interaction structure of the NJL model remains unchanged under the addition of the pseudovector bilinear term and the general form (\ref{njl_gap}) of the self-consistent Hartree approximation of the gap equation is kept intact.
The full fermion finite temperature propagator, which accounts for the axial bilinear modification, takes the form:
\begin{widetext}
\begin{align}
\begin{split}
&S(p_n)= (\slashed{p}_n \!+\mu\gamma^0-m-igB_\nu \gamma_5\gamma^\nu)^{-1} \\[3pt]
 &= 
\mfrac{
(\slashed{p}_n \!+\mu\gamma^0 +m + igB_\nu \gamma_5\gamma^\nu) \,
\{ (i \omega_n +\mu)^2 \!-\!\mathbf{p}^2 \!-\!m^2 \!-\!g^2 B^2
+2igm B_\nu \gamma_5 \gamma^\nu
+2ig [B_0 (i\omega_n +\mu) - \mathbf{B}\cdot\mathbf{p}] \gamma_5
\}
}{
[(i \omega_n +\mu)^2 -\mathbf{p}^2 -m^2 +g^2 B^2 ]^2
+4 g^2 \{
[B_0 (i\omega_n +\mu) - \mathbf{B}\cdot\mathbf{p}]^2
-B^2 [(i\omega_n+\mu)^2 - \mathbf{p}^2]
\}
}
,
\end{split}
\end{align}
and thus
\begin{align}
\label{PT_propagator_trace}
\begin{split}
\text{tr}[S(\omega_n,{\bf p})] =& \,
\mfrac{
4m [(i \omega_n +\mu)^2 -\mathbf{p}^2 -m^2 +g^2 B^2 ]
}{
[(i \omega_n +\mu)^2 -\mathbf{p}^2 -m^2 +g^2 B^2 ]^2
+4 g^2 \{
[B_0 (i\omega_n +\mu) - \mathbf{B}\cdot\mathbf{p}]^2
-B^2 [(i\omega_n+\mu)^2 - \mathbf{p}^2]
\}
}
, 
\end{split}
\end{align}
\end{widetext}
where $p_n= (i\omega_n,{\bf p})$, $\omega_n = (2n+1)\pi T$, and $B^2 = B_\nu B^\nu$, similar to the treatment within the four-momentum cutoff scheme \cite{fbk20,fk21}.

The gap equation of the modified NJL model now follows after evaluating the summation over the Matsubara frequencies.  
It can be performed analogously to the standard NJL model after a partial fraction decomposition of (\ref{PT_propagator_trace}). 
To this end one first determines the roots of the denominator, which is a depressed quartic polynomial,
\begin{equation}
\text{tr}[S(x,{\bf p})] = \frac{4m (x^2-a)}{(x^2-a)^2 - ( b x^2+ cx +d )}, 
\end{equation}
denoting $x = i\omega_n + \mu$, as well as 
\begin{alignat}{2}
\label{abcd}
a &= \mathbf{p}^2 +m^2 -\tilde{g}^2 (1- s^2), \,\,\,
&&b = -4\tilde{g}^2 s^2 , \notag \\[-8pt]
& && \\[-8pt]
\notag 
c &= 8\tilde{g}^2 s \,\lvert\bf{p}\rvert \cos\theta, 
&&d  = -4\tilde{g}^2 \mathbf{p}^2 (1  - s^2 \sin^2 \theta) .
\end{alignat}
For the latter parameters the angle $\theta$ between the spatial vectors $\bf{B}$ and $\bf{p}$ has been introduced, so that 
$\mathbf{B}\cdot \mathbf{p} = \lvert \mathbf{B}\rvert \, \lvert \bf{p} \rvert \, \cos \theta$, which remains an argument of the three-momentum integration in the gap equation (\ref{njl_gap}). 
Furthermore, $\tilde{g} = gB_0$ denotes the scaled bilinear coupling constant and the parameter $s = \lvert \mathbf{B} \rvert/B_0$ quantifies the characteristics of the background field $B^\nu$:
$s>1$ for a spacelike background, $s <1$ for a timelike background and $s=1$ for a lightlike background field.

Following Ferrari's method \cite{c68}, the depressed quartic denominator is expanded to the form 
\begin{equation}
[x^2-a+n]^2 - [ (2n+b) x^2+ cx +(d-2na +n^2) ], 
\end{equation}
where $n$ is then chosen to complete the square in the second term.
That is to say, $n$ satisfies the cubic equation
\begin{equation}
n^3 + n^2 \left(\mfrac{1}{2} b -2a\right) + n \left(d-ab\right) + 
\left(\mfrac{1}{2} bd- \mfrac{1}{8} c^2 \right)
=0 .
\end{equation}
Out of the three solutions determined by the cubic formula,
\begin{widetext}
\begin{equation}
n = \frac{4a-b}{6} - 
\frac{K_1}{3\,\bigl[\tfrac{1}{2} (K_2 + \sqrt{4K_1^3+K_2^2}\,)\bigr]^{1/3}} 
+
\mfrac{1}{3}\,\bigl[\tfrac{1}{2} (K_2 + \sqrt{4K_1^3+K_2^2}\,)\bigr]^{1/3} 
\end{equation}
is selected for simplicity, where 
\begin{equation}
K_1 = -4a^2\!-\!ab-\!\mfrac{1}{4}b^2\!+\!3d, \quad\,
K_2 = 16a^3 +6a^2b-\!\mfrac{3}{2}ab^2\!-\!\mfrac{1}{4}b^3\!+\mfrac{27}{8}c^2
\!-\!18ad -\!9bd.
\end{equation}
\end{widetext}
The denominator of (\ref{PT_propagator_trace}) thus takes the form
\begin{equation}
[x^2-a+n]^2 - \left[x\, \sqrt{2n+b} + \mfrac{c}{2\sqrt{2n+b}} \,\right]^2, 
\end{equation}
which is now factorized straightforwardly into 
\begin{equation}
(x+r_1)\,(x+r_2)\,(x-r_3)\,(x-r_4)
\end{equation}
with 
\begin{align}
\label{PT_roots}
r_{1,2} =& \mfrac{1}{2} \left (
\!\sqrt{2n+b} \mp \sqrt{b\!-\!2n\!+\!4a\!-\mfrac{2c}{\sqrt{2n+b}}} \,
\right),  \notag \\[5pt]
r_{3,4} =& \mfrac{1}{2} \left (\!
\sqrt{2n+b} \pm \sqrt{b\!-\!2n\!+\!4a\!+\mfrac{2c}{\sqrt{2n+b}}} \,
\right).
\end{align}
While not immediately apparent, $r_1$ and $r_2$, as well as $r_3$ and $r_4$, form complex conjugate pairs, as is to be expected, based on the complex conjugate root theorem, since the coefficients (\ref{abcd}) are real-valued expressions. 
\begin{figure}[!t]
\centering
\includegraphics[width=0.45\textwidth]
{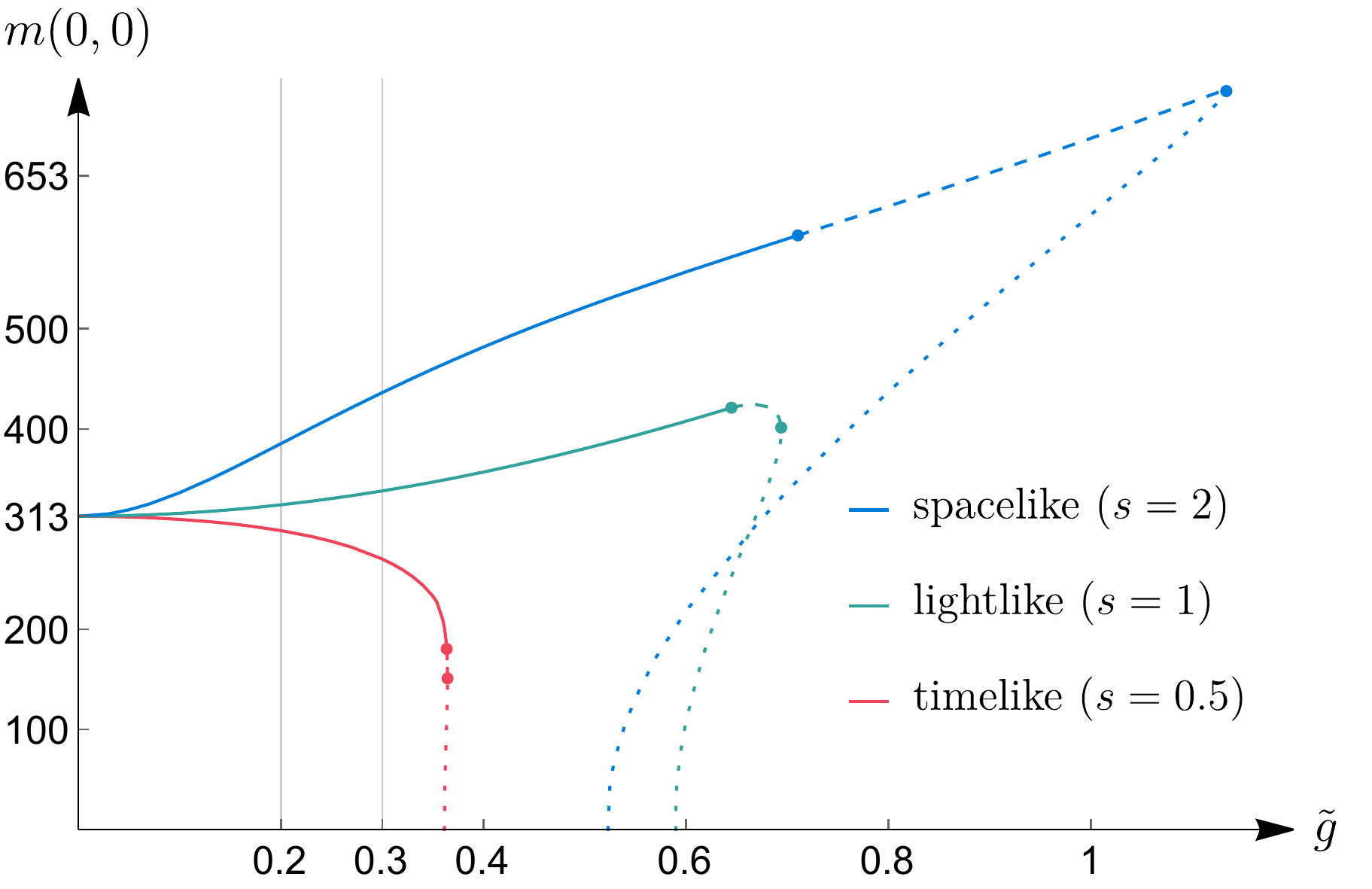}
\caption{
\label{f12}
Effective fermion mass $m$ of the modified NJL model at vanishing temperature and chemical potential as a function of the (scaled) coupling strength $\tilde{g}$ of the extension term for a space-, light-, and timelike background field case.
}
\end{figure}

Using this factorization, the partial fraction decomposition of the trace of the full fermion propagator (\ref{PT_propagator_trace}) has the form
\begin{widetext}
\begin{equation}
\mfrac{1}{4m}
\text{tr}[S(\omega_n,{\bf p})] =\,
-\mfrac{\alpha_1}{i\omega_n +(r_1+\mu)} - \mfrac{\alpha_2}{i\omega_n +(r_2+\mu)}
+\mfrac{\alpha_3}{i\omega_n - (r_3-\mu)}+\mfrac{\alpha_4}{i\omega_n -(r_4-\mu)}
,
\end{equation}
\end{widetext}
with the coefficients
\begin{eqnarray}
\label{PT_root_coeffs}
\alpha_{1,2} &=& \frac{r_{1,2}^2-a}{(r_{1,2}-r_{2,1})(r_{1,2}+r_3)(r_{1,2}+r_4)} , \nonumber \\
\alpha_{3,4} &=& \frac{r_{3,4}^2-a}{(r_{3,4}+r_1)(r_{3,4}+r_2)(r_{3,4}-r_{4,3})} ,
\end{eqnarray}
obtained using the residue theorem.
As a consequence of the complex conjugate pair structure of (\ref{PT_roots}) the coefficients $\alpha_1$ and $\alpha_2$, as well as $\alpha_3$ and $\alpha_4$, are complex conjugate expressions. 

The summation over the Matsubara frequencies $\omega_n$ can now be performed in complete analogy to the standard NJL model, cf.~\cite{k92,m55,fw03}, resulting in the gap equation of the modified system in the chiral limit:
\begin{equation}
\label{PT_gap}
m =  4 GN_c N_f\,  m \sum_{i=1}^4 \int^\Lambda \hspace{-0.2cm} \mfrac{\mathrm{d}^3{\bf p}}{(2\pi)^3}\,
\Bigl[\,
\alpha_i \tanh\Bigl(\frac{r_i+\textrm{sgn}\mu}{2T}\Bigr)\Bigr]
,
\end{equation}
where sgn is $+1$ for $i=1,2$ and $-1$ for $i=3,4$.
Notice that this is a real-valued expression due to the complex conjugate pair structure of (\ref{PT_roots}) and (\ref{PT_root_coeffs}), facilitating the existence of real effective mass solutions.

Before discussing the behavior of the effective fermion mass $m$ at finite temperature and chemical potential, it is instructive to consider briefly the case of vanishing $T$ and $\mu$ and compare the results obtained within the three-momentum cutoff scheme used in this study, with those of the previously employed four-momentum cutoff regularization in \cite{fbk20,fk21}.
In particular, utilizing the parameter $s$ the behavior of both timelike and lightlike cases of the background field $B^\nu$ can be examined, whereas the discussion in \cite{fbk20,fk21} was restricted to the spacelike case only.
Their behavior is visualized in Fig.~\ref{f12} as a function of the scaled coupling strength $\tilde{g}$ for the illustrative values $s = 0.5$ (timelike), $s = 2$ (spacelike) and the lightlike case with $s = 1$.
The dynamical mass generation of the spacelike case at small coupling values 
reproduces the behavior observed within the four-momentum cutoff scheme in \cite{fbk20, fk21}. 
A mass increase relative to the standard NJL model is found furthermore in the lightlike case, as well as timelike cases close to the lightlike case.
However, sufficiently far within the timelike sector (small values of $s$) one instead observes a monotonic decrease of the effective fermion mass with increasing strength of the non-Hermitian extension term. Dynamical mass generation is then no longer possible.

\begin{figure*}[t]
\centering
\subfloat[]{
\includegraphics[width=0.45\textwidth]
{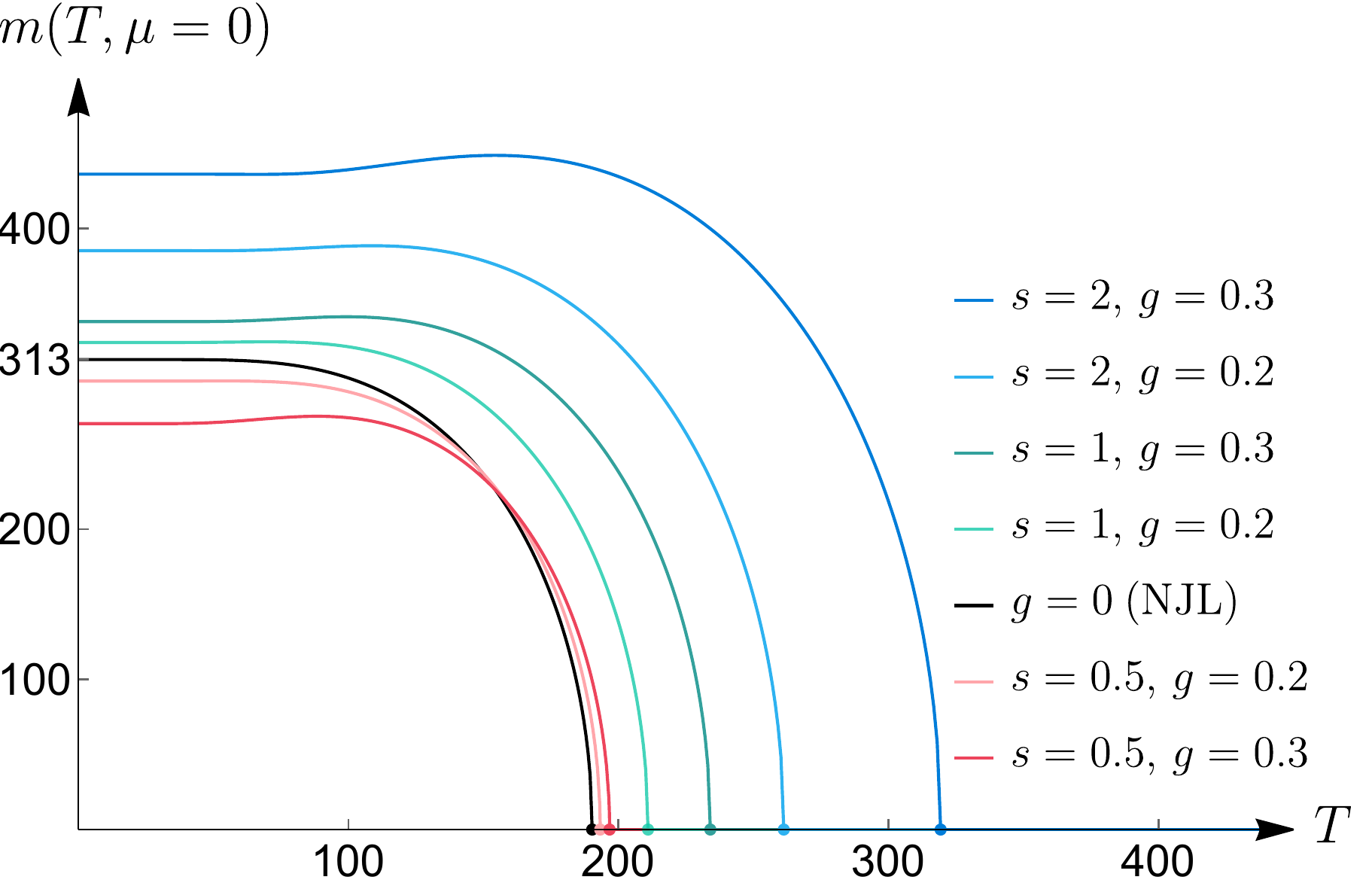}
\label{f13a}
}
\hfill
\subfloat[]{
\includegraphics[width=0.45\textwidth]
{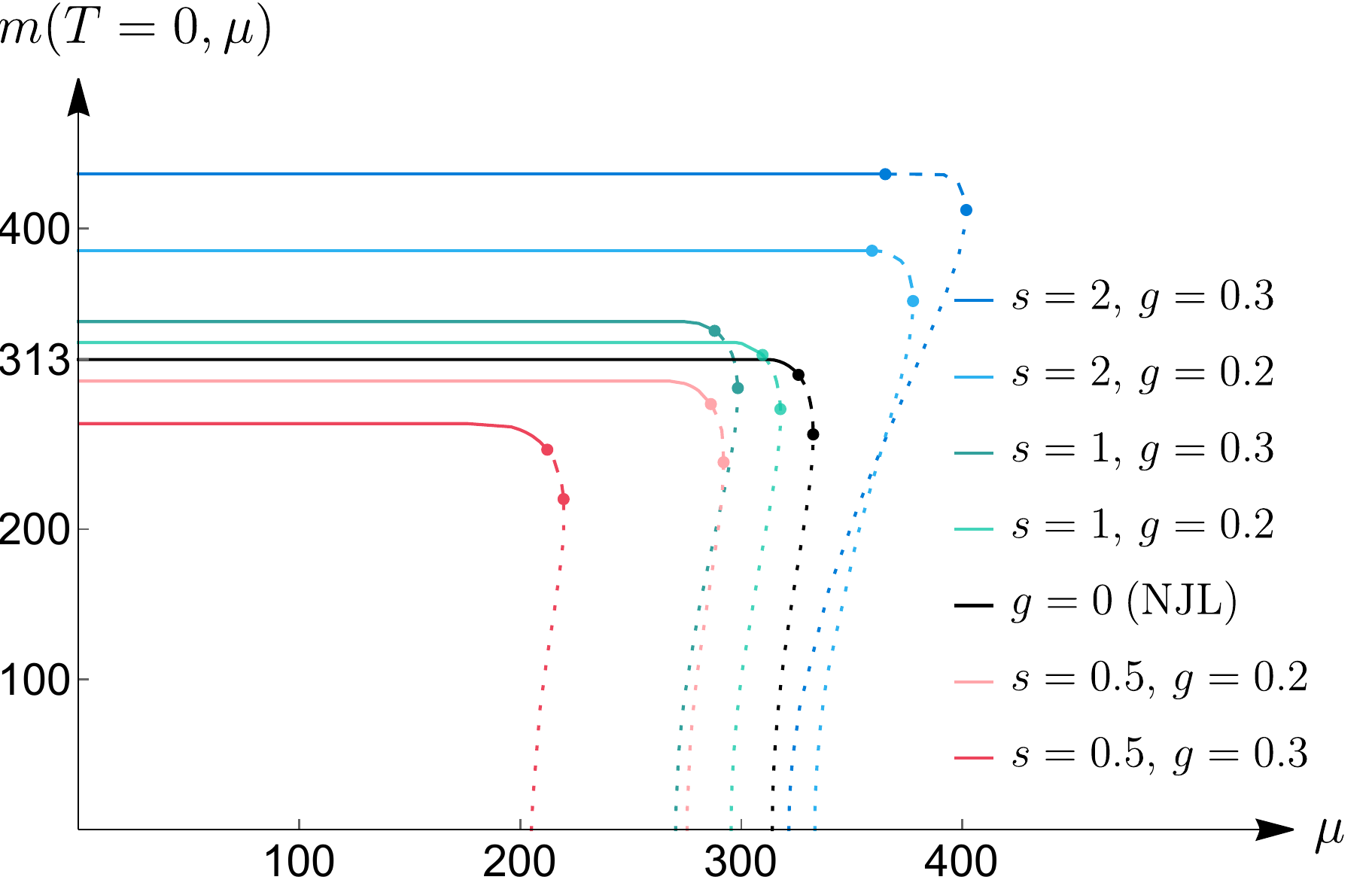}
\label{f13b}
 }
\caption{
\label{f13}
(a) Behavior of the effective fermion mass $m$ within the pseudovector extension of the NJL model in MeV as a function of the temperature $T$ at vanishing chemical potential $\mu$ for the bilinear coupling values $\tilde{g}= 0.2\Lambda$ and $\tilde{g}= 0.3\Lambda$ in a spacelike ($s=2$), the lightlike ($s=1$), and a timelike case ($s=0.5$).
(b)
Behavior of the effective mass $m$ as a function of the chemical at vanishing temperature $T$ potential $\mu$ for various values of the bilinear coupling strength $\tilde{g}$ and the parameter $s$. 
The stable physical solutions associated with the global minimum of $\Omega$ are shown as solid lines, while metastable and unstable solutions of the gap equation are shown as dashed and dotted lines respectively.
}
\end{figure*}

At sufficiently large coupling values $\tilde{g}$ the existence of real (nontrivial) fermion mass solutions, indicative of an unbroken $\cPT\!$-symmetric region, breaks down for any value $s$ and the system is realized in a  regime of spontaneously broken $\cPT$ symmetry instead.
Contrary to the behavior within the four-momentum cutoff scheme, where the system undergoes a continuous, second-order transition, one instead finds the $\cPT$ phase transition to be of first order within the three-momentum cutoff regularization for all parameters $\tilde{g}$ and $s$. 
(The position of the transition as well as the identification of stable (solid lines), metastable (dashed lines), and unstable mass solutions (dotted lines), see Fig.~\ref{f12}, is determined based on the thermodynamic potential $\Omega$, as described for finite $T$ and $\mu$ in the following.)
This difference between regularization schemes, together with the observation that the $\cPT$ transition typically 
occurs at comparably large coupling values relative to the cutoff length $\Lambda$, suggests that for the characterization of the $\cPT\!$-symmetry breaking phase transition of the modified NJL model, the self-consistent Hartree approximation is not robust in this regime. 
From a physical point of view, one might expect dynamical mass changes that are small, so that values $\tilde{g}\ll 1$ are important. Thus, the following discussion focuses on the system at small coupling values of the $\cPT$ extension term ($\tilde{g}= 0.2\Lambda$ and $\tilde{g}= 0.3\Lambda$), the study of its behavior at finite temperature and baryon chemical potential, and an examination of the effects of the $\cPT$ bilinear term on the chiral phase transition. 

Figure~\ref{f13a} presents the  effective fermion mass $m$, as determined by the self-consistent gap equation (\ref{PT_gap}) of the extended model, as a function of the temperature $T$ at vanishing chemical potential $\mu$. Illustrated are a spacelike ($s=2$, blue), a lightlike ($s=1$, green), and a timelike ($s=0.5$, red) case at coupling values $\tilde{g}= 0.2\Lambda$ and $\tilde{g}= 0.3\Lambda$.
As in the standard NJL model, visualized in black, the system undergoes a continuous second-order chiral phase transition, beyond which chiral symmetry is restored and the effective mass vanishes.
The position $T^c(\tilde{g}, s)$ of this transition increases with the coupling $\tilde{g}$ and with increasing values of $s$, see also Table \ref{t03}. 
In all cases, the extension of the system through the inclusion of the $\cPT\!$-symmetric non-Hermitian term results in a raised critical temperature compared to the standard NJL model.
An effective increase of the fermion mass relative to the standard NJL result is found at small finite values of the temperature as well, when $s=2$ and $s=1$, in agreement with the behavior of $m(T=0,\mu=0)$ shown in Fig.~\ref{f12}.
For $s=0.5$, however, an effective mass {\it loss} due to the extension term is observed.
A notable difference to the behavior of the effective mass in the standard NJL model is the fact that it does not decrease monotonically with increasing temperature $T$. Instead, $m$  increases  initially to reach a maximum, before decreasing to vanish at the transition temperature $T^c$.

When the gap equation (\ref{PT_gap}) is evaluated as a function of the chemical potential $\mu$ at vanishing temperature $T$, the behavior of the effective mass qualitatively resembles that obtained within the standard NJL model, see Fig.~\ref{f13b}. Shown are again a spacelike ($s=2$, blue), the lightlike ($s=1$, green), and a timelike ($s=1/2$, red) case at the coupling values $\tilde{g}= 0.2\Lambda$ and $\tilde{g}= 0.3\Lambda$, in addition to the standard NJL model behavior in black.
The effective fermion mass does not decrease to vanish continuously. 
Instead a parametric region with multiple mass solutions is found and the physical stable result has to be identified again using the thermodynamic potential $\Omega(T,\mu,\tilde{g}, s)$.
It can be determined from the thermodynamic average of the interaction energy in analogy to the discussion for the standard NJL model, cf.~(\ref{coupling_integration_method}), by following a coupling-constant integration method. 
As in the case of the pseudoscalar extension, the non-Hermitian pseudovector bilinear term does not affect the approach structurally. But the substitution of the effective mass within the coupling-constant integral relies on the modified gap equation (\ref{PT_gap}), through which the extension enters implicitly:
\vspace*{-0.14cm}
\begin{widetext}
\begin{eqnarray}
 2\int_0^1  \mfrac{\mathrm{d}\lambda}{\lambda} \,(m_\lambda-m_0) \mfrac{\mathrm{d}m_\lambda}{\mathrm{d}\lambda} 
&= &
4 G N_c N_f\, T\, \int^\Lambda  \mfrac{\mathrm{d}^3{\bf p}}{(2\pi)^3}\,
\sum_{j= 1}^4
\int_{x_j(0)}^{x_j(1)} \hspace{-0.2cm} \mathrm{d}x_j \, 
\bigl[4m \,\alpha_j \Bigl(\mfrac{\mathrm{d}r_j}{\mathrm{d}m} 
\Bigr)^{-1} \bigr] 
\tanh(x_j) \nonumber
\\
& =&
4 G N_c N_f\, T\, \int^\Lambda \hspace{-0.2cm} \mfrac{\mathrm{d}^3{\bf p}}{(2\pi)^3}\,
\ln \Bigg(\,
\frac{\cosh[x_1(1)] \, \cosh[x_2(1)] \,\cosh[x_3(1)] \,\cosh[x_4(1)] 
}{
\cosh[x_1(0)] \,\cosh[x_2(0)] \,\cosh[x_3(0)] \,\cosh[x_4(0)]}
\,\Bigg) 
,\nonumber
\end{eqnarray}
\end{widetext}
where $x_1(\lambda)= ({r_1(\lambda)+\mu})/{2T}$, $x_2(\lambda)= (r_2(\lambda)+\mu)/{2T}$, $x_3(\lambda)= (r_3(\lambda)-\mu)/{2T}$, and $x_4(\lambda)= (r_4(\lambda)-\mu)/{2T}$. The $\lambda$-dependence refers to 
the use of the mass result $m_\lambda$, which solves the $\lambda$-dependent equivalent of the gap equation (\ref{PT_gap}), where $G\rightarrow \lambda G$ within the coupling-constant integration method.
One thus obtains the formal equivalent of the relation (\ref{njl_potential_coupling_integrated}), 
which establishes the thermodynamic 
potential $\Omega(T,\mu,\tilde{g},s)$ after subtracting off the contribution 
$\Omega_0 $ of the non-Hermitian free theory obtained at $\lambda =0$:
\begin{align}
\label{PT_potential}
\Omega(T,\mu,\tilde{g}, s)=\,&
\mfrac{(m-m_0)^2}{4G} 
-\frac 12 N_c N_f 
\sum_{i=1}^4\int^\Lambda  \mfrac{\mathrm{d}^3{\bf p}}{(2\pi)^3}\, 
{r_i}
\notag
\\
&\!\!\!\!\!\!
-T\, N_c N_f 
\int^\Lambda \hspace{-0.2cm} \mfrac{\mathrm{d}^3{\bf p}}{(2\pi)^3}\,
\ln\prod_{i=1}^4
\bigl[1\!+\mathrm{e}^{-(r_i+\textrm{sgn}\mu)/T}\,\bigr]
,
\end{align}
where once again sgn is $+1$ for $i=1,2$ and $-1$ for $i=3,4$.
Like the gap equation (\ref{PT_gap}), which is 
recovered from the extremal condition $d\Omega/dm = 0$ in the limit of vanishing bare mass $m_0$, the thermodynamic potential (\ref{PT_potential})
is a real-valued expression due to the complex conjugate pair structure of (\ref{PT_roots}).
Notably, this property is unaffected by a removal of the three-momentum cutoff limit $\Lambda$ in the logarithmic integral contribution for the comparison to the SB limit.

\begin{table*}[]
\centering
\setlength\tabcolsep{6.8pt}
\renewcommand{\arraystretch}{1.4}
\small    
\begin{tabular}{ |c||c|c|c|c|c|c|c| }  
 \cline{2-8}
 \multicolumn{1}{c||}{} & 
 \shortstack{\\$\tilde{g}=0$ \\ (NJL)} & 
 \shortstack{$s = 0.5,$\\$\tilde{g} = 0.2\Lambda$} &  
 \shortstack{$s = 0.5,\,$\\$\tilde{g} = 0.3\Lambda$} & 
 \shortstack{$s = 1,\,$\\$\tilde{g} = 0.2\Lambda$} & 
 \shortstack{$s = 1,\,$\\$\tilde{g} = 0.3\Lambda$} & 
 \shortstack{$s = 2,\,$\\$\tilde{g} = 0.2\Lambda$} & 
 \shortstack{$s = 2,\,$\\$\tilde{g} = 0.3\Lambda$} \\ 
 \hline
 \hline 
$T^c (\mu=0)$ &
$190$~MeV & $193 $~MeV & $197 $~MeV & $211 $~MeV & $234 $~MeV & $261 $~MeV & $ 319$~MeV\\ 
 \hline
$\mu^c (T=0)$ &
$326$~MeV  & $ 286$~MeV & $212 $~MeV &
$310 $~MeV  & $288 $~MeV &  $359 $~MeV & $365 $~MeV \\
\hline
\end{tabular}

\caption{
Phase transition temperatures $T^c (\mu=0, \tilde{g},s)$ at vanishing chemical potential and transition chemical potentials $\mu^c (T=0, \tilde{g},s)$ at vanishing temperature for various coupling strengths $\tilde{g}$ and parameters $s$ of the non-Hermitian extension term.
}
\label{t03}
\end{table*}


As in the previous sections, the stable physical fermion mass result is determined as the global minimum of the thermodynamic potential $\Omega(T,\mu,\tilde{g}, s)$ under variation of the fermion mass $m$. Meanwhile, local minima characterize metastable solutions and maxima correspond to unstable solutions of the gap equation.
These properties of the fermion mass are shown in Fig.~\ref{f13b} as solid, dashed, and dotted lines respectively. The position $\mu^c$ of the first-order chiral phase transition, accompanied by an abrupt transition to a vanishing fermion mass, is visualized as a dot and its values are listed in Table~\ref{t03}. 
Notice that with increasing coupling strength $\tilde{g}$ of the non-Hermitian modification term, this transition moves to larger values of the chemical potential in the spacelike case with $s=2$, while decreasing in the lightlike case, where $s=1$, and the timelike case with $s = 0.5$. 
Similar to Fig.~\ref{f13a}
an effective mass increase relative to the standard NJL result is found when $s=2$ and $s=1$, while an effective mass loss arises for $s=0.5$,
in agreement with the behavior of $m(T=0,\mu=0)$ shown in Fig.~\ref{f12}.

\begin{figure}[]
\captionsetup[subfigure]{labelformat=empty}
\centering
\subfloat[]{
\includegraphics[width=0.42\textwidth]
{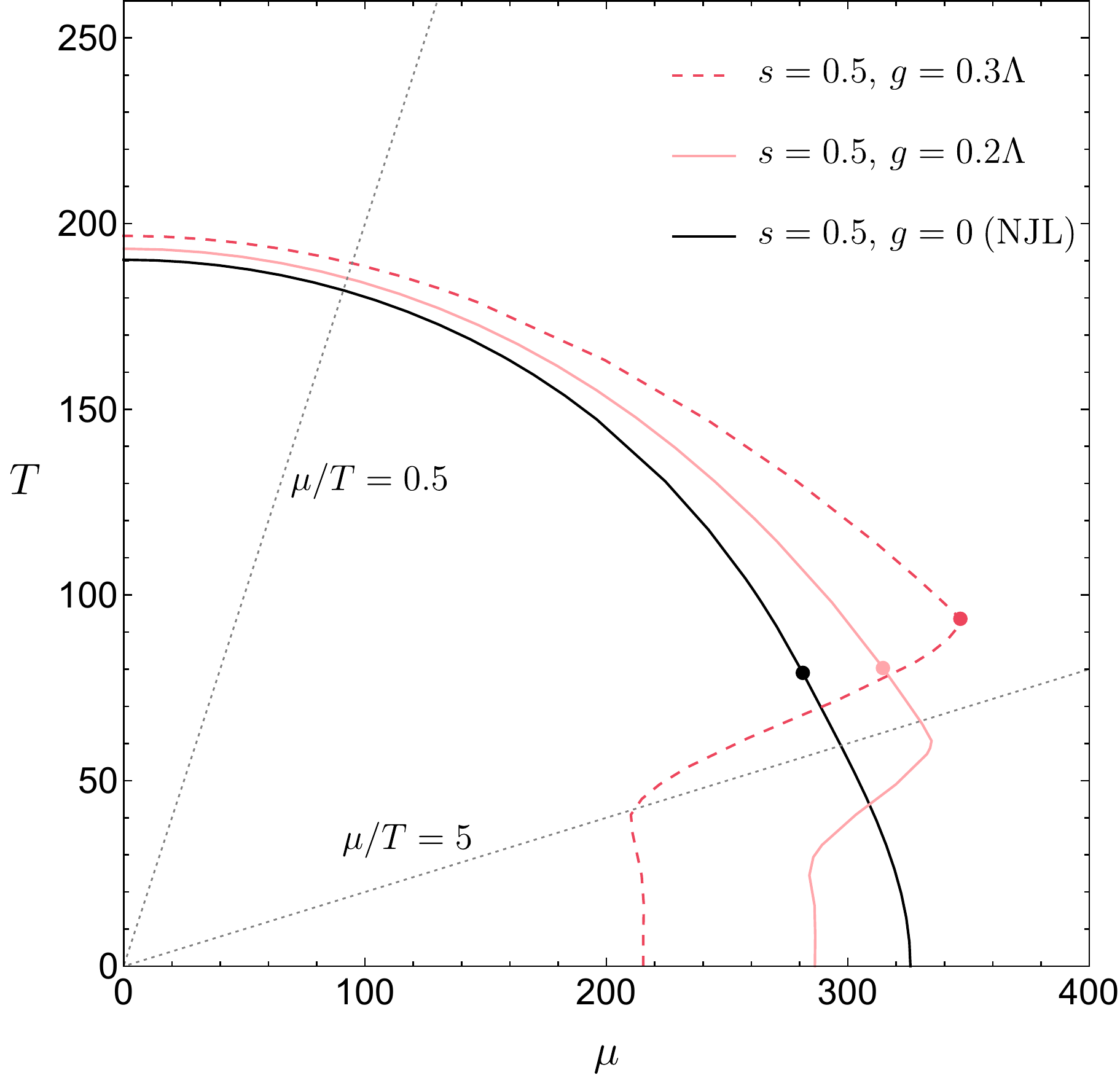}
\label{f14a}
}
\hfill
\subfloat[]{
\includegraphics[width=0.42\textwidth]
{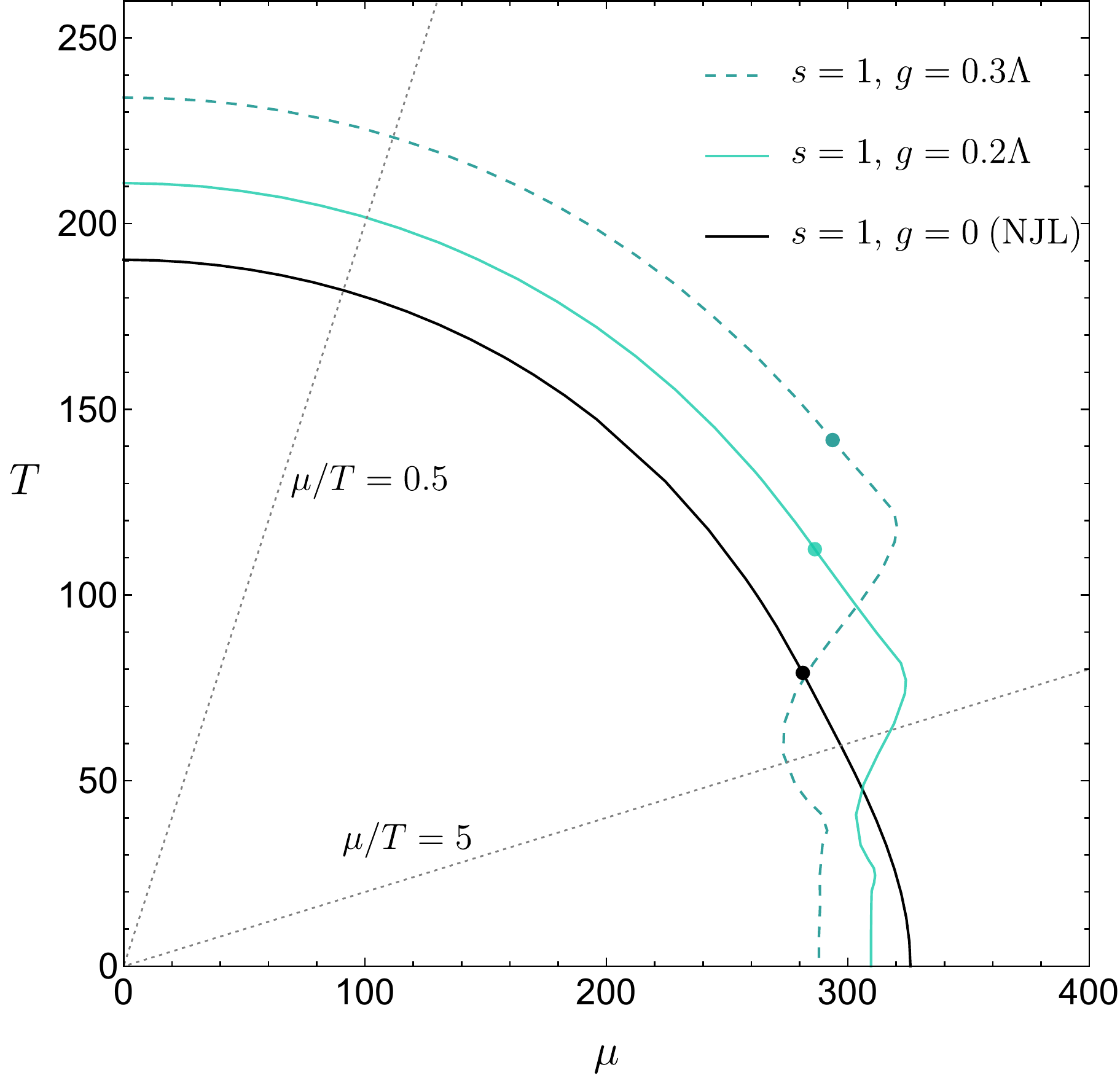}
\label{f14b}
}
\hfill \\
\vspace*{-0.55cm}
\subfloat[]{
\includegraphics[width=0.41\textwidth]
{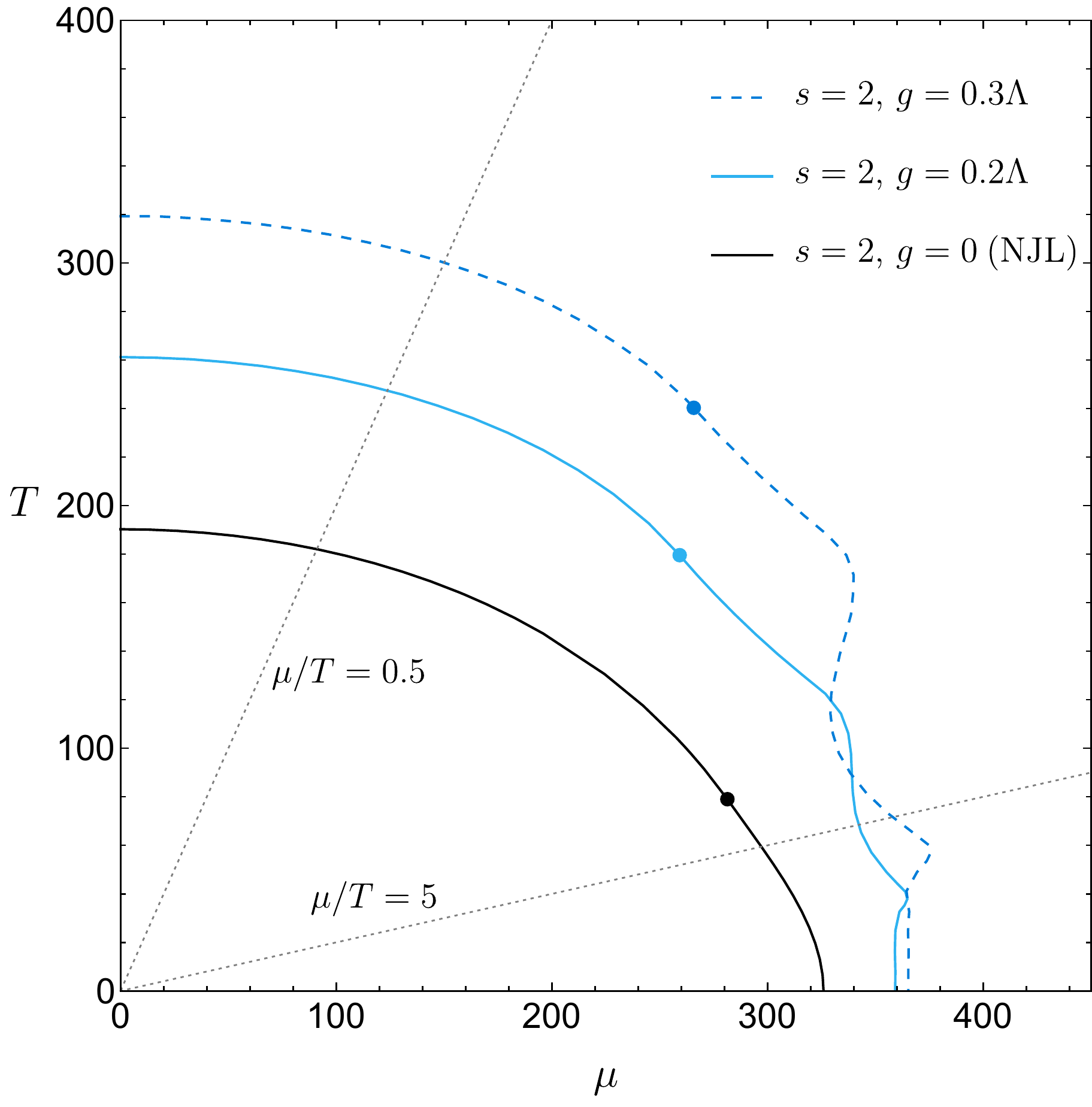}
\label{f14c}
}
\caption{
\label{f14}
Phase diagrams of the modified NJL model in the $T$-$\mu$--plane for coupling values $\tilde{g} = 0.2 \Lambda$ and $\tilde{g} = 0.3 \Lambda$ in the timelike case with $s = 0.5$, the lightlike case ($s=1$), and the spacelike case with $s=2$.
}
\end{figure}

The overall behavior of the chiral phase transition within the $T$-$\mu$--plane is visualized in Figs.~\ref{f14} for the timelike case with $s=0.5$,  the lightlike case ($s=1$), and  the spacelike case with $s=2$.
Shown is the boundary between the spontaneously broken and restored regions for coupling strengths $\tilde{g}= 0.2\Lambda$ (solid line) and $\tilde{g}= 0.3\Lambda$ (dashed line) of the non-Hermitian extension, as well as the standard NJL model case in black.
The phenomenological behavior along the $T$ and $\mu$ axes, as shown in Fig.~\ref{f13}, is generally continued into the $T$-$\mu$--plane:
For small chemical potentials, the model undergoes a second-order chiral phase transition at sufficiently large temperatures; the transition temperature increases with the coupling $\tilde{g}$ and with the value $s$ quantifying the space- or timelikeness of the non-Hermitian background.
At small temperatures, a first-order chiral phase transition is found at sufficiently large chemical potentials; the transition chemical potential decreases with increasing $\tilde{g}$ for $s= 0.5$ and $s=1$, but increases in the spacelike case with $s=2$.
The respective critical end-points, marking the change from a second-order to a first-order transition behavior, are illustrated as dots and their position is listed in Table~\ref{t3}.
The CEP moves to higher values of the temperature and chemical potential with an increasing bilinear coupling $\tilde{g}$; for increasing value $s$, the  chemical potential $\mu_{\text{CEP}}$ decreases, while the temperature $T_\text{\text{CEP}}$ increases.

\begin{table*}[]
\centering
\renewcommand{\arraystretch}{1.4}
\small    
\begin{tabular}{ |c||c|c|c|c|c|c|c| }  
 \cline{2-8}
 \multicolumn{1}{c||}{} & 
\shortstack{\\$\tilde{g}=0$ \\ (NJL)} & 
 \shortstack{$s = 0.5,$\\$\tilde{g} = 0.2\Lambda$} &  
 \shortstack{$s = 0.5,\,$\\$\tilde{g} = 0.3\Lambda$} & 
 \shortstack{$s = 1,\,$\\$\tilde{g} = 0.2\Lambda$} & 
 \shortstack{$s = 1,\,$\\$\tilde{g} = 0.3\Lambda$} & 
 \shortstack{$s = 2,\,$\\$\tilde{g} = 0.2\Lambda$} & 
 \shortstack{$s = 2,\,$\\$\tilde{g} = 0.3\Lambda$} \\  
 \hline
 \hline 
$\mu_{\text{CEP}}$ &
$281$~MeV & $315$~MeV & $347$~MeV& $286$~MeV & $294$~MeV& $259$~MeV & $266$~MeV
\\ 
 \hline
$T_{\text{CEP}}$ &  
$79$~MeV & $80$~MeV & $94$~MeV & $112$~MeV & $141$~MeV & $180$~MeV & $240$~MeV 
\\
\hline
$(\mu/T)_{\text{CEP}}$ &
$3.56$ &  $3.92$ & $3.70$ & $ 2.55$& $2.07 $&  $ 1.44$&  $ 1.11 $
\\
\hline
\end{tabular}

\caption{
Temperature and chemical potential of the critical end-point of the modified NJL model for various coupling strengths $\tilde{g}$ and parameters $s$ of the non-Hermitian extension term.
}
\label{t3}
\end{table*}

Beyond the identification of the physical fermion mass at finite temperature and chemical potential, the thermodynamic potential (\ref{PT_potential}) of the modified NJL model allows for the study of the thermodynamic observables.
Again the approach parallels that within the standard NJL model, leading to the quark number density
\begin{eqnarray}
\label{PT_quarknumber}
n(T,\mu,\tilde{g}, s) &=&  -\frac{\partial \,\Omega(T,\mu,\tilde{g}, s)}{\partial \mu} \,\Bigr\rvert_T 
\\
&= & \,
N_c N_f \int^\Lambda \hspace{-0.2cm} \mfrac{\mathrm{d}^3{\bf p}}{(2\pi)^3}\, 
\mfrac{1}{2} \sum_{i=1}^4\Bigl[\,
\textrm{sgn}\tanh\Bigl(\frac{r_i+\textrm{sgn}\mu}{2T}\Bigr)\Bigr]
, \nonumber
\end{eqnarray}
the entropy density
\begin{align}
\label{PT_entropy}
\begin{split}
s(T,\mu,\tilde{g}, s) =& -\frac{\partial \,\Omega(T,\mu,\tilde{g}, s)}{\partial T} \,\Bigr\rvert_\mu
\\
=& \, 
2 N_c N_f \!\!\int^\Lambda \hspace{-0.2cm} \mfrac{\mathrm{d}^3{\bf p}}{(2\pi)^3}\,
\Bigl\{
\mfrac1{4T} {\sum_{i=1}^4r_i} \\
&+
\mfrac{1}{2} 
\ln\Bigl(\prod_{i=1}^4
\bigl[1+\mathrm{e}^{-(r_i+\textrm{sgn}\mu)/T}\,\bigr]\!
\Bigr)
\\[4pt]
&
\hspace*{0.1cm}
-
\mfrac{1}{2} 
\sum_{i=1}^4\Bigl[\,
\mfrac{r_i+\textrm{sgn}\mu}{2T} \tanh\Bigl(\mfrac{r_i+\textrm{sgn}\mu}{2T}\Bigr) 
\Bigr]
\Bigr\}
,
\end{split}
\end{align}
and the pressure density
\begin{equation}
\label{PT_pressure}
p(T,\mu,\tilde{g}, s) \!=\! - \bigl[ \Omega(T,\mu,\tilde{g}, s) \!-\! \Omega(0,0, \tilde{g}, s) \bigr].
\end{equation}
The energy density and interaction measure are then determined through the well-established relations
\begin{align}
\label{PT_energy}
\begin{split}
\epsilon(T,\mu, \tilde{g}, s) =&
-p(T,\mu, \tilde{g}, s) + T s(T,\mu, \tilde{g}, s) +\mu n(T,\mu, \tilde{g}, s)
\end{split}
\end{align}
and
\begin{align}
\label{PT_anomaly}
\begin{split}
 I(T,\mu,  \tilde{g}, s) =
 \epsilon(T,\mu, \tilde{g}, s)-3\,p(T,\mu, \tilde{g}, s)
 .
\end{split}
\end{align}
As with the thermodynamic potential  (\ref{PT_potential}) itself, these quantities form real-valued expressions, due to the complex conjugate pair structure of (\ref{PT_roots}).

Moreover, the bilinear coupling $g$ ultimately always enters in combination with the temperature as $g/T$ in (\ref{PT_quarknumber}) to (\ref{PT_anomaly}),
since the terms $r_1$ to $r_4$ in (\ref{PT_roots}) enter as $r/T$, with $g$ contained in the parameters (\ref{abcd}). 
The large temperature limit 
therefore remains unchanged by the inclusion of the non-Hermitian $\cPT\!$-symmetric pseudovector bilinear extension
along lines of fixed $\mu/T$ and when removing the cutoff scale $\Lambda \rightarrow \infty$ (except for the UV-divergent term in $\Omega$ and the related pressure and energy densities as well as the interaction measure).
The same SB limits of an ideal massless fermion gas  are found as in the standard NJL model, see (\ref{njl_SBlimit_potential}) -- (\ref{njl_SBlimit_entropy}).

\begin{figure*}
\centering
\subfloat[]{
\includegraphics[width=0.48\textwidth]
{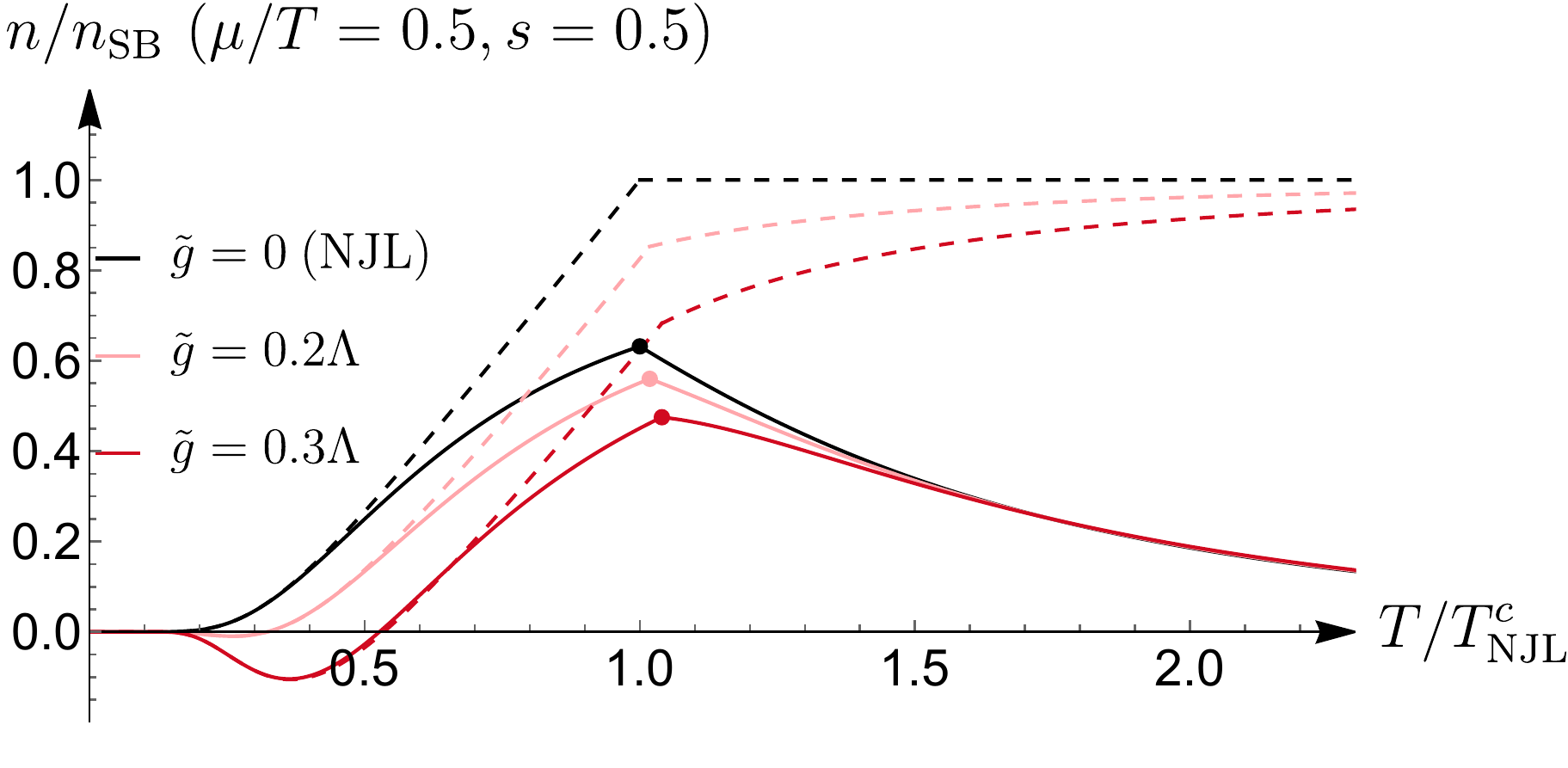}
\label{f15a}
}
\subfloat[]{
\includegraphics[width=0.48\textwidth]
{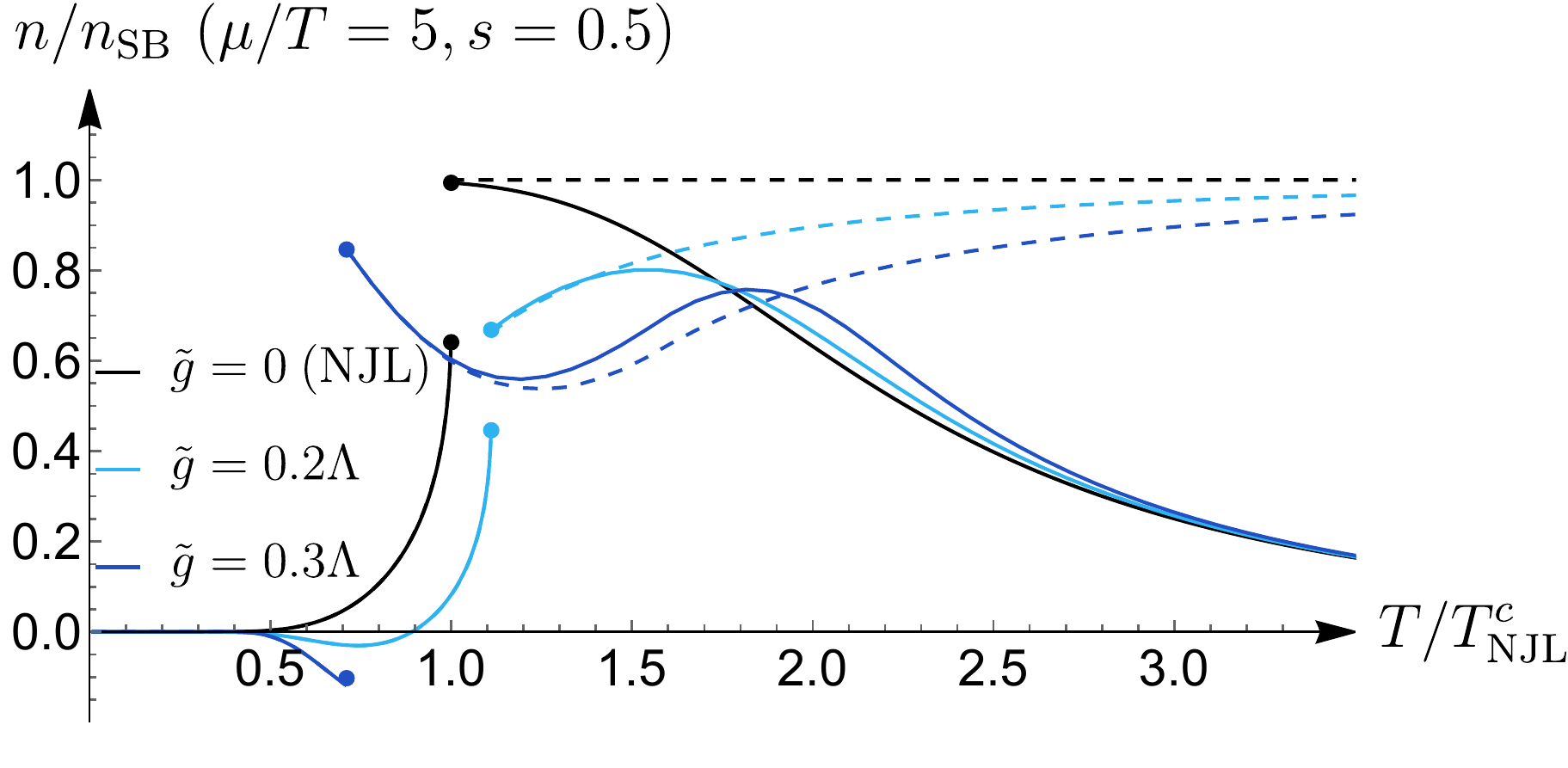}
\label{f15a2}
}
\hfill\\
\vspace*{-0.5cm}
\subfloat[]{
\includegraphics[width=0.48\textwidth]
{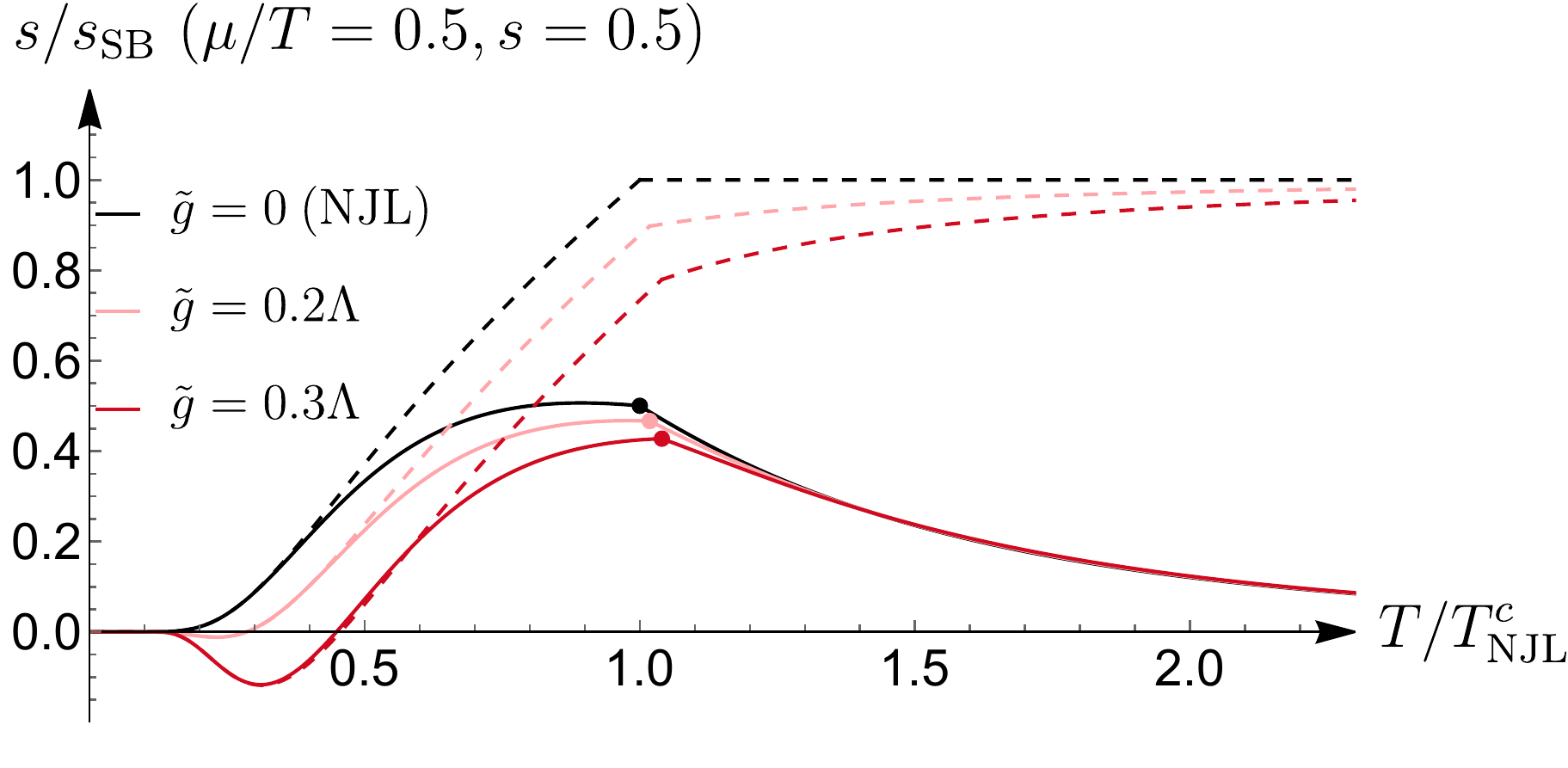}
\label{f15b}
}
\subfloat[]{
\includegraphics[width=0.48\textwidth]
{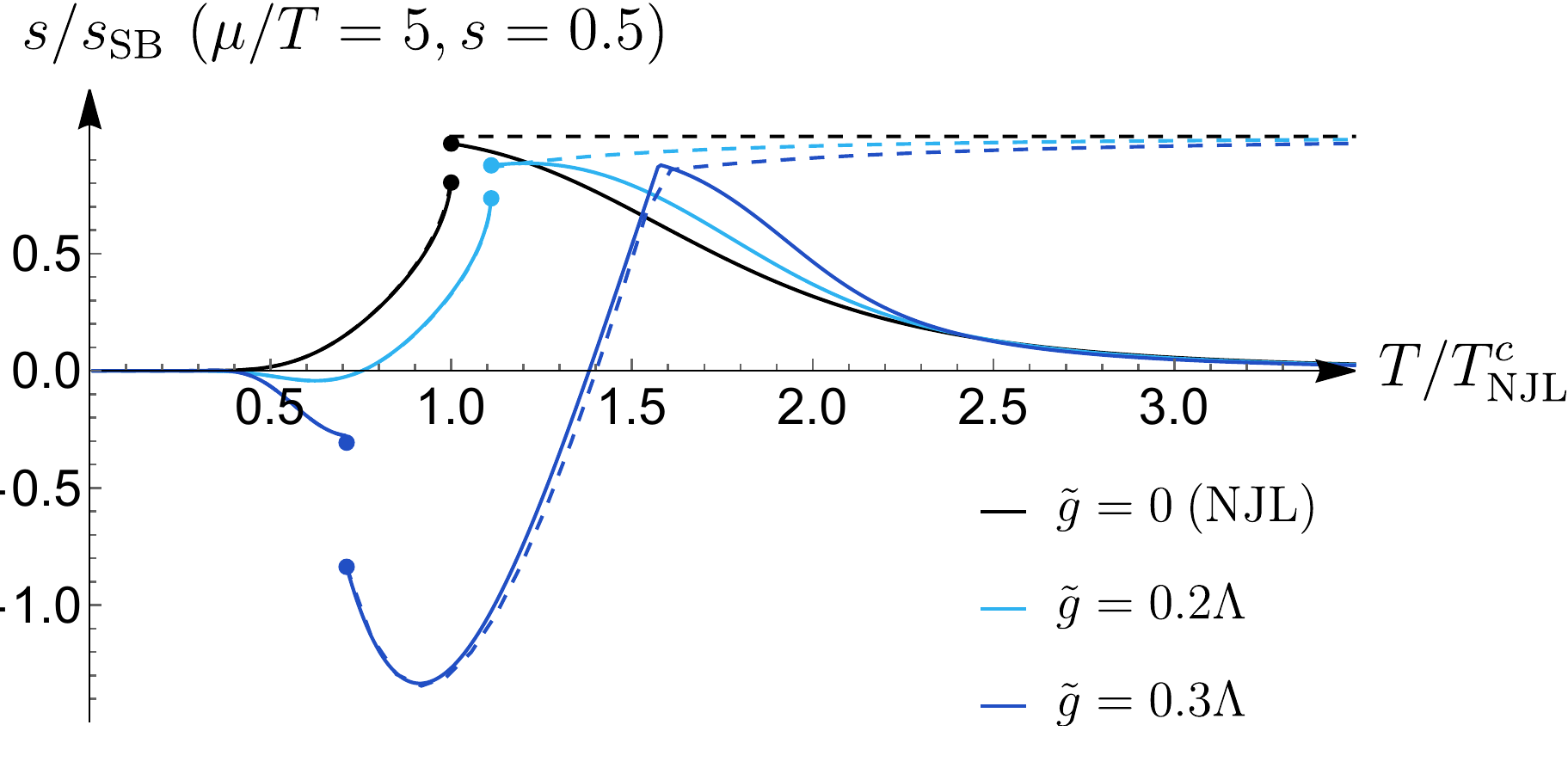}
\label{f15b2}
}
\hfill\\
\vspace*{-0.5cm}
\subfloat[]{
\includegraphics[width=0.48\textwidth]
{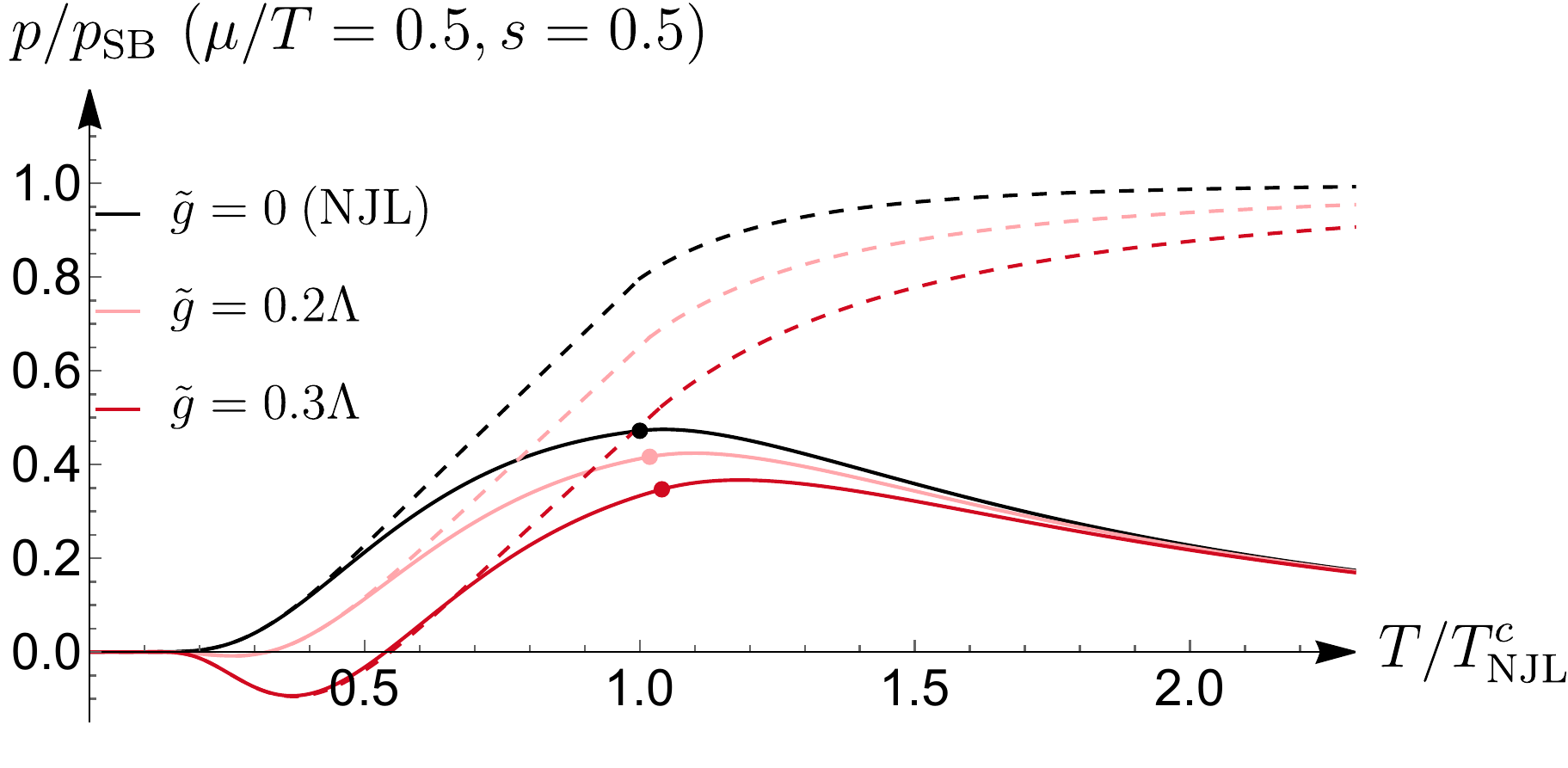}
\label{f15c}
}
\subfloat[]{
\includegraphics[width=0.48\textwidth]
{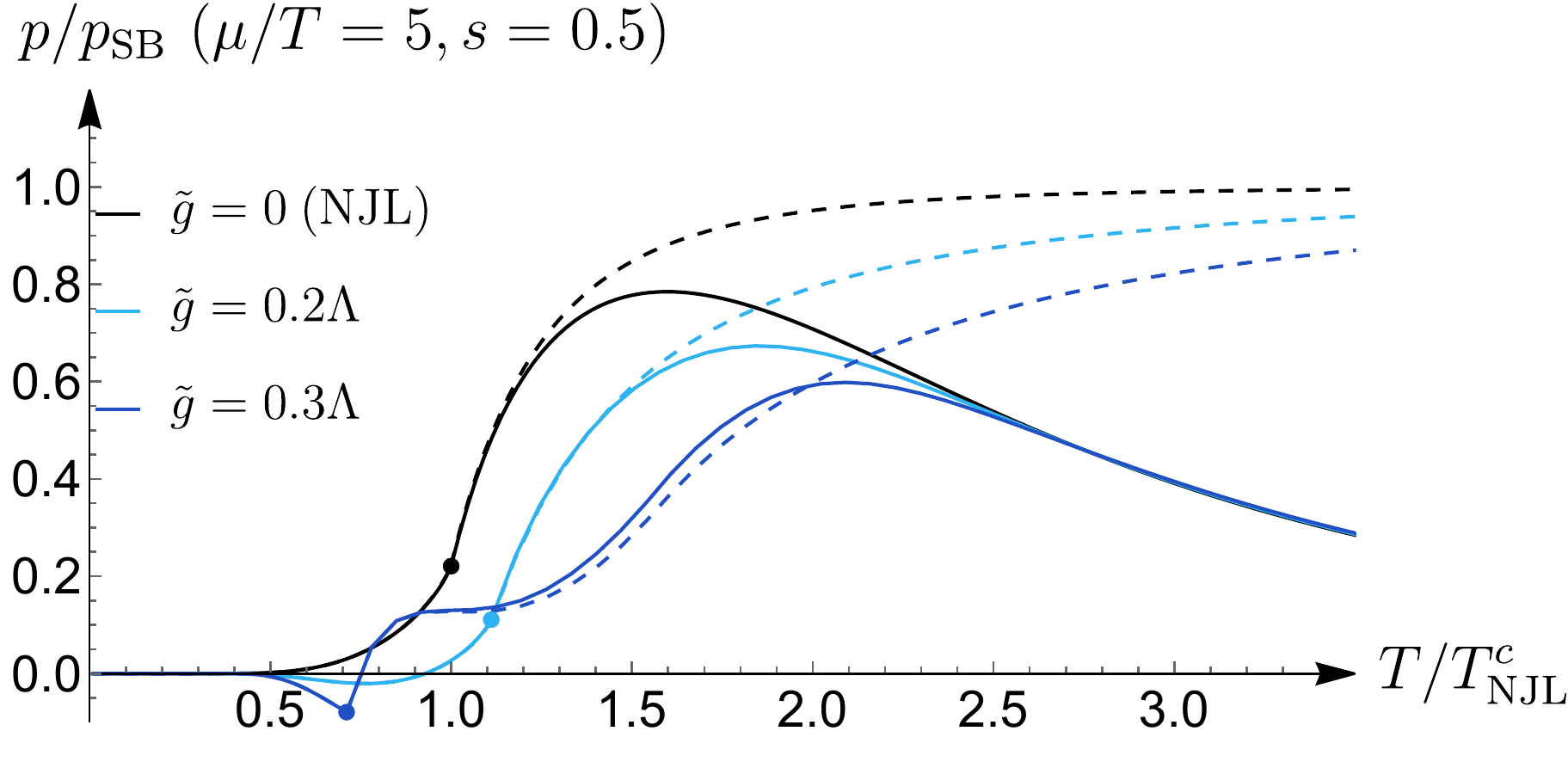}
\label{f15c2}
}
\hfill\\
\vspace*{-0.5cm}
\subfloat[]{
\includegraphics[width=0.48\textwidth]
{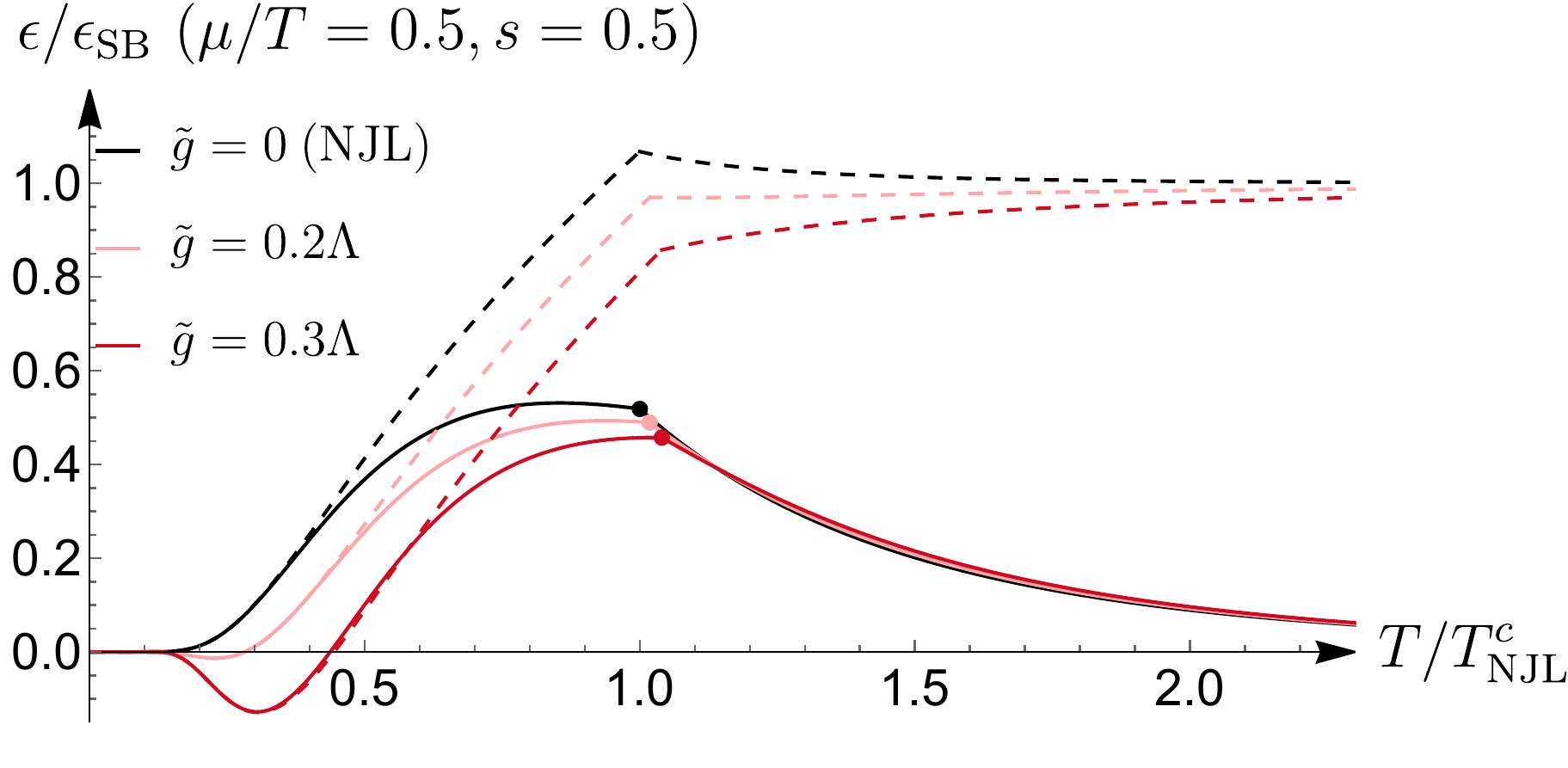}
\label{f15d}
}
\subfloat[]{
\includegraphics[width=0.48\textwidth]
{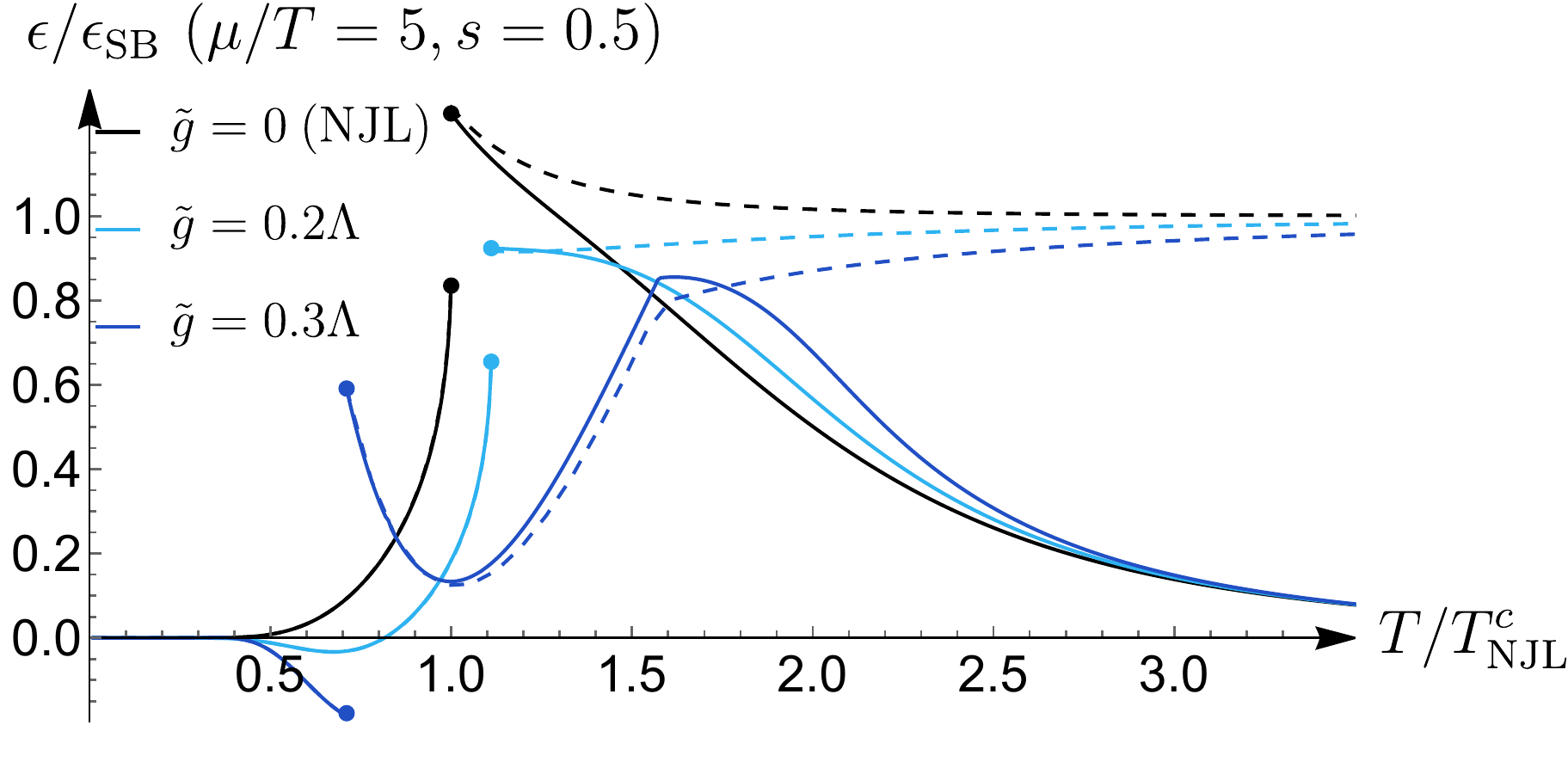}
\label{f15d2}
}
\hfill\\
\vspace*{-0.5cm}
\subfloat[]{
\includegraphics[width=0.48\textwidth]
{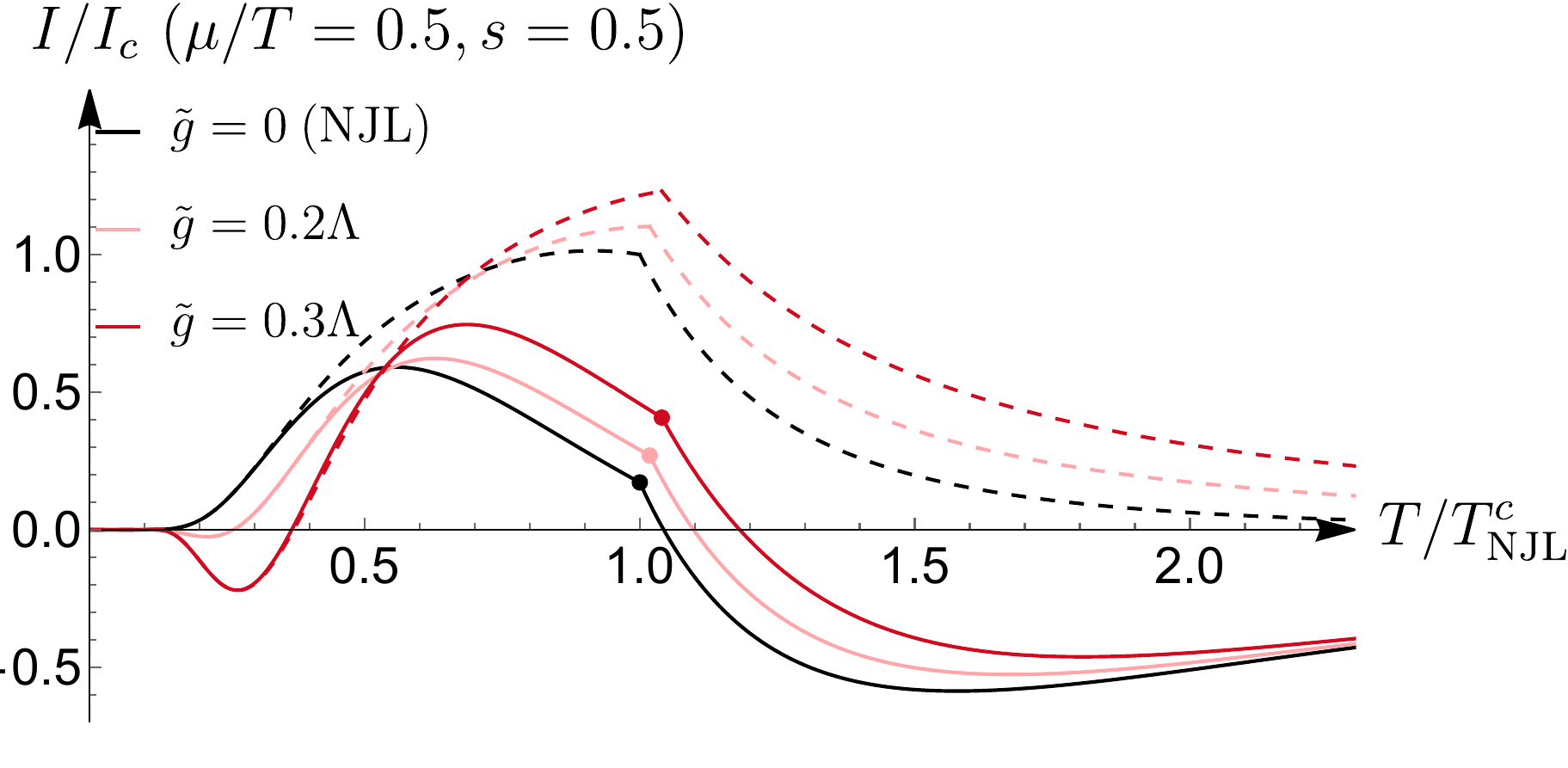}
\label{f15e}
}
\subfloat[]{
\includegraphics[width=0.48\textwidth]
{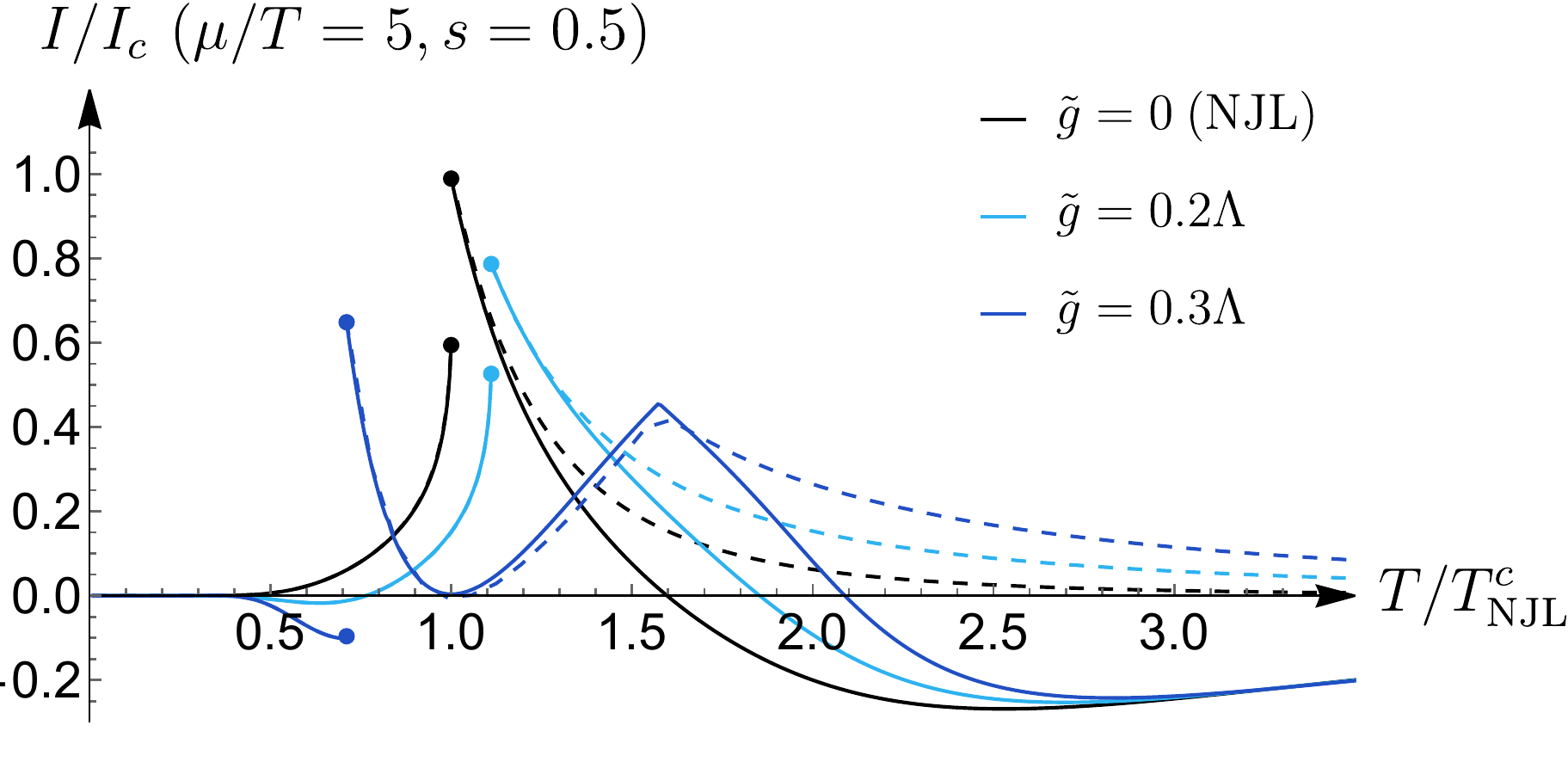}
\label{f15e2}
}
\caption{
\label{f15}
Thermodynamic functions for $s= 0.5$ (timelike) along $\mu/T = 0.5$ (red) and $\mu/T = 5$ (blue) at couplings $\tilde{g} = 0.2\Lambda$ (light colors) and $\tilde{g}= 0.3 \Lambda$ (dark colors). Standard NJL case in black.
}
\end{figure*}

\begin{figure*}
\centering
\subfloat[]{
\includegraphics[width=0.48\textwidth]
{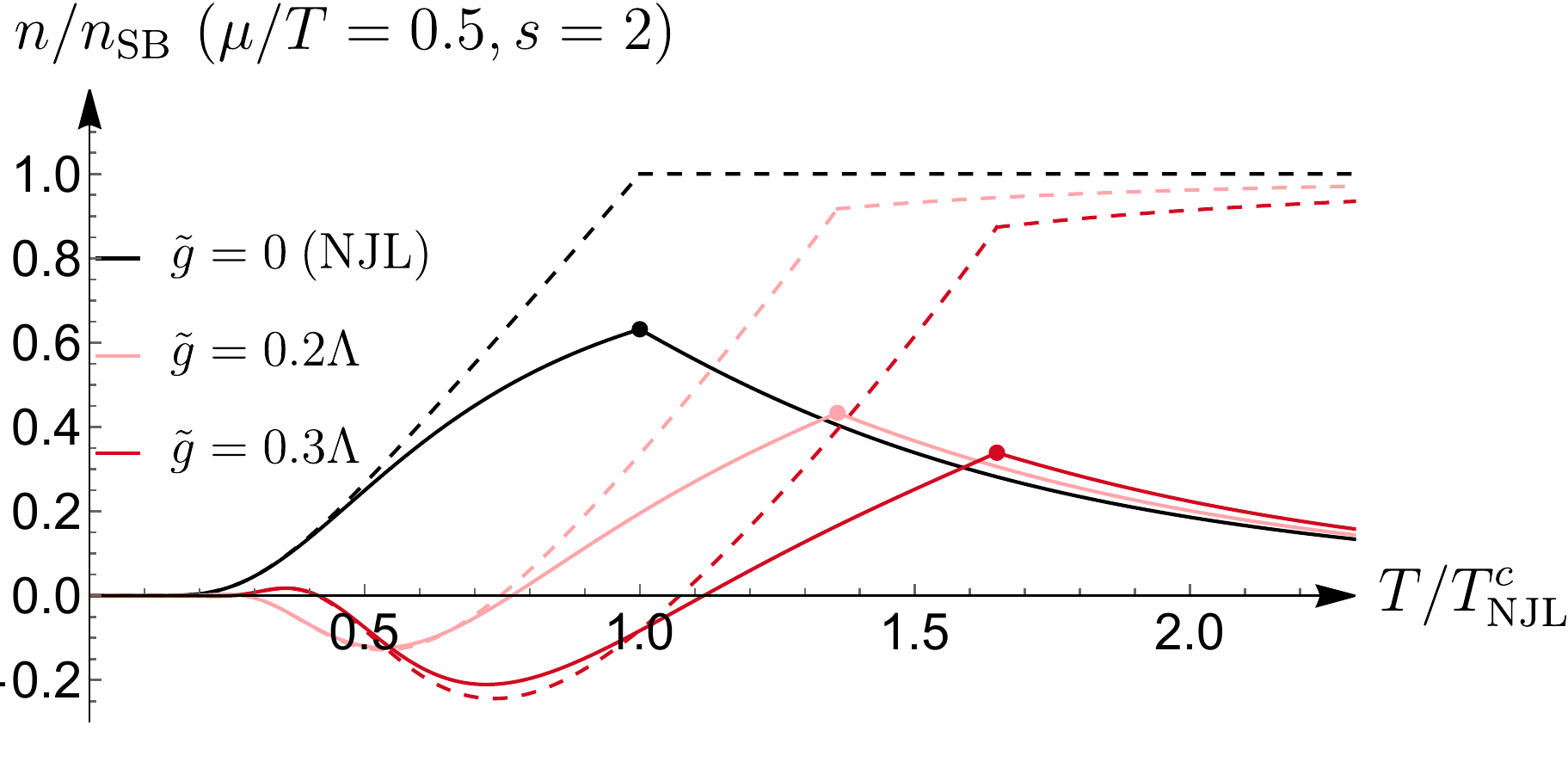}
\label{f16a}
}
\subfloat[]{
\includegraphics[width=0.48\textwidth]
{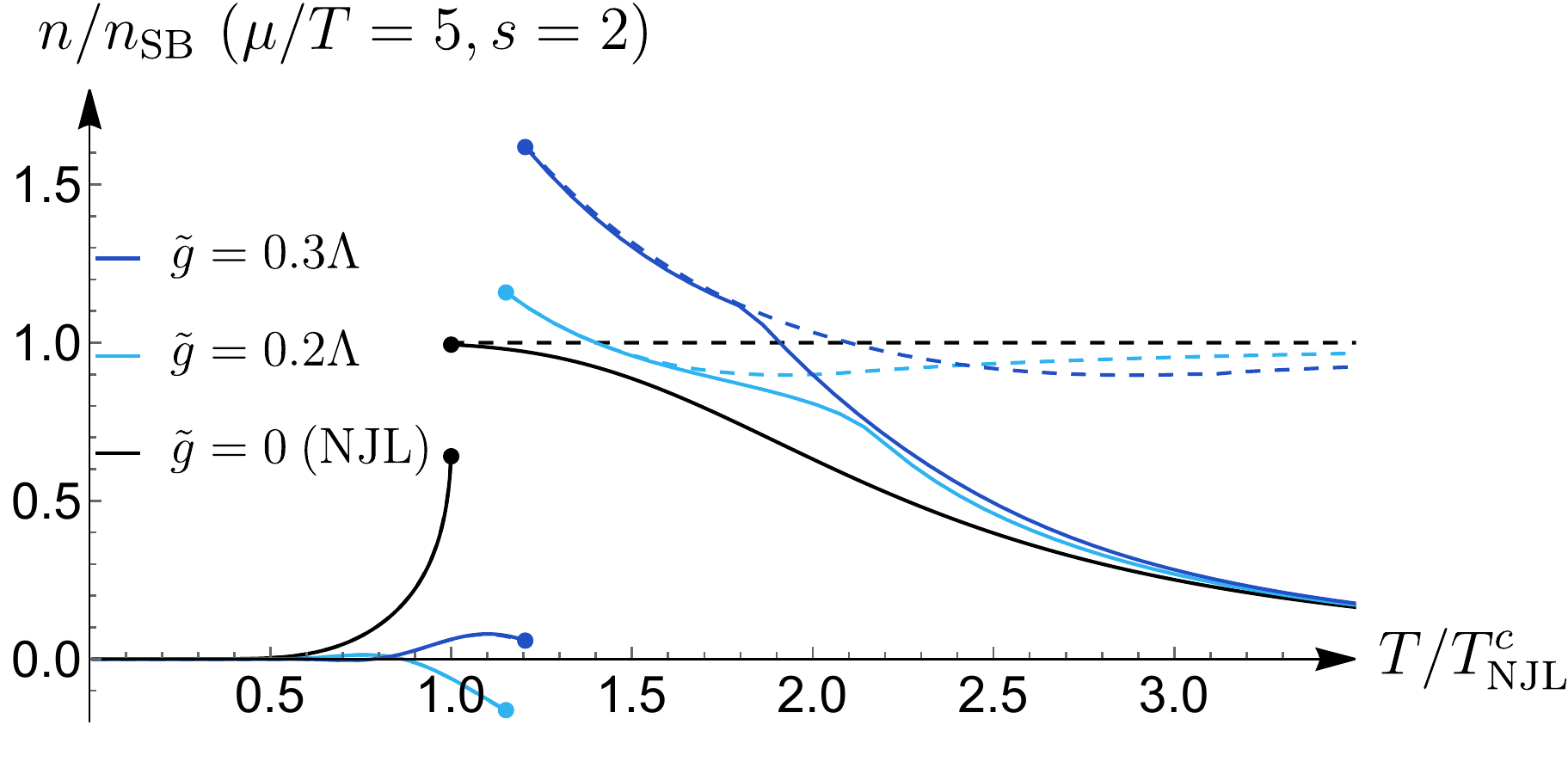}
\label{f16a2}
}
\hfill\\
\vspace*{-0.5cm}
\subfloat[]{
\includegraphics[width=0.48\textwidth]
{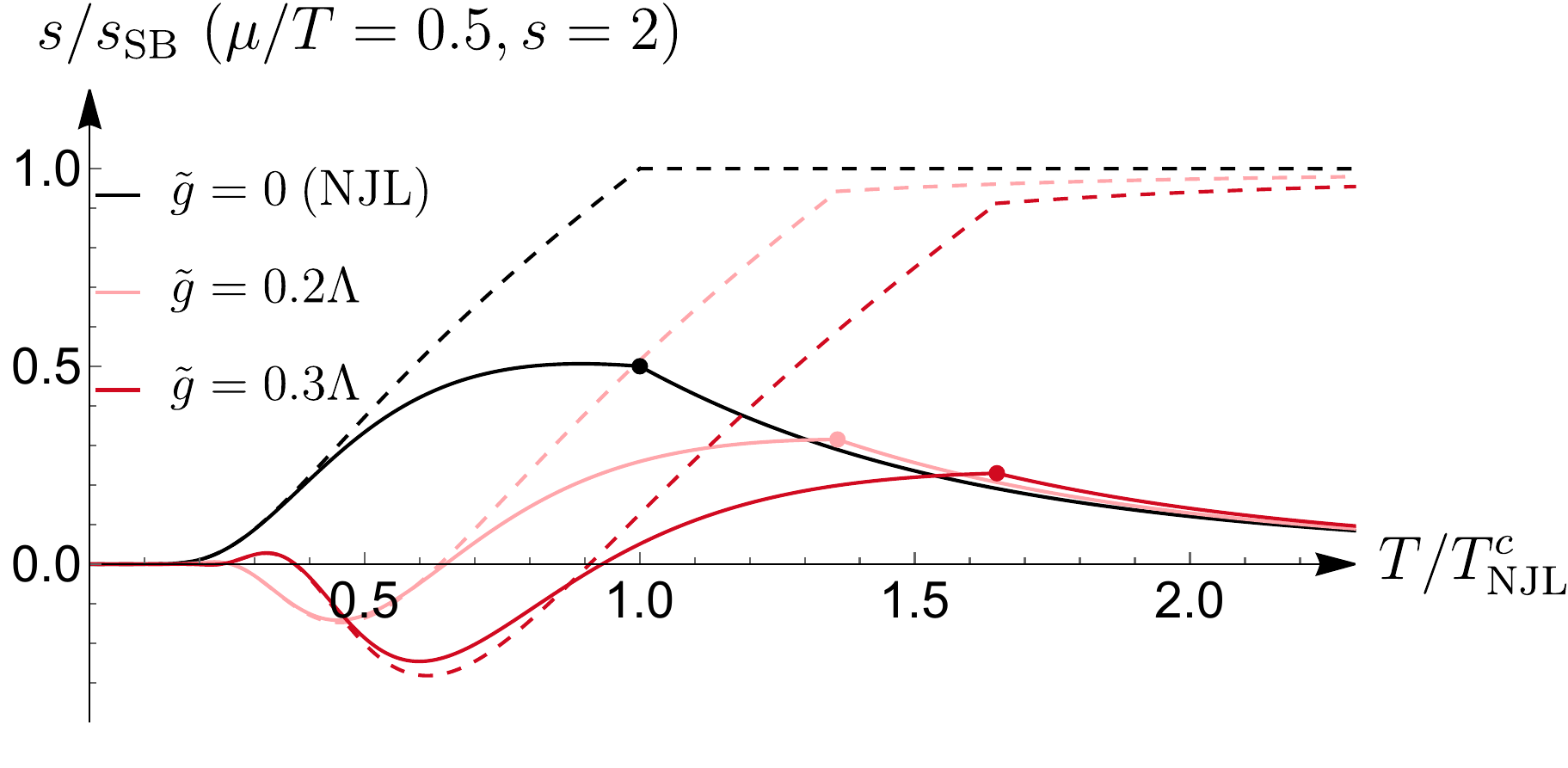}
\label{f16b}
}
\subfloat[]{
\includegraphics[width=0.48\textwidth]
{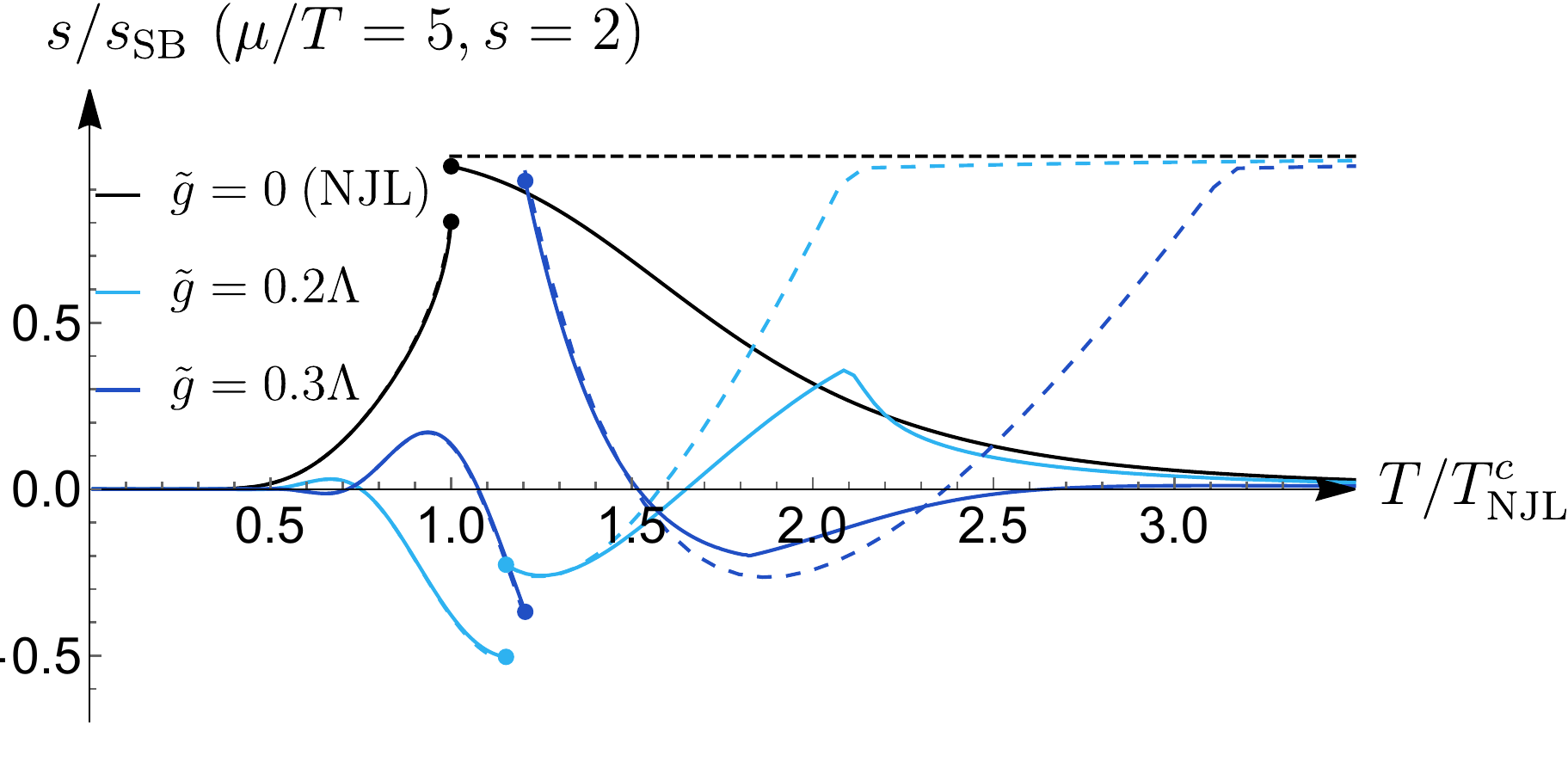}
\label{f16b2}
}
\hfill\\
\vspace*{-0.5cm}
\subfloat[]{
\includegraphics[width=0.48\textwidth]
{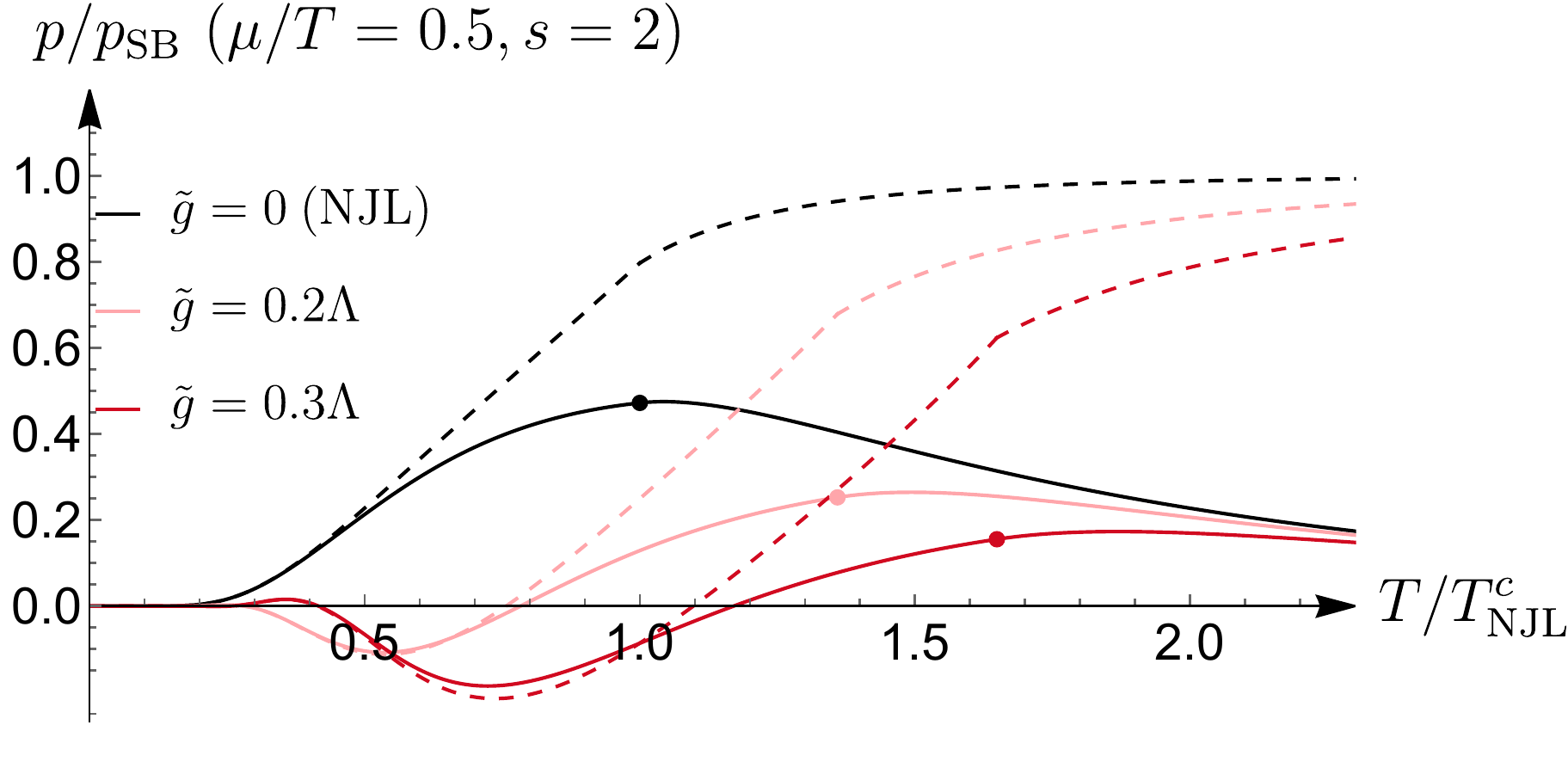}
\label{f16c}
}
\subfloat[]{
\includegraphics[width=0.48\textwidth]
{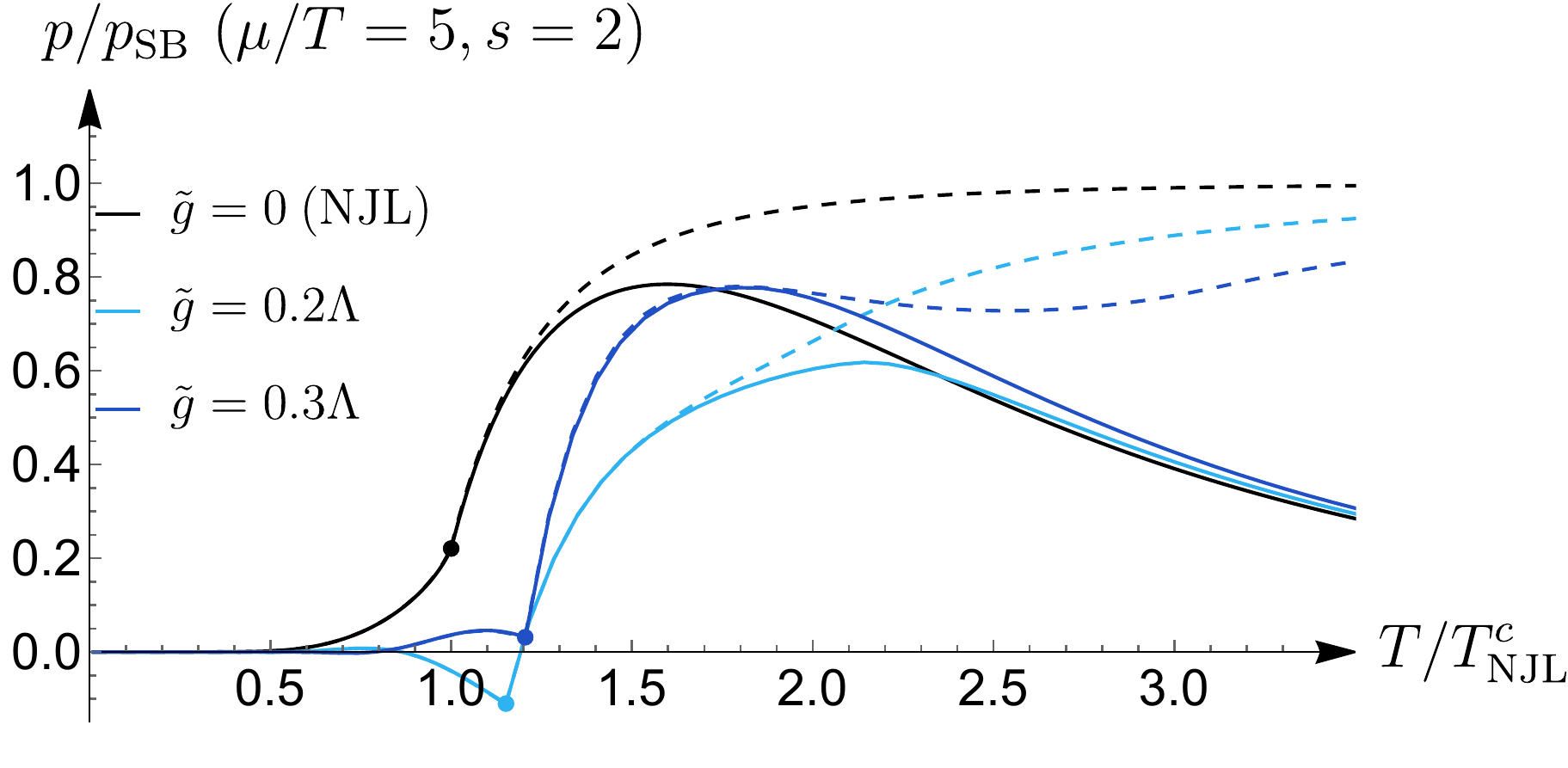}
\label{f16c2}
}
\hfill\\
\vspace*{-0.5cm}
\subfloat[]{
\includegraphics[width=0.48\textwidth]
{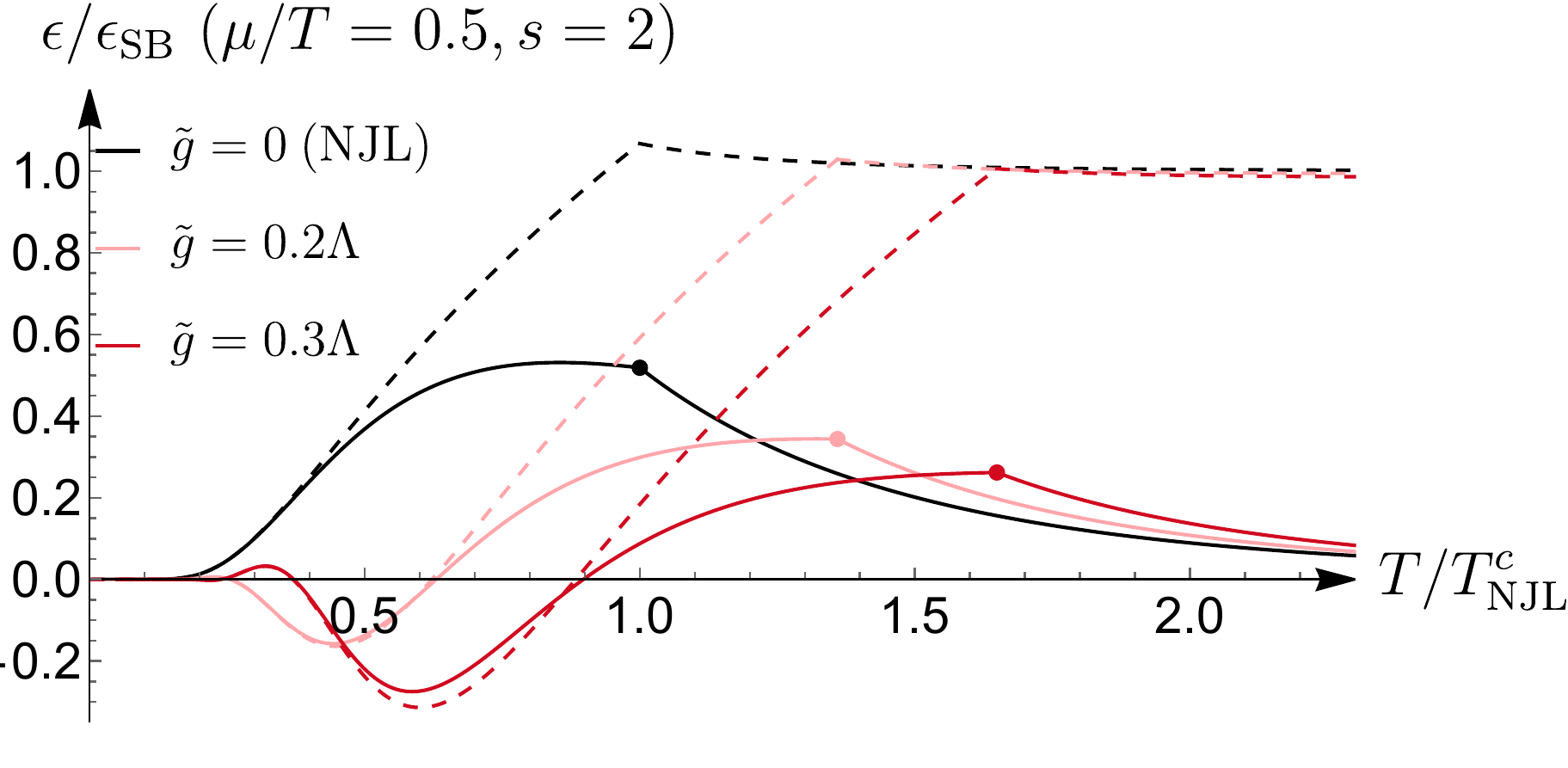}
\label{f16d}
}
\subfloat[]{
\includegraphics[width=0.48\textwidth]
{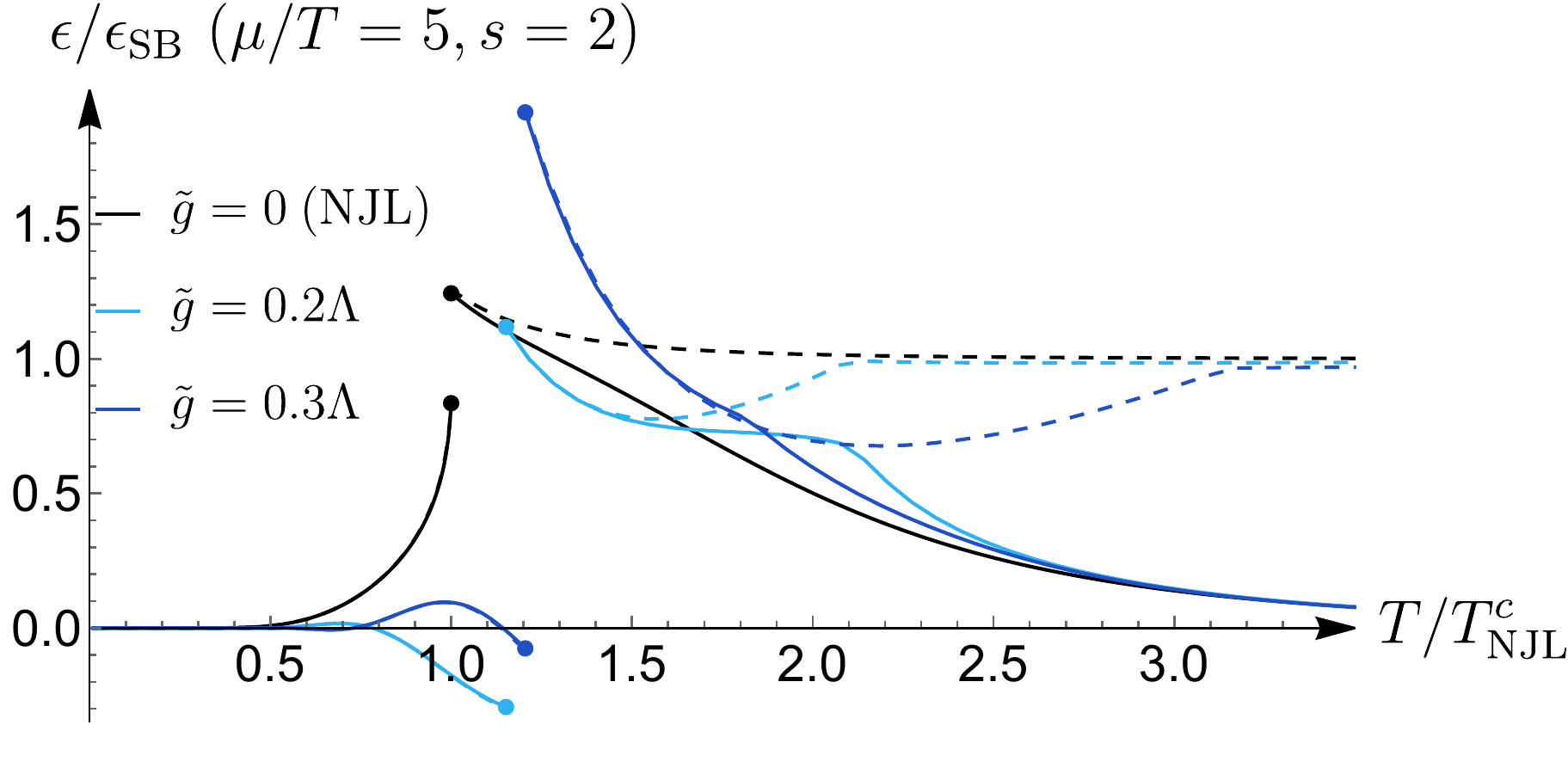}
\label{f16d2}
}
\hfill\\
\vspace*{-0.5cm}
\subfloat[]{
\includegraphics[width=0.48\textwidth]
{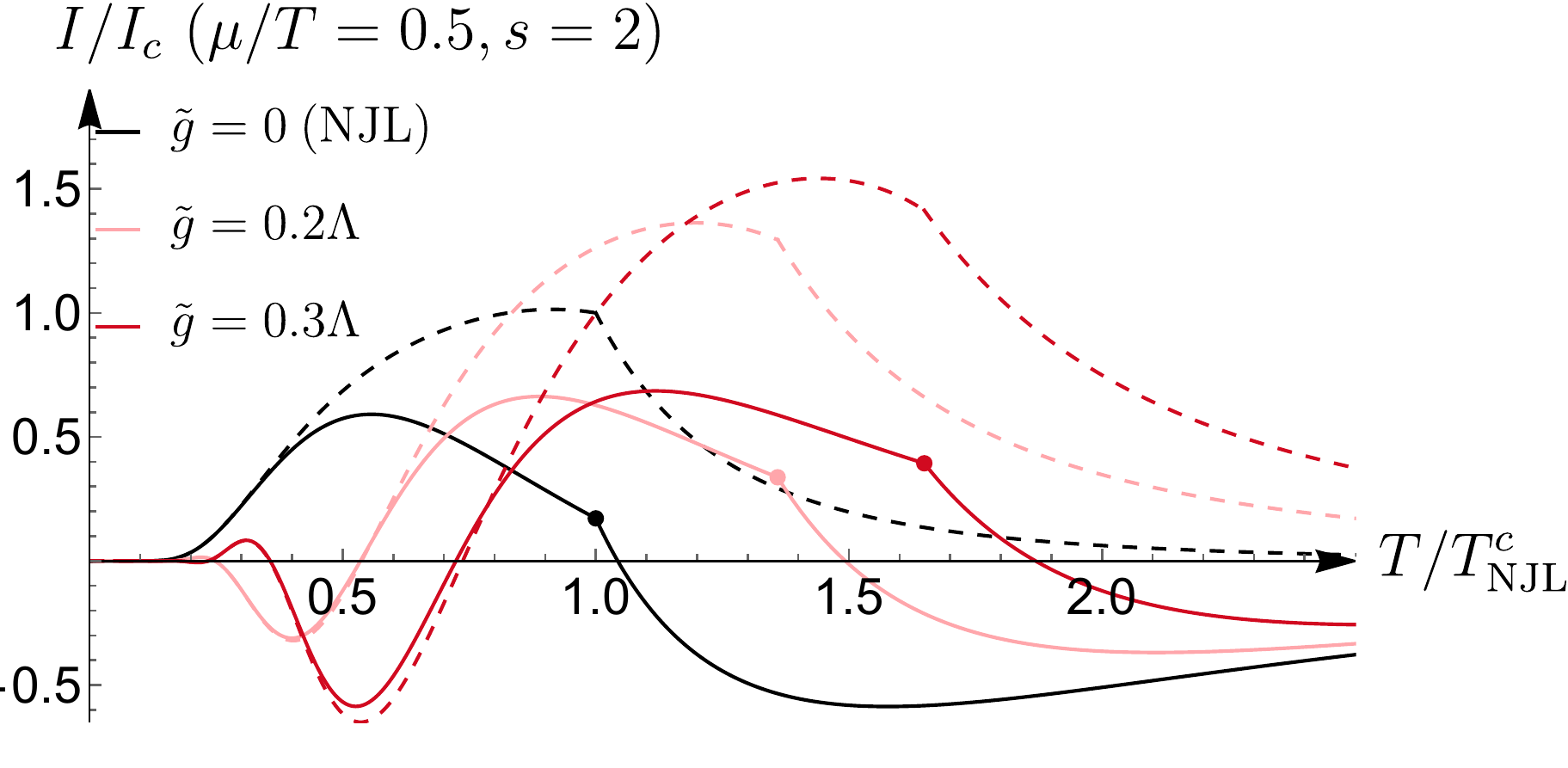}
\label{f16e}
}
\subfloat[]{
\includegraphics[width=0.48\textwidth]
{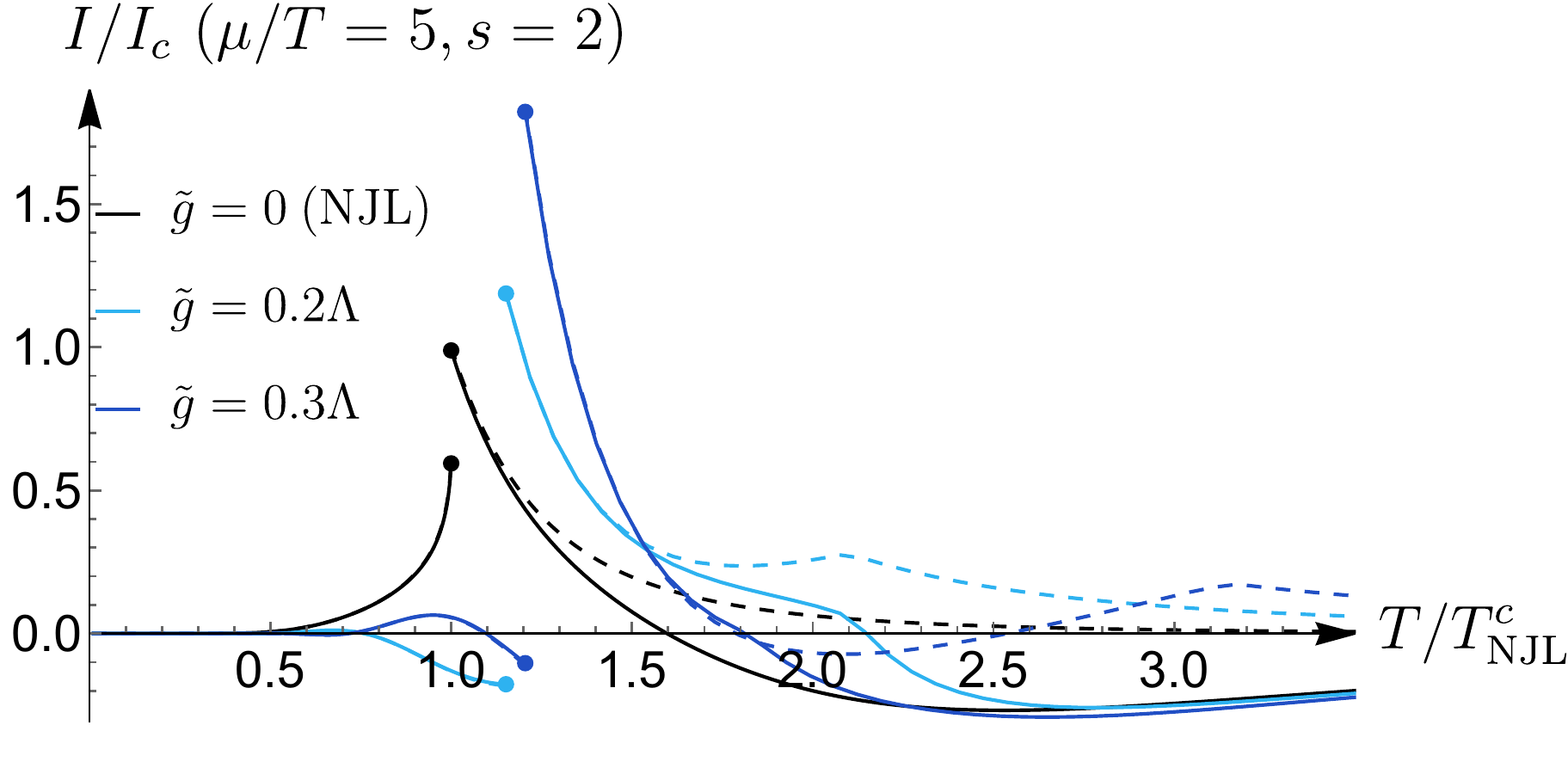}
\label{f16e2}
}
\caption{
\label{f16}
Thermodynamic functions for $s= 2$ (spacelikelike) along $\mu/T = 0.5$ (red) and $\mu/T = 5$ (blue) at couplings $\tilde{g} = 0.2\Lambda$ (light colors) and $\tilde{g}= 0.3 \Lambda$ (dark colors). Standard NJL case in black.
}
\end{figure*}

The effect of the $\cPT$ extension on the thermodynamic observables at finite values of the temperature and chemical potential is illustrated in Figs.~\ref{f15} and \ref{f16}. 
Shown is the behavior of the expressions (\ref{PT_quarknumber}) to (\ref{PT_energy}), scaled to their respective SB limit, as a function of the scaled temperature $T/T^c_\text{NJL}$. 
The timelike case with $s= 0.5$ is shown in Fig.~\ref{f15}, while Fig.~\ref{f16} shows the spacelike case with $s=2$.
In both cases the thermodynamic observables are evaluated along lines in the $T$-$\mu$--plane with constant ratio $\mu/T=0.5$ (red), where the phase transition remains of second order for all cases, and for the ratio $\mu/T=5$ (blue), 
for which the system undergoes a first-order phase transition.
Coupling values of $\tilde{g}= 0.2 \Lambda$ and $\tilde{g}= 0.3 \Lambda$ are shown as light and dark color variants respectively.
Solid lines denote the behavior with a fixed cutoff length $\Lambda$,  
while dashed lines show the behavior when the cutoff is removed.
A list of the critical temperatures $T^c(\tilde{g},s)$ for the illustrated cases can be found in Table~\ref{t4}.

For the second-order transition case with $\mu/T=0.5$, a \emph{decrease} of all thermodynamic functions with increasing coupling strength $\tilde{g}$ is observed compared to the corresponding standard NJL model observables throughout the spontaneously broken and the restored chiral symmetry phases for both a finite cutoff and for $\Lambda\rightarrow\infty$.
When considering for instance the behavior of the quark number density $n$ in Figs.~\ref{f15a} and \ref{f16a}, one notes that,
contrary to the monotonically increasing behavior toward the phase transition within the standard NJL model,
$n$ decreases to a minimum at \emph{negative} values within the non-Hermitian NJL model before subsequently increasing when approaching the phase transition at $T^c$.
This behavior remains present when removing the three-momentum cutoff, $\Lambda\rightarrow\infty$, and is therefore not a cutoff artifact.
In the case of a finite cutoff the quark number vanishes asymptotically beyond the phase transition, while it approaches the SB limit when the cutoff is removed. In this, the behavior of the modified system qualitatively agrees with that of the standard NJL model, but a notable deviation from the massless ideal fermion gas behavior remains present until the temperature well exceeds the transition value $T^c$.
For a finite cutoff $\Lambda$ the behavior in the spacelike case with $s=2$ differs from that in the timelike case with $s=0.5$ only in so far as that the decrease of the thermodynamic functions compared to the NJL model is more pronounced. 
For $\Lambda\rightarrow\infty$, on the other hand, the decrease of the thermodynamic functions is initially more pronounced at small temperatures, but in the vicinity of the phase transition the difference to the standard NJL model behavior becomes less prominent when $s=2$, than in the timelike case with $s=0.5$.
In addition, the position of the minimum within the spontaneously broken chiral symmetry region increases to higher temperatures in the spacelike case with $s=2$, but remains decidedly below the phase transition temperature $T^c$.
A comparable phenomenology is found for the entropy, pressure and energy density.

\begin{table*}[]
\centering
\setlength\tabcolsep{7.8pt}
\renewcommand{\arraystretch}{1.4}
\small     
\begin{tabular}{ |c||c|c|c|c|c| }  
 \cline{2-6}
 \multicolumn{1}{c||}{} & 
\shortstack{\\$\tilde{g}=0$ \\ (NJL)} & 
 \shortstack{$s = 0.5,$\\$\tilde{g} = 0.2\Lambda$} &  
 \shortstack{$s = 0.5,\,$\\$\tilde{g} = 0.3\Lambda$} & 
 \shortstack{$s = 2,\,$\\$\tilde{g} = 0.2\Lambda$} & 
 \shortstack{$s = 2,\,$\\$\tilde{g} = 0.3\Lambda$} 
 \\ 
 \hline
 \hline 
$\mu/T = 0.5$ &
$ 182 $~MeV & $ 185$~MeV & $ 189$~MeV& $ 247$~MeV & $ 300$~MeV
\\
\hline
\hline
   $\mu/T = 5$ &
   $59 $~MeV&  $ 66$~MeV & $ 42$~MeV&  $ 68$~MeV& $ 72$~MeV
 \\
\hline
\end{tabular}

\caption{
Phase transition temperature $T^c(\tilde{g},s)$ along lines of constant $\mu/T = 0.5$ (second-order transition region) and $\mu/T = 5$ (first-order transition region), cf. Fig.~\ref{f14}
}
\label{t4}
\end{table*}

In the first-order transition case with $\mu/T=5$ of the timelike ($s=0.5$) system for $\Lambda\rightarrow\infty$, shown as dashed blue lines in Fig.~\ref{f15},
one again observes a decrease of the thermodynamic functions relative to the standard NJL model. 
Similar to the second-order transition case, the quark number density $n$, see Fig.~\ref{f15a2}, admits an initial decrease to negative values.
A notable difference to the previous case is the fact that the range of temperatures in which a decrease toward a minimum is found, lies beyond the phase transition when the  coupling constant is sufficiently large:
For $\tilde{g}= 0.2 \Lambda$ such a minimum occurs in the spontaneously broken symmetry phase, with $n$ increasing monotonically thereafter when approaching the phase transition and following the behavior of $n_{\text{NJL}}$ qualitatively. 
For $\tilde{g}= 0.3 \Lambda$, however, $n$ decreases up to the phase transition, 
undergoing the characteristic discontinuous jump at $T^c$, and then continues to decrease toward a minimum within the restored chiral symmetry regime before asymptotically approaching the SB limit.
The behavior of the entropy, pressure, and energy density follows a comparable trend with respect to the standard NJL model behavior.
For a finite three-momentum cutoff $\Lambda$ an additional asymptotic decay at high temperatures is found, as in all models discussed prior.

In the spacelike case with $s=2$, see Fig.~\ref{f16}, the first-order transition behavior for $\mu/T=5$ at $\Lambda\rightarrow\infty$ also admits a region, in which the thermodynamic functions decrease toward a minimum, similar to the timelike case.
But these regions here occur at even higher temperatures, deep within the  restored  chiral symmetry region. 
With an increase of the coupling constant $\tilde{g}$, the minimum again shifts toward even higher temperatures.
As such, the decrease of the thermodynamic functions due to the non-Hermitian extension within the spontaneously broken symmetry region at low temperatures is found to be less pronounced for 
$\tilde{g}= 0.3 \Lambda$ than for $\tilde{g}= 0.2 \Lambda$.
Another notable difference to the timelike case with $s=0.5$ is the increased jump at the discontinuous phase transition. 
Due to this increase the quark number density $n$ and the energy density $\epsilon$ even increase in the restored symmetry region beyond $T^c$ when being close to the phase transition, exceeding the SB limit.
Nevertheless this is followed by a rapid decrease toward the aforementioned minimum at high temperatures and a successive asymptotic approach of the SB limit.
A finite three-momentum cutoff $\Lambda$ introduces an additional asymptotic decay toward a vanishing limit at high temperatures instead.

The effects of the non-Hermitian but $\cPT\!$-symmetric pseudovector extension $igB_\nu \,\bar{\psi}\gamma_5\gamma^\nu \psi$ can be interpreted by considering the particle and antiparticle contributions to the 
fermion wavefunction $\psi$.
Due to the structure of the Dirac matrices (\ref{iic1s0e4}), the $B_0$ component of the bilinear extension introduces a mixing between fermionic and antifermionic contributions, while the $B_k$, $k\in[1,3]$ components do not.
Instead they modify the system comparable to a mass term, resulting in either an increase ($s=2$) or a decrease ($s=0.5$) of the effective fermion mass, cf. Fig.~\ref{f13}.
In direct contrast to the pseudoscalar extension discussed in section~\ref{s3}, a decrease of the quark number density, $n = n_{q} - n_{\bar{q}}$, compared to the standard NJL model and the occurrence of \emph{negative} values of $n$ describes an emphasis on the antifermionic component within the pseudovector-extended theory, rather than the fermion excess found in the pseudoscalar-extended system.
One nevertheless observes an increase toward the same SB limit of an ideal massless fermion gas as in the pseudoscalar extension case and the standard NJL model at high temperatures, because of the temperature-suppressed influence of the extension term.

Figures~\ref{f15e},\ref{f15e2}, \ref{f16e}, and \ref{f16e2} show the behavior of the interaction measure (\ref{PT_anomaly}), scaled to the value $I_c$ at the phase transition of the NJL($\infty$) case since its high temperature limit vanishes.
As in all previous cases, the pressure and the energy density are affected notably by the momentum cutoff $\Lambda$,
so that the behavior of $I/I_c$ when accounting for a finite cutoff (solid lines) has to be considered largely artificial. 
For $\Lambda\rightarrow\infty$ (dashed lines) the interaction measure shows two competing trends in both the second-order ($\mu/T=0.5$) and the first-order ($\mu/T=5$) transition region and in both the spacelike ($s=0.5$) and timelike ($s=2$) case:
A localized decrease toward a minimum in accordance with the 
corresponding behavior within the other thermodynamic observables, in particular the quark number density;
and an overall increase of the interaction measure compared to the standard NJL model behavior, is found throughout all temperatures.
As before, the localized decrease toward a minimum arises at higher scaled temperatures for larger coupling constant values $\tilde{g}$, for the first-order transition region with $\mu/T= 5$ compared to the second-order region with $\mu/T= 0.5$, and in the spacelike case with $s=2$ compared to the timelike case with $s=0.5$.
Notably, the interaction measure becomes negative within this region, marking a notable change in behavior of the modified non-Hermitian system compared to the 
standard NJL model.
Contrary to the pseudoscalar extension discussed in section~\ref{s3}, 
this deviation from the standard NJL model arises typically at comparatively low temperatures and within the spontaneously broken chiral symmetry regime.

Overall, the inclusion of the non-Hermitian, but $\cPT\!$-symmetric and chiral symmetry preserving, pseudovector bilinear term $igB_\nu \,\bar{\psi}\gamma_5\gamma^\nu \psi$ presents an intriguing complement to the non-Hermitian pseudoscalar extension.
The dynamical generation of effective fermion mass within the spontaneously 
broken chiral symmetry region, previously described for a spacelike background 
$B^\nu$ at vanishing $T$ and $\mu$ \cite{fbk20,fk21}, remains a prominent and robust feature at finite values of the temperature and chemical potential. 
However, this property does depend on the space- or timelikeness of the background; an effective mass loss can be found for a timelike $B^\nu$ instead.
The extent of the spontaneously broken chiral symmetry regime and the position of the chiral phase transition within the $T$-$\mu$--plane are affected by the extension term as well, ranging toward higher temperatures for low chemical potential values, i.e., in the second-order transition region, similar to the effects of the pseudoscalar extension. 
At low temperatures (second-order transition region), the position of the phase transition decreases toward lower chemical potential values in the time- and lightlike cases of the background field - again similar to the $g \bar{\psi}\gamma_5 \psi$ modification. But for a sufficiently spacelike background it increased to higher values of $\mu^c$ instead.
A notable departure from the behavior of both the standard NJL model and the pseudoscalar extension of the system is found in all cases within the quark number, entropy, pressure, and energy densities, displaying a marked decrease compared to the standard NJL model behavior and even extending to negative values. Instead of the fermion excess within the pseudoscalar extension of the system, the $\cPT\!$-symmetric pseudovector modification shows an emphasis on the antifermionic component within the theory.

\section{Concluding remarks}
\label{s5}

Due to the presence of real effective fermion masses in non-Hermitian extensions of the NJL model at vanishing temperature and density \cite{fbk20, fk21}, there seems to be no reason for discarding such systems. 
In this study we have generalized the established finite temperature and chemical potential approach of the NJL model to investigate the effects of the non-Hermitian bilinear extension terms $g\bar{\psi}\gamma_5 \psi$ and $igB_\nu \,\bar{\psi}\gamma_5\gamma^\nu \psi$ on the thermodynamic behavior of the system in search for general characteristic signals of non-Hermitian fermionic quantum field theories. 

In both extensions of the NJL model a dynamical generation of effective fermion mass due to the non-Hermitian contribution can be observed in the spontaneously broken (approximate) chirally symmetric regime; in the case of the pseudovector modification, however, this property depends on the characteristics of the background field $B^\nu$, resulting in an effective mass loss for sufficiently timelike cases instead.
The position of the chiral phase transition in the $T$-$\mu$--plane moves to higher temperatures at small fixed chemical potentials, that is in the second-order transition region, in both non-Hermitian systems.
In the first-order transition region at small fixed temperatures, on the other hand, the transition chemical potential decreases for increasing coupling strength of a non-Hermitian pseudoscalar bilinear. 
In the pseudovector modified model the change of the transition chemical potential depends on the characteristics of the background field again, increasing for sufficiently spacelike cases, but decreasing otherwise.
The position of the critical end-point marking the boundary between first- and second-order chiral phase transitions moves toward higher critical temperatures $T_\text{CEP}$ in both modified NJL models; for the pseudoscalar extension the critical chemical potential $\mu_\text{CEP}$ decreases, while it increases for the inclusion of a pseudovector bilinear term.

Further deviations from the standard NJL model become apparent in the behavior of the quark number, entropy, pressure, and energy densities.
When the system is extended through the inclusion of the term $g\bar{\psi}\gamma_5 \psi$, these thermodynamic observables remain initially unchanged compared to the standard NJL model behavior in the spontaneously broken approximate chiral symmetry region at low temperature and chemical potential despite the dynamical fermion mass generation. But in the vicinity of the phase transition and throughout the restored symmetry phase a notable fermion excess arises, increasing beyond the high-temperature SB limit.
This is contrasted by the behavior of the $\cPT\!$-symmetric pseudovector modification, where the thermodynamic observables decrease due to the extension term, reaching even negative values. This non-Hermitian extension reflects an emphasis on the antifermionic component of the theory instead. 
These trends may provide a first indication of curious potential mechanisms for producing non-Hermitian baryon asymmetry.

Moreover, negative values of the interaction measure $I = \epsilon-3p$ are found in both non-Hermitian extensions of the NJL model, arising within the restored approximate chiral symmetry region for the pseudoscalar bilinear, but in the spontaneously broken symmetry region and the vicinity of the phase transition for the pseudovector term.
This feature builds an interesting connection to recent discussions of the constraints of neutron star masses and extended theories of general relativity.


\end{document}